\documentclass[twocolumn,tighten]{aastex63}

\newcommand\beq{\begin{equation}}
\newcommand\eeq{\end{equation}}
\newcommand\beqn{\begin{eqnarray}}
\newcommand\eeqn{\end{eqnarray}}

\newcommand\Msol{$\mathrm{M}_{\odot}$}
\newcommand\kms{km s$^{-1}$}

\newcommand\slos{$\sigma_{los}$}
\newcommand\stwo{$\sigma_{200}$}

\newcommand\sloanr{$r^\prime$}
\newcommand\sloani{$i^\prime$}
\newcommand\sloang{$g^\prime$}

\newcommand\sloangr{$(g^{\prime} - r^{\prime})$}

\newcommand{\hval}{$\mathrm{h}_{70}^{-1}$}
\newcommand{\hvaltwo}{$\mathrm{h}_{70}^{-2}$}
\newcommand{\masstwo}{$\mathrm{M}_{200}$}

\newcommand{\rvirial}{$\mathrm{r}_{vir}$}
\newcommand{\rtwo}{$\mathrm{r}_{200}$}
\newcommand{\oii}{{\rm [O\,}{{\sc ii}]}}
\newcommand{\oiii}{{\rm [O\,}{{\sc iii}]}}

\newcommand{\hh}{$^{h}$}
\newcommand{\hm}{$^{m}$}
\defcitealias{verdugo2020}{Paper\,I}
\defcitealias{gioia1998}{G98}

\accepted{June 14, 2021}

\submitjournal{ApJ}

\shorttitle{Dissecting the Strong Lensing Galaxy Cluster MS\,0440.5$+$0204}
\shortauthors{Carrasco et al.}
\begin{document}

\title{DISSECTING THE STRONG LENSING GALAXY CLUSTER MS\,0440$+$0204\\ 
II. NEW OPTICAL SPECTROSCOPIC OBSERVATIONS IN A WIDER AREA AND CLUSTER DYNAMICAL STATE\\}

\correspondingauthor{E. R. Carrasco}
\email{rodrigo.carrasco@noirlab.edu}

\author[0000-0002-7272-9234]{Eleazar R. Carrasco}
\affil{Gemini Observatory/NSF’s NOIRLab, Casilla 603, La Serena, Chile}

\author[0000-0003-4062-6123]{Tom\'as Verdugo}
\affiliation{Observatorio Astron\'omico Nacional, Instituto de Astronom\'ia, Universidad Nacional Aut\'onoma de M\'exico, Ensenada, B.C., M\'exico}

\author[0000-0003-4446-7465]{Ver\'onica Motta}

\author[0000-0001-9271-4155]{Gael Fo{\"e}x}
\affiliation{Instituto de F\'{\i}sica y Astronom\'ia, Facultad de Ciencias, Universidad de Valpara\'iso, Avda. Gran Breta\~na 1111, Valpara\'iso, Chile.}

\author[0000-0002-6481-7162]{E. Ellingson}
\affiliation{Center for Astrophysics and Space Astronomy, 389 UCB, University of Colorado, Boulder, CO 80309-0389, USA}

\author[0000-0003-0408-9850]{Percy L. Gomez}
\affiliation{W. M. Keck Observatory, 65-1120 Mamalahoa Highway, Kamuela, HI 96743, USA.}

\author[0000-0002-7061-6519]{Emilio Falco}
\affiliation{Harvard-Smithsonian Center for Astrophysics, 60 Garden St., Cambridge, MA 02138, USA.}

\author[0000-0001-6636-4999]{Marceau Limousin}
\affiliation{Aix Marseille Univ, CNRS, CNES, LAM, F-13013 Marseille, France.}




\begin{abstract}
We present an optical study of the strong lensing galaxy cluster \object{MS\,0440.5$+$0204} at $z=0.19593$, based on CFHT/MegaCam \sloang, \sloanr-photometry and GMOS/Gemini and CFHT/MOS/SIS spectroscopy in a broader area compared to previous works. We have determined new spectroscopic redshifts for the most prominent gravitational arcs surrounding the central galaxy in the cluster. The new redshifts and the information provided by the photometric catalog yield us to perform a detailed weak and strong lensing mass reconstruction of the cluster. The large number of member galaxies and the area covered by our observations allow to estimate more accurately the velocity dispersion and mass of cluster and examine in detail the nature of the cluster and surroundings structures. The dynamical mass is in good agreement with the mass inferred from the lensing analysis and X-ray estimates. About $\sim$68\% of the galaxies are located in the inner $\lesssim$0.86 \hval\,Mpc region of the cluster. The galaxy redshift distribution in the inner region of the cluster shows a complex structure with at least three sub-structures along the line-of-sight. Other sub-structures are also identified in the galaxy density map and in the weak lensing mass map. The member galaxies in the North-East overdensity are distributed in a filament between \object{MS\,0440.5$+$0204} and \object{ZwCL\,0441.1$+$0211} clusters, suggesting that these two structures might be connected. \object{MS\,0440$+$0204} appears to be dynamically active, with a cluster core that is likely experiencing a merging process and with other nearby groups at projected distances of $\lesssim$1 \hval\,Mpc ~that could be being accreted by the cluster.\end{abstract}


\section{Introduction}\label{sec:intro}

Galaxy clusters are the largest gravitationally bound systems in the Universe.
They serve as unique laboratories to study the large-scale structure formation
and galaxy evolution. Clusters are excellent locations to probe the mass
distribution, of baryons and dark matter (DM), using a variety of
multi-wavelength data (UV/optical, near-infrared, X-ray) and over a wide range
of scales. Moreover, in the UV and optical wavelength regimes, gravitational
lensing became an important tool for measuring the mass distribution at
different scales without any assumption of the  dynamical state and the nature 
of the intervening matter. In the last decade, improved lensing models
\citep[e.g.,][]{lim2007, zitrin2013, new13, diego2015,Jauzac2015} allowed to
better understand the mass profile on scale from $\sim 10$ \hval~kpc
(strong-lensing regime) up to $\sim 5$ \hval~Mpc (weak-lensing regime).
Indeed, it has been evident the necessity to combine complementary probes of 
the gravitational potential in order to recover in a better way the mass density
profiles \citep[e.g.,][]{Limousin2013,new13,ver2011,Monna2015,verdugo2016}, 
in particular to constrain the mass in the external regions of the clusters, 
since the strong lensing is accurate only in the inner regions 
\citep[see ][for a discussion]{verdugo2016}.

An interesting new approach to measure the mass profile in galaxy clusters at
scales between SL and WL regimes is to look at ``regular'' galaxy clusters, i.e.
clusters where no obvious sub-structures are detected, and for which a single DM
halo modeling can be applied. In a recent works, \citet{ver2011} and \citet{verdugo2016}
show that it is possible to obtain a complete understanding of the mass profile 
at the these scales by combining SL modeling with dynamical measurements. Although this
method constrains large-scale properties inaccessible to SL models, this method
requires observational constraints and assumptions about the dynamical state of
the system. This approach requires detailed spectroscopic analysis of the
clusters based on the measurement of the redshift for a large number of galaxies
in the cluster and its environment. From the observational point of view, this
could be challenging, because requires access to a large amount of telescope
time. With the advent of new observational techniques that allow 
to obtain spectroscopic and photometric redshift of hundred of galaxies in 
large areas of the sky,  systematic determination of galaxy redshifts in clusters 
only recently began \citep[e.g.,][]{rines2013, geller2014, balestra2016, miniJPAS2020}.

\begin{figure*}[t!]
\centering
\includegraphics[width=0.90\hsize]{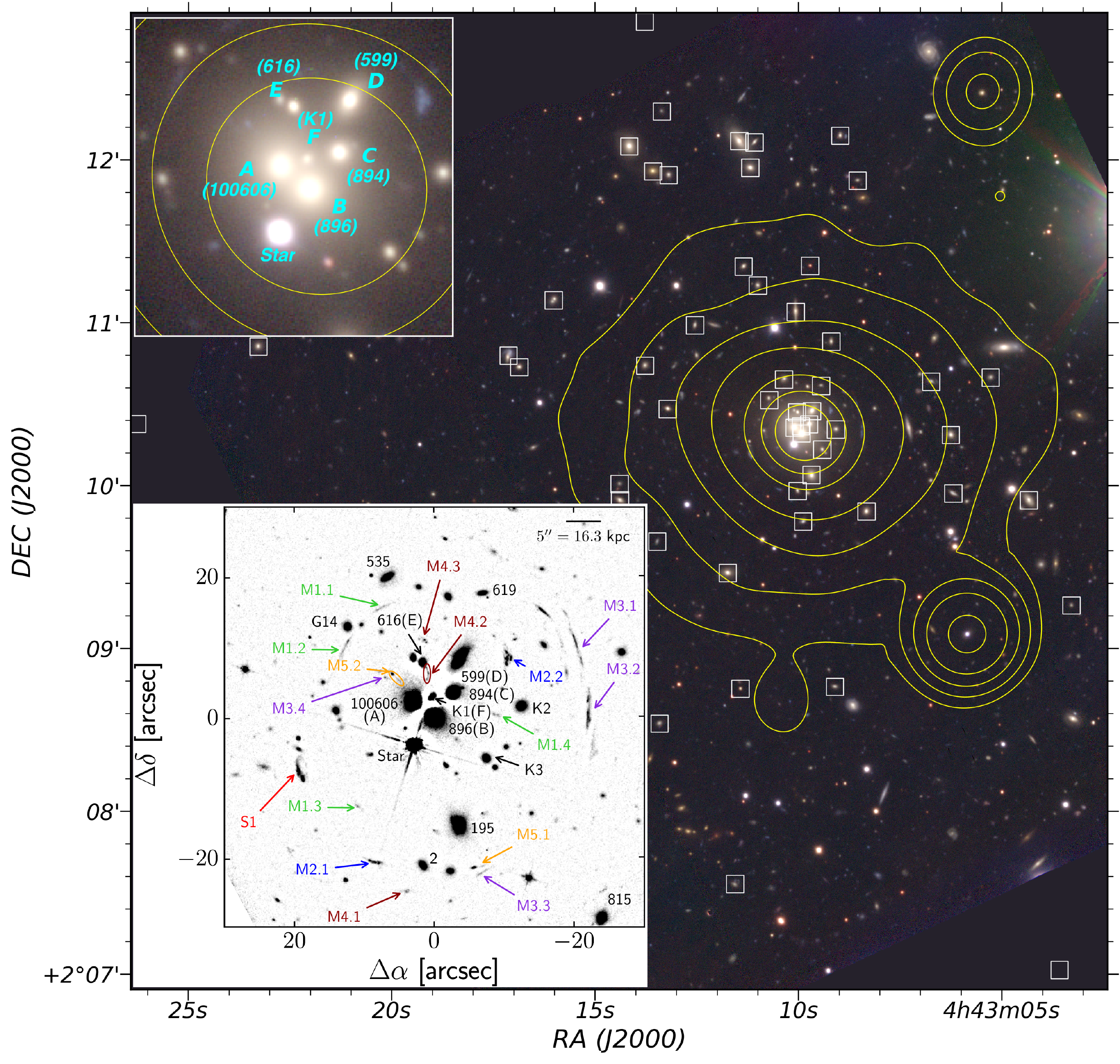} \label{fig:colorfig}
\caption{Color \sloang, \sloanr, \sloani~composite image of the central $\sim
5\arcmin\!\times\!5\arcmin$ ($1\,\times\,1$ h$^{-2}_{70}$ Mpc$^{2}$ at 
$z=0.196$) region of the MS0440.5$+$0204 cluster observed with GMOS at Gemini
South. The location of cluster member galaxies are depicted (small white
squares). The solid lines shows the iso-contours derived from the adaptively
smoothed Chandra X-ray image (0.2-6.0 KeV energy). The top-left inset
shows the central 30\arcsec $\times$ 30\arcsec~ region of the cluster 
($\sim 100$ \hval~kpc $\times$ $\sim 100$ \hval~kpc) with  
the six distinct nuclei embedded in a common symmetric envelope. 
The bottom-left inset shows the HST/WFPC2/F702W local
median average subtracted image of the central 60\arcsec $\times$
60\arcsec~region of the cluster. Member galaxies are identified with a numbers
and letters (section \ref{sec:galred} and Table \ref{tab:galcat}), along with
the different gravitational arc systems and sub-components, including the radial
arcs M4.2 and M5.2 (see Table \ref{tab:tabarcs}). North is up, and East is
left.}
\end{figure*}

To extend the method presented in \citet{ver2011} and \citet{verdugo2016}, we test 
analogous and improved study to the strong lensing galaxy cluster 
MS\,0440$+$0204 located at $z \sim 0.19$. \object{MS\,0440.5$+$0204} is a relatively massive 
high X-ray luminosity galaxy cluster \citep[$L_{X} \sim 6 \times 10^{44}$ erg s$^{-1}$, ][]{hicks2006}. 
It was discovered in X-ray in the Extended Medium  Sensitivity Survey by \citet{gioia1990}. 
This system  is a good example of so called a ``regular'' cluster, where no obvious 
sub-structures are presented. This assumption is based on a previous X-ray and kinematic
analysis of the cluster by \citet[][hereafter \citetalias{gioia1998}]{gioia1998}
based only in 40 confirmed members. In its center, the cluster is dominated
by a multi nucleus cD galaxy \citep[hereafter the BCG,][]{luppino1993}, with six 
distinct nuclei merging within a common symmetric envelope of $\sim 48$\arcsec ($\sim 157$ 
\hval~kpc) in diameter. The BCG in \object{MS\,0440.5$+$0204} is quite similar to the
central galaxy in Abell 3827, a super-giant elliptical that appears to be in the process 
of cannibalizing at least four other galaxies \citep[][and references therein]{carrasco2010}.
Surrounding the BCG are at least five gravitational lensed arc systems and two radial arcs 
\citep[][\citetalias{gioia1998}]{luppino1993}, corresponding to lensed sources 
located at different redshifts. The different gravitational
arcs systems and its sub-components, within the central 1\arcmin $\times$
1\arcmin~region of cluster are shown in Figure \ref{fig:colorfig}
(bottom-left inset). The mass profile of this cluster  has been studied previously from the 
ground and space by combining weak lensing and X-ray analysis
\citep[\citepalias{gioia1998}][]{hicks2006,hoekstra2012,mahdavi2013,hoekstra2015}. 
Despite the extensive multi-wavelength studies presented in the literature, the number of
galaxies with confirmed redshift is small. The same is extended to the lensed
arcs. Only one gravitational arc system (S1) has its redshift confirmed via
spectroscopic observations (see Fig. \ref{fig:colorfig}). Therefore, no detailed
SL analysis of this cluster exist except for an early attempt by \citetalias{gioia1998},
who used different lensing models to estimate the redshifts of the lensed sources 
based on HST images.  To provide a reliable measurement of the mass profile at ``medium'' 
scales, it is imperative to determine, with a good precision, the spectroscopic redshift of the 
lensed arc systems and the galaxies belonging to the cluster in a wider area.


This is the second paper in a series of two,  with the aim of put forward a 
detailed dynamical measurements of the galaxy cluster \object{MS\,0440.5$+$0204}
based on analysis of new redshift obtained for a large number of galaxies ($\sim 100$) 
up to $\sim 2 \times$ \rtwo~on the sky. The large number of member galaxies 
allow us to study the presence of sub-structures and confirm (or reject) the 
relaxed nature of the cluster. The results allow also to characterize the mass 
profile from the SL region of the cluster up to medium scales. A detailed 
explanation of the lensing modeling of \object{MS\,0440.5$+$0204}, and 
the comparison with measured masses reported previously in the literature 
are presented in \citet[][hereafter \citetalias{verdugo2020}]{verdugo2020}.

The paper is organized as follow. Section \S \ref{sec:obsred} provides details
about the observations, the data reduction and the redshift estimation of the
lensed sources and the galaxies. In section \S \ref{sec:lensmod} we provide a
brief explanation of the lensing models and highlight the main result obtained
from the lensing analysis presented in \citetalias{verdugo2020}. In \S
\ref{sec:analysis} we present the dynamical analysis of the cluster and discuss
our finding.  Finally, in section \S \ref{sec:summary}, we summarize our results
and present our conclusions.

Throughout this work we adopt the following  cosmological parameters: 
$\Omega_{M} = 0.3$, $\Omega_{\Lambda} = 0.7$ and  $H_{0}=70$ $h_{70}$ 
km s$^{-1}$ Mpc$^{-1}$. With this cosmology 1\arcsec~corresponds to 3.24 
\hval~kpc at the redshift of the cluster  ($z=0.196$). All magnitudes presented 
in this paper are quoted in the AB-system.

\section{Observations}\label{sec:obsred}

\subsection{Gemini data}\label{sec:gemdata}

The images and spectroscopic data were obtained with the Gemini Multi-Object
Spectrograph \citep[hereinafter GMOS, ][]{hook2004} mounted at the Gemini South
telescope in Chile, in queue mode.

The central 5\farcm5 $\times$ 5\farcm5 region of \object{MS0440.5$+$0204} galaxy
cluster was  observed with the \sloang~($3 \times300$ second exposures),
\sloanr~($3 \times200$ second exposures) and \sloani~filters ($3 \times200$
seconds exposures) on 2011 November 12 UT (program ID: GS$-$2011B$-$Q$-$59),
during dark time, in photometric conditions, and with median seeing values of
0\farcs51, 0\farcs50 and 0\farcs57  in \sloang, \sloanr~and \sloani~filters,
respectively. All images were processed with the reduction package {\tt THELI}
\citep{esd05,schirmer13}. The science frames were bias/overscan-subtracted,
trimmed and flat-fielded. The resulting processed images were registered to a
common pixel and sky coordinate positions using the program {\tt SCAMP}
\citep{ber06} called from {\tt THELI}. A common astrometric solution for the
three filters was derived using an external catalog  generated from archival
Canada-France-Hawaii Telescope (CFHT) images observed with MegaCam (see 
section \S \ref{sec:cfhtdata} for details). The astrometric calibrated images were 
sky subtracted using constant values, re-sampled to a common position and then
stacked by filter using the program {\tt SWARP} \citep{ber2010a} called from
{\tt THELI}. The final co-added images were normalized to 1-second exposure.
Although the GMOS images were obtained under photometric conditions, no
photometric standard stars were observed during the night. Thus, to calibrate
the photometry, we used the average magnitude zero points listed in GMOS public
web page\footnote{http://www.gemini.edu/node/10445} of 28.15, 28.23 and 27.93 mag
for the \sloang, \sloanr\ and \sloani~ filters, respectively.

We used program {\tt Source Extractor} version 2.19.5 \citep{ber96} and the GMOS
image in the \sloanr~ filter for object detection and photometry (our primary
filter). The photometric parameters in the \sloang~ and \sloani~filters were
extracted for those objects detected in the \sloanr~image. The MAG$\_$AUTO
parameter was adopted as the value for the total magnitude of the objects.
Colors  were derived by measuring fluxes inside a fixed circular aperture of
1\farcs2 in diameter. The star-galaxy separation was done using the
{\tt CLASS\_STAR} and {\tt FLUX\_RADIUS} parameters. All objects with 
{\tt CLASS\_STAR} $<0.3$ and {\tt FLUX\_RADIUS} $>2.5$ pixels were selected 
and flagged as galaxies. The number counts for the objects classified as 
galaxies reach their maximum at \sloang$\sim24.8$ mag and at \sloanr$\sim24.3$ 
The galaxy classification was checked visually and by plotting pair of 
parameters \citep[see ][and references therein]{carrasco2006}. In both
cases the classifications are consistent for $\gtrsim95$\% of the objects
classified as galaxies down to \sloanr$\sim 24.3$ mag. Using the number count 
information given above and the  uncertainties in the galaxy classification, 
we have adopted the value of \sloanr$=24$ mag as our limiting magnitude.

The final GMOS photometric catalog contains total magnitudes, colors and 
structural parameters for 355 objects classified as galaxies brighter than 
\sloanr~$=24$ ($M_{r\prime}=-15.91$ at the distance of the cluster,
using a distance modulus of 39.91). The color composite (\sloang, \sloanr, \sloani) 
image of the central 5\arcmin $\times$ 5\arcmin region of the cluster covered by GMOS 
is shown in Fig.
\ref{fig:colorfig}. The top-left inset in the figure shows the 
BCG with the six distinct nuclei embedded in a common symmetric envelope. 
The X-ray contours in the figure are from the \textit{Chandra} smoothed images. 
The arcs and arclets belonging to the gravitational arcs systems M1, M2, M3, M4, 
M5 and S1 are shown  in the bottom-left inset image of Fig.\ref{fig:colorfig}.

\begin{figure}[ht!]
\centering
\includegraphics[width=0.9\hsize]{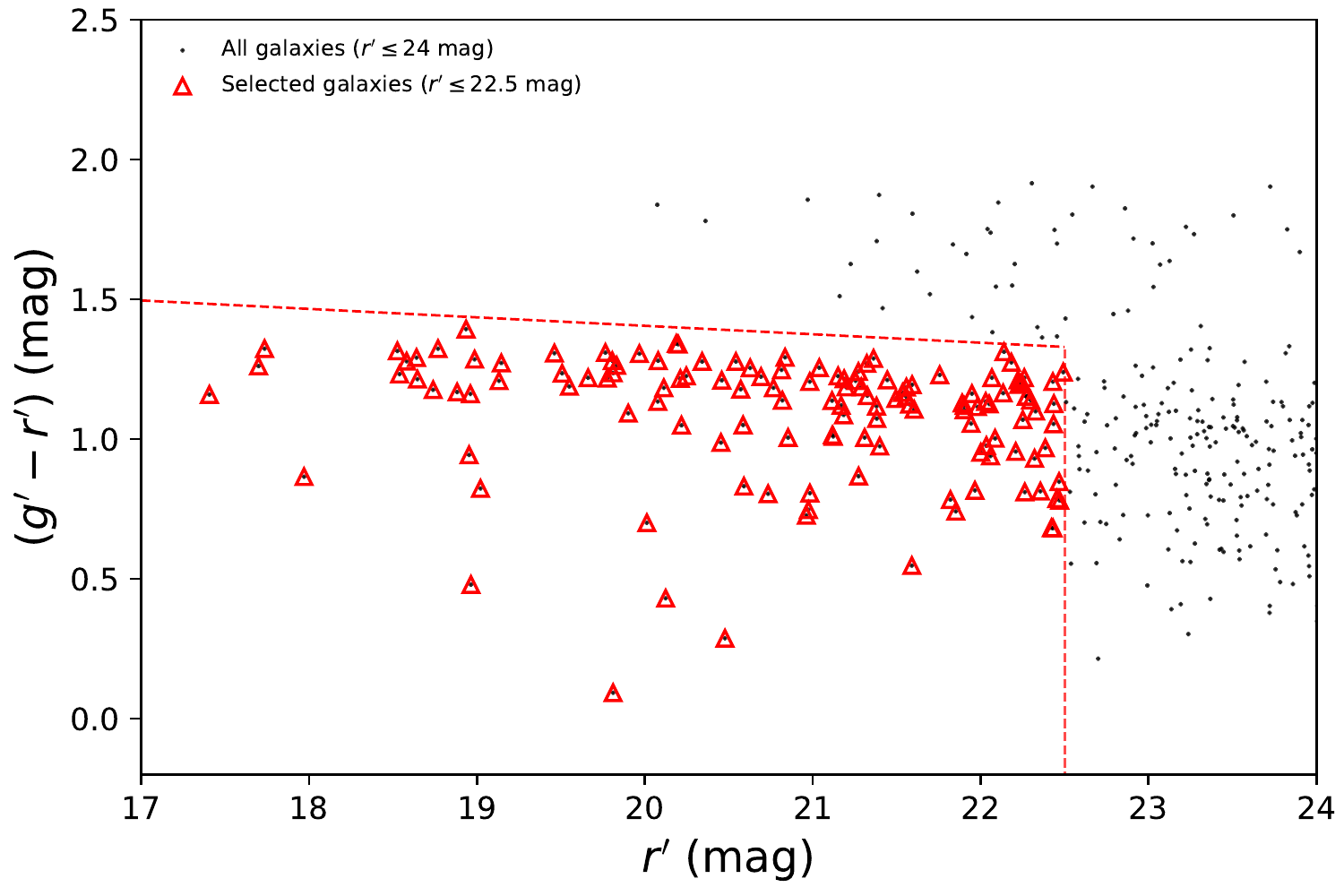}
\caption{Color-magnitude diagram of all objects classified as galaxies (black dots) 
within the area covered by GMOS observations  (see text). A total of 134 galaxies 
brighter than \sloanr$= 22.5$ mag and located in the region between the two dashed 
lines were selected as a potential targets for spectroscopy (red triangles).
\label{fig:galsel}}
\end{figure}

The GMOS photometric catalog is mainly used to select the galaxies for spectroscopic 
follow-up. The galaxies 
were selected by their \sloanr~magnitudes and \sloangr~colors. 
Galaxies brighter than \sloanr $=22.5$ mag and inside the area delimited by the 
two dashed lines in
Fig. \ref{fig:galsel} (red triangles) were selected as a potential targets for
spectroscopy . Ninety eight out of 134 galaxies ($\sim 73$\% of the 
sample) were included in four GMOS masks. It is worth noting that, despite the
remarkable of the gravitational arc system in \object{MS\,0440.5$+$0204}, no
accurate photometric nor spectroscopic redshifts for the lensed sources exist.
Only the distorted arclike galaxy S1 have a secure spectroscopic redshift
obtained by \citetalias{gioia1998}. To model in detail the arc system and hence
constrain the mass of the cluster, redshift of the lensed sources have to be
determined. To secure the redshifts, the brightest lensed sourced presented in
the gravitational systems M1, M2, M3, M4, M5, and the source S1 (lower-left
inset in Fig. \ref{fig:colorfig}) were included in two of the designed masks.

The GMOS multi-object observations (MOS) were carried out a year later, 
between 2013 January 13 and 2013 February 4 UT (Program ID: 
GS-2012B-Q-53), during dark time and under photometric conditions.  All spectra
were acquired using the R400$+$ grating, 1\arcsec\ slitest and $2\times2$
binning. The two masks, with the fifteen strongly lensed features and some of
the faintest galaxies in the cluster, were observed with the
\textit{nod-and-shuffling} technique in \textit{band shuffling} mode
\citep{glaz2001}, using the OG515 blocking filter, and centered at 7500\AA. The
two other masks were observed using a central wavelength of 6000\AA. Offsets of
70\AA\ and 50\AA\ toward the blue and the red were applied between exposures
to avoid the loss of any emission and/or absorption lines that could be, by
chance, lie in the gaps between CCDs. The spectroscopic observation log is
summarized in Table \ref{tab:specobslog}. For each mask, spectroscopic flats and
CuAr comparison lamps spectra were taken before or after each science exposure.
To calibrate in flux the science spectra, the spectrophotometric standard stars
LTT 9239 and  LTT 3864  were observed with the same instrument setup. However,
the spectra were obtained in a different night (August 29, 2012 UT) and under 
different observing conditions, providing only a relative flux calibration of
the science spectra.

\begin{deluxetable}{ccccc}
\tabletypesize{\scriptsize}
\tablewidth{0pc}
\tablecolumns{5}
\tablecaption{GMOS spectroscopy observations log\label{tab:specobslog}}
\tablehead{
\colhead{Mask} & \colhead{Obs. Date (UT)} & \colhead{ Exposure} & \colhead{Seeing} & \colhead{Airmass} \\
\colhead{(1)} & \colhead{(2)} & \colhead{(3)} &\colhead{(4)} & \colhead{(5)}} 
\startdata
1  & 2013-01-14 & $3 \times 1800$ sec & 0\farcs78 & 1.196 \\
    & 2013-02-03 & $1 \times 1800$ sec & 0\farcs70 & 1.239 \\
2  & 2013-02-04 & $4 \times 1800$ sec & 0\farcs92 & 1.185 \\
3  & 2013-01-13 & $4 \times 1800$ sec & 0\farcs79 & 1.184 \\
4  & 2013-02-02 & $4 \times 1800$ sec & 0\farcs92 & 1.229 \\
\enddata
\end{deluxetable}

All spectra were reduced with the Gemini GMOS package version
1.14\footnote{https://www.gemini.edu/node/11823}, following the standard
procedures for MOS and \textit{nod and shuffle} MOS observations. The science
exposures, spectroscopic flats and CuAr comparison lamps were over-scan and
bias-subtracted and trimmed. Spectroscopic flats were processed by removing the
calibration unit and GMOS spectral response, normalized and leaving only the
pixel-to-pixel variations. The two-dimensional science spectra were then flat
fielded, wavelength calibrated, rectified (S-shape distortion corrected), and
extracted to one-dimensional format. The obtained \textit{rms} for the
wavelength solution varied between $\sim$0.10 \AA\, and 0.20 \AA.  For the
1\arcsec~ slits, the final spectra have a resolution of $\sim 7.1$\AA\ (measured
from the sky lines FWHM), a dispersion of $\sim 1.37$ \AA~ pixel$^{-1}$, and an
average wavelength coverage of $\sim 4000$~\AA~ -- $\sim 9500$~\AA~ (the
coverage depends on the slit position in the GMOS field-of-view).

\subsection{CFHT data}\label{sec:cfhtdata}

The \object{MS\,0440.5$+$0204} cluster was observed with MegaCam at CFHT as part
of the Canadian Cluster Comparison Project \citep[CCCP --][]{hoekstra2012,mahdavi2013}, 
using the \sloang~ and \sloanr~filters, under photometric
conditions. The images were re-processed using the {\tt Elixir}
pipeline\footnote{http://www.cfht.hawaii.edu/Instruments/Elixir/}. The
photometric calibration of the individual images was obtained using photometric
standard star obtained during the night of observation. The calibrated images
were then co-added into a single image per filter using the {\tt MegaPipe} image
stacking pipeline\footnote{http://www.cadc-ccda.hia-iha.nrc-cnrc.gc.ca/en/megapipe/}
\citep{gwyn08} at the Canadian Astronomy Data Center (CADC). The final MegaCam
co-added images  have sky level of 0 ADU and scaled to have a photometric
zero-point of 30 mag in the AB system. The final co-added images have an
effective exposure times of 2440 and 6060 seconds with an average seeing of  
0\farcs89 and 0\farcs85 in \sloang~and \sloanr~filters, respectively. 

The object detection and photometry were performed following the same recipe 
used with the GMOS images. In summary, the MegaCam \sloanr-band image was used as 
the primary filter for object detection with {\tt Source Extractor} 
\citep{ber96}. The photometric parameters in the \sloang-band were determined only 
for those objects detect in common with the \sloanr-band image (dual mode). 
The parameter MAG$\_$AUTO was adopted as the value for the total magnitude of the objects.  
The colors of the objects were estimated by measuring the flux inside a fixed circular
aperture of 1\farcs2 in diameter. Objects with {\tt CLASS\_STAR} $<0.3$ and {\tt
FLUX\_RADIUS}  $>2.8$ pixels in \sloanr-filter were selected as galaxies.
We checked the objects classified as galaxies by plotting pairs of parameters 
and by visual inspection, as we did for the GMOS images in section \S \ref{sec:gemdata}. 
We found that $\gtrsim$90\% of the galaxies were classified identically by the three 
methods down to \sloanr$\sim 24.3$ mag (\sloang$\sim 24.8$ mag). The galaxy counts 
calculated using the objects classified 
as galaxies reach their maximum at \sloanr$\sim 24.5$ (\sloang$\sim 25$ mag).
Using this information and the uncertainties in the galaxy classification above
\sloanr$\sim 24.3$ mag, we have adopted a conservative value of \sloanr$=24$ mag
for our limiting magnitude.

The final catalog contains the magnitudes, colors and structural parameters of 
5547 objects classified as galaxies brighter than $\sim 24$ mag in \sloanr-filter 
(our completeness limit). It is worth clarifying that we use the photometric 
catalog constructed from the CFHT/Megacam archival images to analysis the projected
distribution of the galaxies in the cluster beyond the area covered by GMOS 
(see section \S \ref{sec:analysis}) and to build the catalog used in the weak 
lensing analysis presented in \citetalias{verdugo2020} and summarized in 
section \S \ref{sec:weaklen}.

\subsection{Redshifts of the lensed sources}\label{sec:redarcs}

\begin{figure}[ t!]
\centering
\includegraphics[width=0.9\hsize]{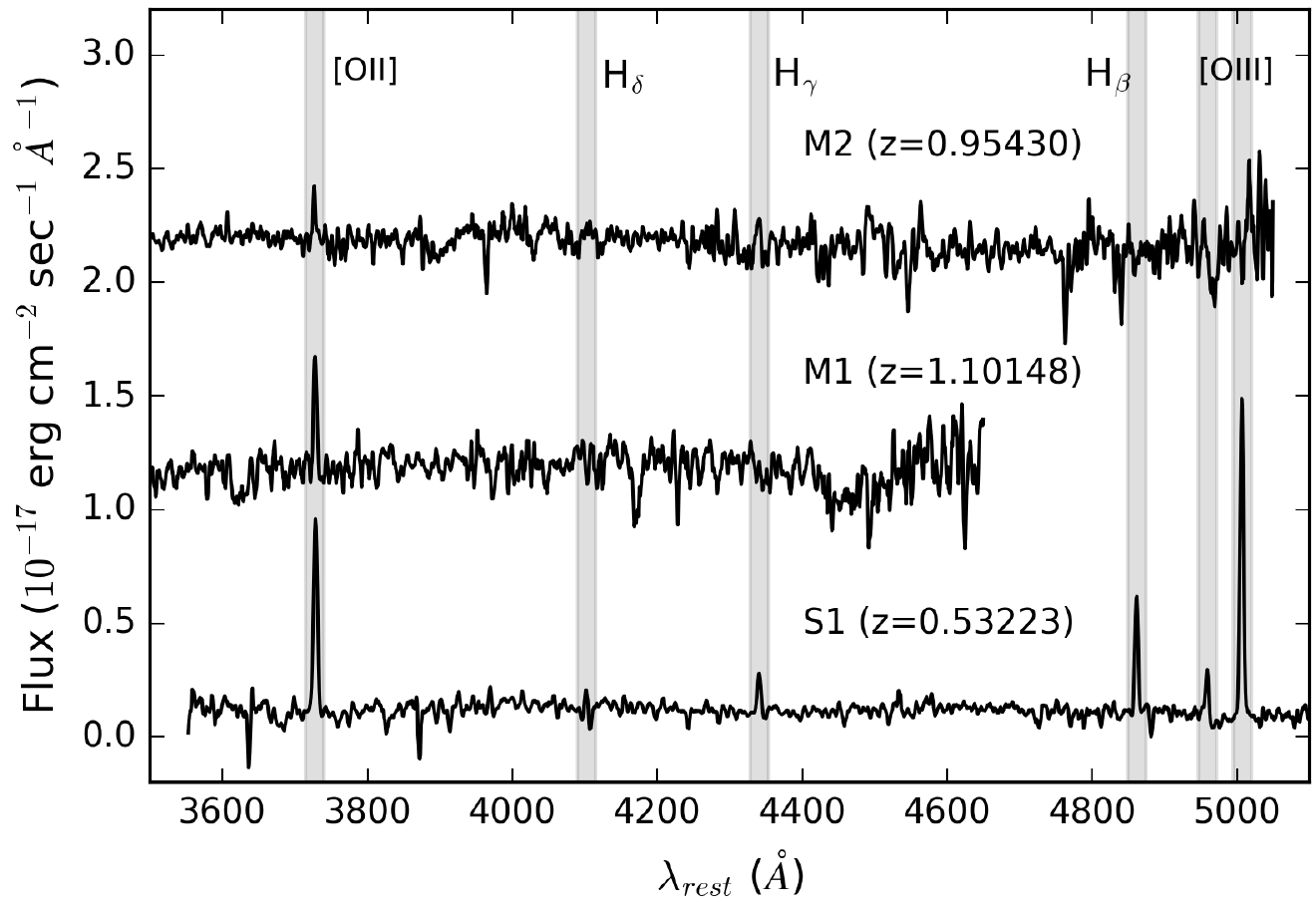}
\caption{Relative flux calibrated spectra, at the optical rest frame, of the
lensed sources with secure spectroscopic redshifts.  For clarity, the M1 and M2
lensed source spectra are shifted by 0.8 dex and 2.0 dex in flux, respectively.
The most prominent emission lines are clearly seen in the spectra.\label{fig:arcspectra}}
\end{figure}

We attempted to determine the redshift of the lensed sources in gravitational
arc systems M1, M2, M3, M4, M5 and S1 (see bottom left inset in Fig.
\ref{fig:colorfig}) by searching for clear spectroscopic features in their
spectra. The redshift of the lensed sources S1, M1.1 - M1.4 and M2.1 - M2.2 
were determined using the emission lines in the arc spectra. The optical
spectra of those lensed with spectroscopically secure redshifts are shown
in Fig. \ref{fig:arcspectra}. The redshifts were measured by employing a
line-by-line Gaussian fitting with the program {\tt RVIDLINES} implemented in
the {\tt IRAF RV} package. From the \oii~ $\lambda$3727, H$_{\gamma}$
$\lambda$4340,H$_{\beta}$ $\lambda$4863 and \oiii\,$\lambda \lambda$4959,5007
emission lines presented in the spectrum of the S1 arclike source, we have
obtained a redshift of 0.53223. The redshift is in good agreement with the 
value of 0.53230 obtained by \citetalias{gioia1998}, after the redshift is corrected by
a zero-point offset (see section \S \ref{sec:galred}). The spectra of the lensed sources 
M1.1, M1.2, M1.3 and M1.4 show a strong emission line located at $\sim$ 7834\AA, 
which may be related to \oii~ $\lambda$3727 at $z\sim 1.1$.  Another possibility is
that this is a strong Ly{\sc $\alpha$}\,$\lambda$1216 emission line at
$z\sim5.4$. However, this high redshift solution is unlikely, since the previous
model \citepalias{gioia1998} and our own estimation locates the source between 
redshift 0.53 and 1.1.  For the system M1 we obtained an average redshift of 
$1.10148$. The spectra of the M2.1 and M2.2 arcs show several emission 
lines at 7284\AA, 9501\AA, 9686\AA~and 9785\AA, corresponding to  
\oii\,$\lambda$3727, H$_{\beta}$ $\lambda$4863 and \oiii\,$\lambda \lambda$
4959,5007 respectively. From these lines, we obtained an average redshift for 
the M2 arc system of $0.9543$.

\begin{figure}[ht!]
\centering
\includegraphics[width=0.75\hsize]{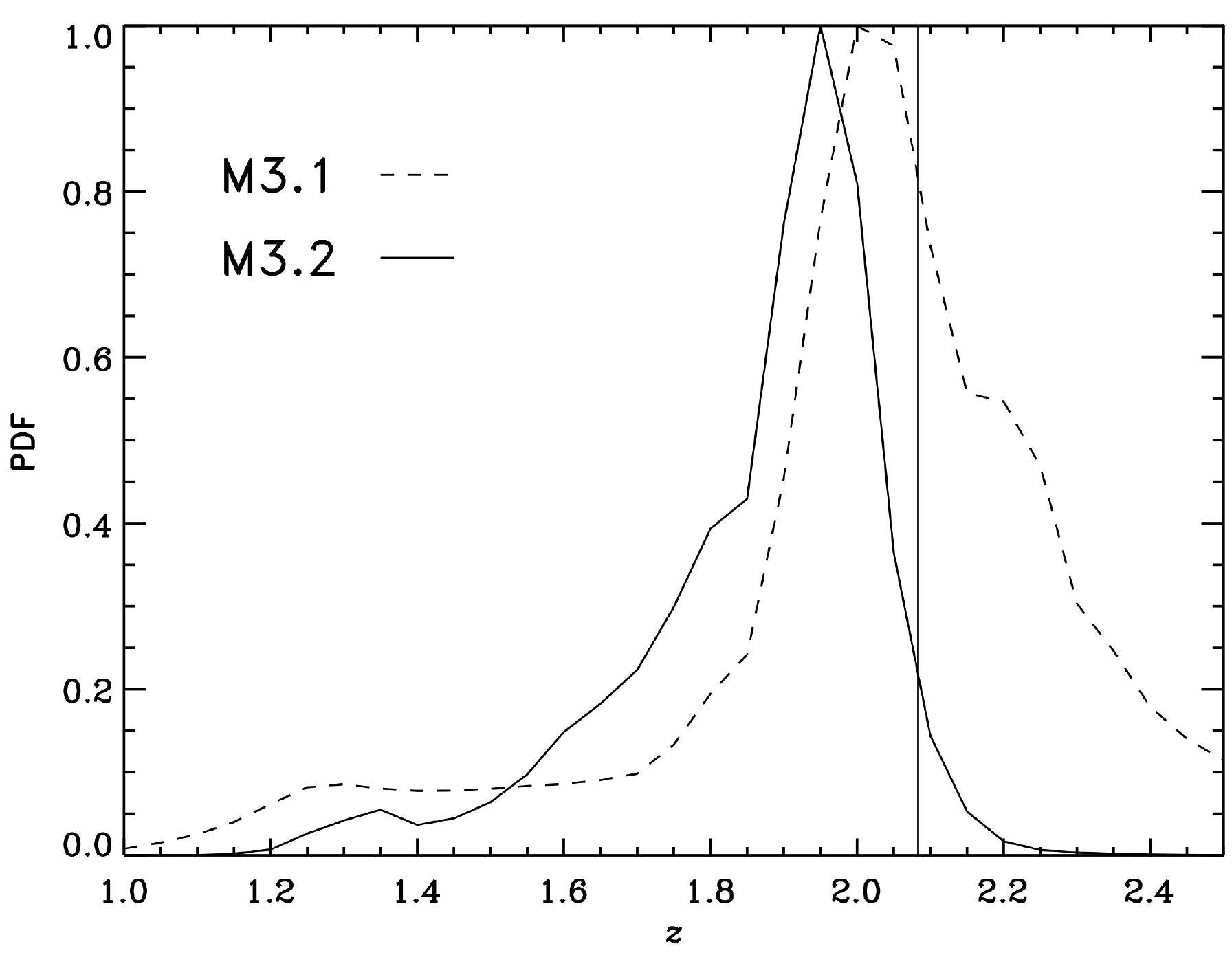}
\includegraphics[width=0.95\hsize]{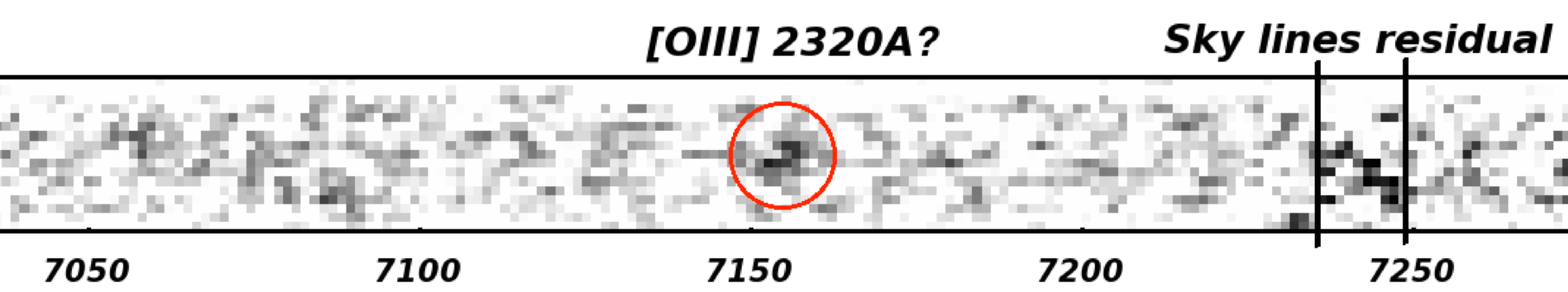}
\includegraphics[width=0.90\hsize]{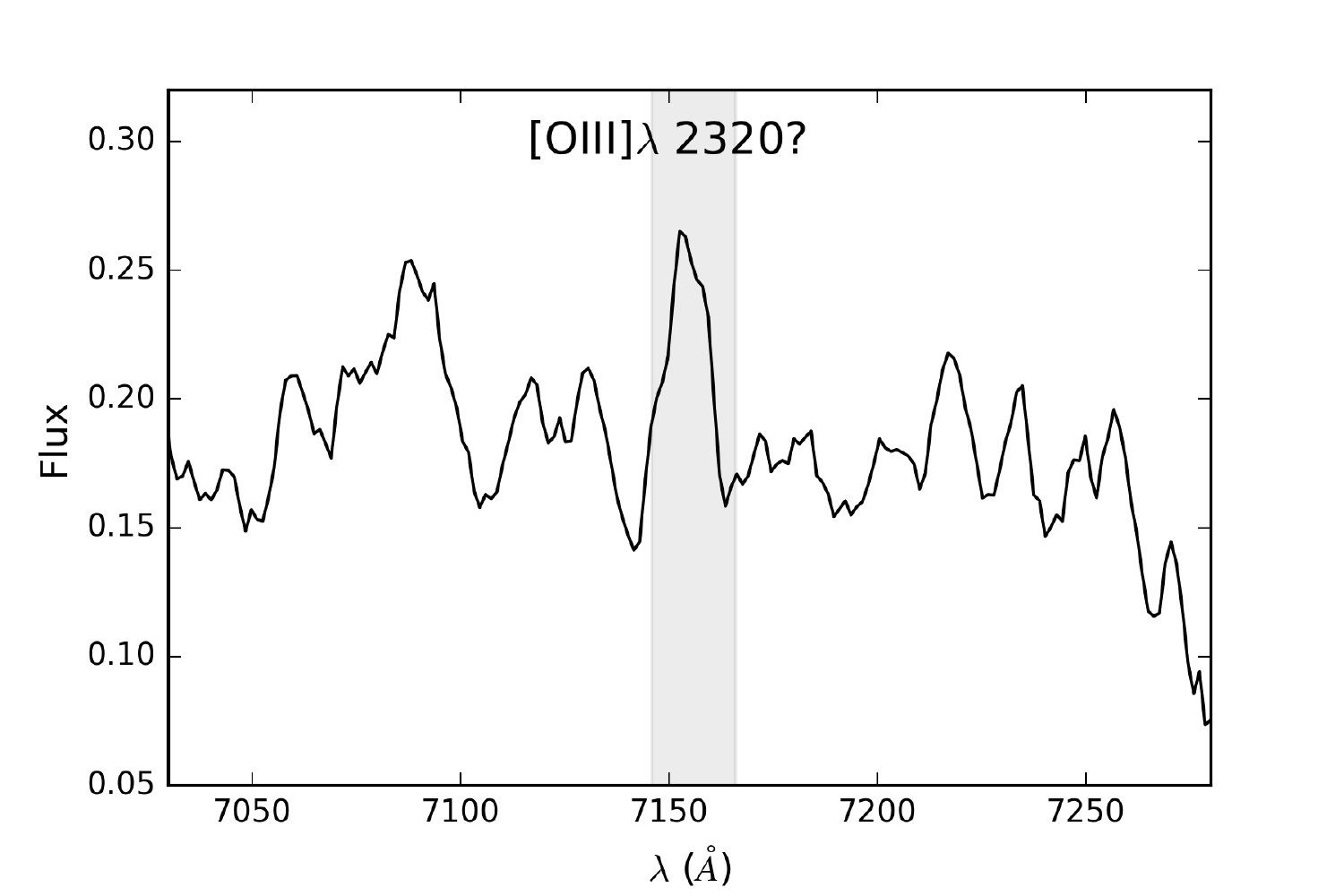}
\caption{\textit{Top panel: } Photometric redshift PDF from HyperZ for M3.1
(dashed line) and M3.2 (continuous line) arcs respectively. The vertical solid
line correspond to the spectroscopic redshift of the arcs assuming an emission
line from \oiii\,$\lambda$ 2321. \textit{Middle panel:} summed 2D spectrum of
arcs M3.1 and M3.2 with the emission line probably associated with
\oiii\,$\lambda$ 2320. \textit{Bottom panel:} summed 1D relative flux calibrated
spectrum (in units of $10^{-17}\,\mathrm{erg}\,\mathrm{cm}^{-2}\,\mathrm{sec}^{-1}$\AA) 
of arcs M3.1 and M3.2.\label{fig:photred3}}
\end{figure}

No spectroscopic features were found in the spectra of the M4.1, M4.2, M4.3,
M5.1 and M52 arcs. However, there is a weak emission line presented in the
spectra of M3.1 and M3.2 arcs (no spectroscopic features are seen in the spectra
of M3.3 arc). The emission line, located at 7156\AA~may correspond to
\oiii\,$\lambda$ 2321 at $z\sim2.1$. Assuming an emission from \oiii\,$\lambda$
2321 and using a Gaussian fitting,  we obtain an average redshift of
$z=2.0834$. The 1D and 2D summed spectrum of M3.1 $+$ M3.2 arcs 
is shown at the bottom of Fig. \ref{fig:photred3}. Our spectroscopic redshift
calculation is supported by \textit{phot-z} estimated using the GMOS magnitudes
in \sloang, \sloanr\,and \sloani\, and ESO/SOFI magnitude in J-band using the 
{\sc HyperZ} software \citep{bolzonella2000}. The infrared data were 
obtained with the SofI camera on the New Technology Telescope at the European 
Southern Observatory (ESO), La Silla Observatory in  Chile \citep{moorwood1998}. 
The observation in the J-band was carried on in 1998 October 29 (PROG= 62.O-0143, 
ID= 4271), with a total exposure time of 180 seconds. The data were reduced 
using the Eclipse data analysis software package \citep{devillard1997}. 
From the fitting results, we estimate a \textit{phot-z} of $2.02 \pm 0.08$ and 
$1.95 \pm 0.07$ for M3.1 and M3.2 arcs respectively. The resulting probability 
distribution functions (PDF) from HyperZ for
the arcs are shown in Fig. \ref{fig:photred3} (top panel). We are aware
about the uncertainties introduced using only 4 photometric bands in the
\textit{phot-z} determination. However, if we include the magnitudes estimated
by \citet{luppino1993} in B, V, R and I, our results do not change.

The redshift of the gravitational arc systems with secure values are presented
in Table \ref{tab:tabarcs}. The residual of the average redshift shifts of all
measurements provided by  {\tt RVIDLINES} was used to estimate the errors.
The errors in the redshift determination varied between $0.00016$
and $0.00044$. The final redshifts used in the lensing model are 
listed in the last column of the table.

\begin{deluxetable*}{ccccccc}[t!]
\tabletypesize{\scriptsize}
\tablewidth{0pc}
\tablecolumns{6}
\tablecaption{Redshifts of the gravitation arc systems \label{tab:tabarcs}}
\tablehead{
\colhead{System} & \colhead{Arc} & \colhead{RA} & \colhead{DEC} &  \colhead{$z_{spec}$} & \colhead{$z_{G98}$} &  \colhead{$z_{final}$}\\
\colhead{(1)} & \colhead{(2)} & \colhead{(3)} & \colhead{(4)} &  \colhead{(5)} &  \colhead{(6)} & \colhead{(7)}} 
\startdata
0 &                 &                        &                          &               &    & 0.53223\\
   & S1            &  04 43 11.17  & $+$02 10 10.49 & 0.53223 & $0.53230^{(a)}$ &  \\
I &                  &                     &                            &                & $ [0.53-1.10]$ & 1.10148 \\
  &M1.1 (A8)  & 04 43 10.44 & $+$02 10 34.54 & 1.10125 &  &  \\  
  &M1.2 (A9) & 04 43 10.74 & $+$02 10 29.85 & 1.10139 &   & \\
  &M1.3 (A12) & 04 43 10.66 & $+$02 10 06.56 & 1.10176 &  & \\
  &M1.4 (A24) & 04 43 09.30 & $+$02 10 19.43 & 1.10147 &  & \\
II & &  & &  &  $[0.60-1.60]$ & 0.95430\\
  &M2.1 (A6) & 04 43 10.49 & $+$02 09 58.53 &  0.95434 &  &  \\
  &M2.2 (A5) & 04 43 09.23 & $+$02 10 27.22 &  0.95426 &  & \\
III & &  & &  &  $[0.59-\infty]$ & 2.08340 \\
  &M3.1 (A3) & 04 43 08.57 & $+$02 10 29.69 &  $2.08362^{(b)}$ & &  \\
  &M3.2 (A2) & 04 43 08.45 & $+$02 10 18.81 &  $2.08318^{(b)}$ & & \\
  &M3.3 (A20) & 04 43 09.45 & $+$02 09 56.88 &  \nodata  &  & \\
  &M3.4 (---) & 04 43 10.29 & $+$02 10 23.82 & \nodata & & \\
IV & &  & &  & $[0.59-1.50]$ & \nodata \\
  &M4.1 (A18) & 04 43 10.20 & $+$02 09 54.35 & \nodata &  &  \\
  &M4.2 (A17) & 04 43 09.97 & $+$02 10 25.18 & \nodata &  & \\
  &M4.3 (A19) & 04 43 10.00 & $+$02 10 30.25 & \nodata &  & \\
V & &  & &  &  $[0.59-1.50]$ & \nodata\\
  &M5.1 (A7) & 04 43 09.55 & $+$02 09 57.69 & \nodata &  & \nodata \\
  &M5.2 (A16) & 04 43 10.31 & $+$02 10 25.30 & \nodata &  & \\
\enddata

\tablecomments{Col. (1) and (2) : System and Arc IDs used in this work followed
by the original names used  in \citetalias{gioia1998} (between parenthesis); Col. (3) and (4):
Right Ascension and  Declination (J2000.0). The units of Right Ascension are
hours, minutes and seconds, and the units of Declination are degrees, arcminutes
and arcseconds; Col. (5): spectroscopic redshifts; Col (6) redshifts of lensed sources
from  \citetalias{gioia1998}; Col. (7): final redshift values, when applicable, used in
the lensing model; $(a)$ - Spectroscopic redshift reported by \citetalias{gioia1998}; 
$(b)$ -There is a weak emission line present in the spectrum of both M3.1 and M3.2 at an 
observed wavelength of 7156\AA~that may be  [OIII] $\lambda$2321 at $z=2.0834$.}
\end{deluxetable*}

\subsection{Redshifts of the galaxy sample}\label{sec:galred}

We determined the redshifts of the galaxies observed with GMOS using the
programs implemented inside the IRAF RV package.The spectra were
visually inspected to search for obvious absorption and/or emission features 
characteristics of an early- and late-type populations. For non-early type 
spectra, the routine {\tt RVIDLINES} was used, employing a line-by-line 
Gaussian fit. The errors of the measurements were estimated using the 
residual of the average redshift shifts of all measurements provided by the program.
For normal early-type spectra, the cross-correlation algorithm \citep{tonry1979}
implemented in the program {\tt FXCOR} was used. The observed spectra
were correlated with four high singal-to-noise (S/N) templates. 
The errors of the redshifts were estimated based on the \textit{R} statistic 
value of \citet{tonry1979}. We were able to measure the redshifts for 96
of the 98 galaxies targeted in the four masks ($\sim$98\% success rate) plus 
one additional galaxy which was found by chance in one slit.

In order to increase the number of galaxies with confirmed spectroscopic 
redshifts to be used in the dynamical analysis, the 97 measured redshifts presented 
above were combined with the 57 spectroscopic redshifts reported by \citetalias{gioia1998} and 
with the 113 un-published spectroscopic redshifts obtained by \citet{yee1996} as part 
of the CNOC cluster Redshift Survey \citep{calberg1994}.
The spectroscopic observations were conducted at CFHT using the MOS/SIS multi-object 
spectrograph \citep[hereafter CFHTMOS,][]{lefevre1994}, covering an area of 
$\sim$0.45\degr~$\times$0.14\degr~($\sim 5.3\times 1.6$ \hvaltwo~Mpc$^2$ at the 
cluster rest frame). We refer the reader to \citet{yee1996} for details about
the data reduction and the criteria  used to select the candidate cluster members. 
The large number of spectroscopic redshifts and the area covered by the CFHTMOS 
observations allows us to study more accurately the dynamical state of the cluster, 
the spatial distribution of the member galaxies and analyse if there are other 
structures at large scale that could be potentially connected to \object{MS\,0440.5$+$0204} 
(see section \S \ref{sec:substructure} for details).

The combined spectroscopic sample contains 267 measured redshifts for 204 galaxies. 
Fifty one galaxies have more than once redshift estimation among the three data sets: 39 
galaxies with 2 measurements and 12 galaxies with 3 measurements. One of the galaxy has
two redshift determinations in the CHFTMOS sample. The large number 
of galaxies with multiple redshift measurements observed in common allows us to 
obtain a more accurate redshifts, obtain an assessment of the general quality of 
the data, identify discordant redshifts and find a potential zero-point shifts 
between data sets. If the difference in the measured redshifts for a given galaxy 
and the zero-points of different instruments differed by significant amounts, 
we could have introduced serious systematic errors in the resulting velocity 
dispersion and in the dynamical analysis of the cluster.

\begin{figure}[t!]
\centering
\includegraphics[width=0.95\hsize]{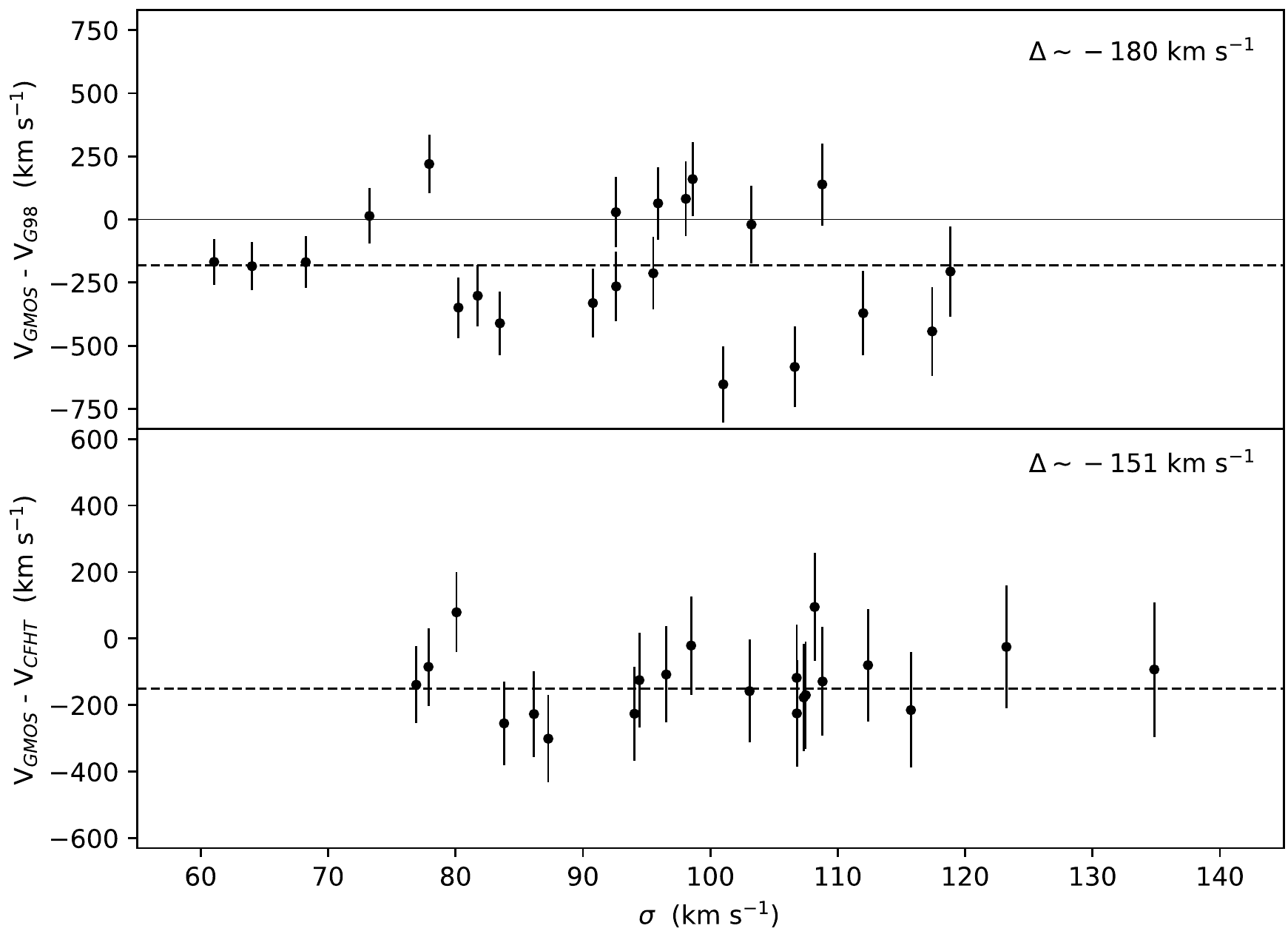}
\caption{Residual of the velocities as a function of the internal quadratic errors for the 
galaxies with observed in common with \citetalias{gioia1998} (top panel) and CFHTMOS (lower panel), 
respectively. The mean residual shifts (dashed lines) are consistent between the two data sets 
(see text). For clarity, only galaxies with $|\Delta V| \lesssim 600$ \kms~are shown in the
figure (see text for details). \label{fig:compfig}}
\end{figure}

To build a coherent catalog to be used in the dynamical analysis, 
we compare the galaxy redshifts observed in common between the three different 
samples. Figure \ref{fig:compfig} shows the residual of the velocities as a function 
of the internal quadratic errors for the galaxies observed in common with 
\citetalias{gioia1998} (top panel) and CFHTMOS (lower panel). A zeroth-order 
polynomial fitting of the data with a 3$\sigma$-clip (to remove the outliers) 
gives a mean residual of $\sim180$ \kms~ (\textit{rms} of 151 \kms) and $\sim151$ 
\kms~(\textit{rms} of 76 \kms) for the galaxies observed in common with 
\citetalias{gioia1998} and CFHTMOS, respectively. The derived mean residual shifts 
are similar, which is expected since both \citetalias{gioia1998} and CFHTMOS data sets 
were obtained using the same telescope and instrument. The results are also 
consistent with the mean shifts obtained when comparing the 14 galaxies observed in 
common between CFHTMOS and \citetalias{gioia1998}.  The comparison gives a mean 
residual shift of $\sim 15$ \kms (\textit{rms} $=77$ \kms), which is negligible 
compared to the \textit{rms}, and consistent with the fact that the two samples 
were obtained using the same observational capabilities.

\begin{figure}[t!]
\centering
\includegraphics[width=1.0\hsize]{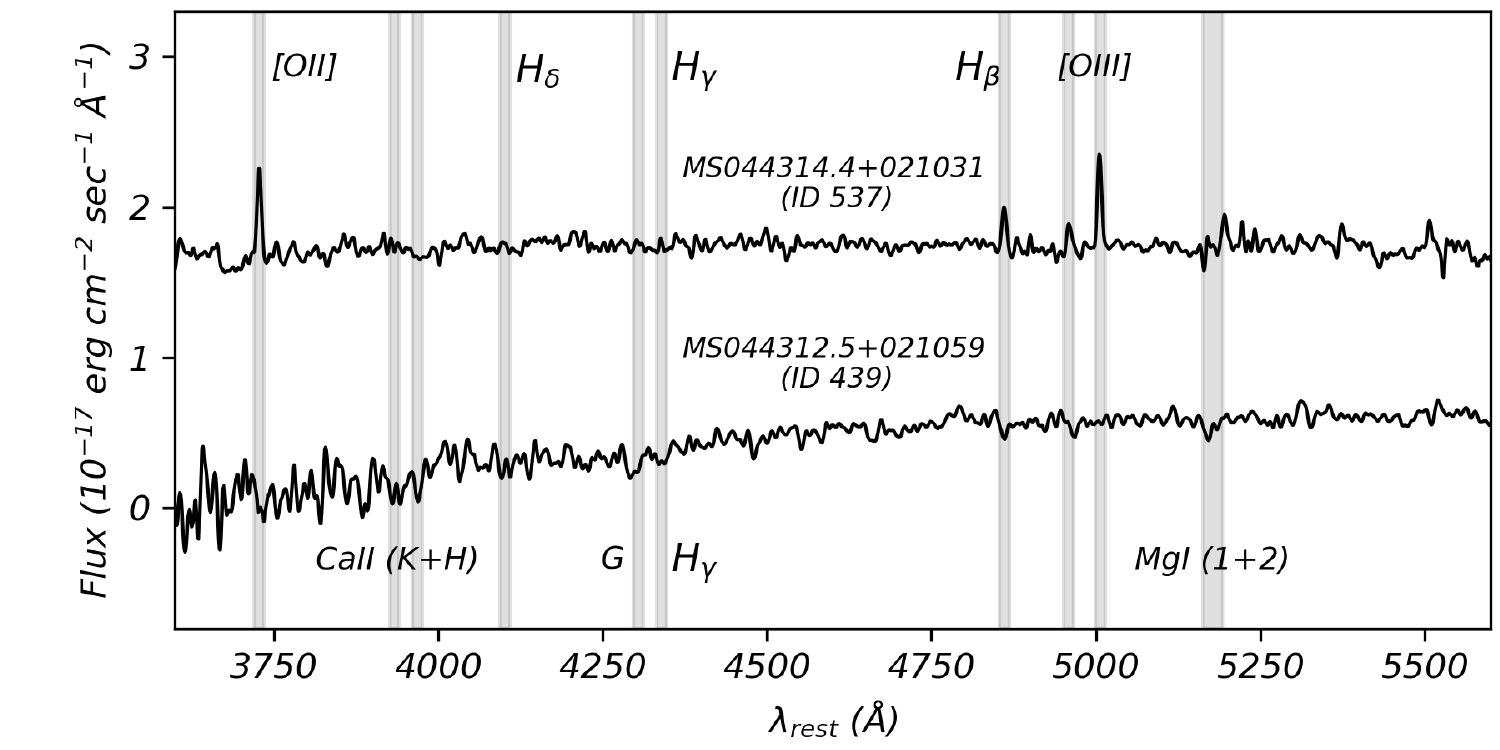}
\caption{Flux calibrated GMOS spectra, at the optical rest-frame, of two galaxies with large 
differences in redshift ($\Delta z = 0.20343$ and $\Delta z = 0.10209$) compared to the 
values reported  by \citetalias{gioia1998}. For clarity, the top spectrum has been shifted 
by $\sim 2$ dexs in flux. \label{fig:spectra}}
\end{figure}

Even though the derived mean residual shifts are similar, the comparison between
GMOS and \citetalias{gioia1998} shows a large dispersion (the \textit{rms} value is similar 
to the mean residual shift). Indeed, after applying the zero-point shifts 
derived above to both data sets, we find that 21 
out of 36 galaxies observed in common with \citetalias{gioia1998} and 2 out of 23 galaxies 
observed in common with CHFTMOS have velocity differences larger than 250 \kms~. It is worth 
mentioning that galaxies redshifts which differed by $\Delta V > 250$ \kms\,from the values 
determined using GMOS are discarded before combining (see below for details 
about how the combination of multiple measurements for a given galaxy is performed). Some
galaxies observed in common with \citetalias{gioia1998} show very large differences
(up to 80,000 \kms!). To determine the origin of these large differences in redshift,
we first checked the GMOS data. We used the 
MEGACAM and the GMOS deep images to eliminate any possibility of coordinate mismatch. 
When this possibility was ruled out (none of the galaxies present coordinate problems),
we visually inspected the GMOS spectra of the problematic galaxies. The GMOS spectra
of these galaxies have high S/N and show clear absorption and/or emission line features
characteristics of an early- and and late-type population of galaxy. The excellente
quality of the GMOS spectra provides a high confidence that our measured redshifts
are correct and reliable. Unfortunately, \citetalias{gioia1998} do not provide the 
necessary information about how the redshifts  and associated errors were estimated. 
Given the absence of information, we can only speculate about the erroneous estimations 
provided by \citetalias{gioia1998} for 
these galaxies. Possible explanations are: absorption/emission lines miss-identification, 
low signal-to-noise of the spectra, which could impact the cross-correlation results, 
velocity miss-typing, among others. 
For the two problematic galaxies observed in common with  CFHTMOS, the derived R-values 
\citep{tonry1979} in the CFHTMOS sample are smaller compared to R-values obtained with
GMOS, indicating that the measured redshifts are less reliable due to, possible, 
poor cross-correlation results. For these two galaxies, we used the redshfits derived
from GMOS spectra. Figure \ref{fig:spectra} shows the GMOS spectra of two galaxies
observed in common with \citetalias{gioia1998}  which show large differences in the redshift 
estimation (after applied the zero-point
correction derived above). The galaxy MS044314.4$+$021031 (top spectrum) shows 
clear emission lines and the measured redshift from the GMOS spectrum is $z=0.40109$. 
The galaxy MS044312.5$+$021059 (bottom spectrum) shows clear absorption lines. The 
measured redshifts derived from GMOS and CFHTMOS sepectra are $z=0.19628$ and 
$z=0.19728$ ($\Delta V =  -150$ \kms), respectively. The redshifts estimated by 
\citetalias{gioia1998} for these galaxies are $z=0.19766$ ($\Delta V = 61,169$ \kms) 
and $z=0.09419$ ($\Delta V = 30,786$ \kms), respectively.  
According to \citetalias{gioia1998} 
the first galaxy (MS044314.4$+$021031) is located at the distance of the cluster and 
the second galaxy (MS044314.4$+$021031) is located at the cluster foreground, which is 
clearly incorrect. The same problem is seen in the remaining 19 galaxies with 
large differences in redshifts and observed in common with \citetalias{gioia1998}.

To construct the final spectroscopic catalog, first we applied the zero-point 
shifts of $-181$ \kms~ and $-150$ \kms~ to both \citetalias{gioia1998} and 
CFHTMOS data sets. Then, for galaxies with multiple measurements, a mean weighted 
redshift is derived using the relation:

\beq 
\overline{\mathrm{Z}} = \frac{\sum_{i} w_{i}z_{i}}{\sum_{i} w_{i}}
\label{eq1} 
\eeq

\noindent where $z_{i}$ is the redshift of the ${i}$ galaxy, $w_{i}=1/\Delta
z_{i}^{2}$ is the weighting factor, and $\Delta z_{i}$ is the redshift
uncertainty. The weight factor $w_{i}$ came from our internal uncertainties,
from CFTHMOS and from the uncertainties published in \citetalias{gioia1998}.  
The errors  in the combined redshifts are calculated following the criteria described in 
\citet{quintana2000}. In order to minimize the errors in the estimation
of the dynamical parameters of the cluster, galaxies with multiple redshifts 
observed in common between the three data sets with $|\Delta Z| > 0.0008$ 
($|\Delta V| > 250$ \kms) are discarded before combining.

\begin{deluxetable*}{lrcccccclccl}[ht!]
\tabletypesize{\scriptsize}
\tablecolumns{12}
\tablecaption{Galaxy redshifts\label{tab:galcat}}
\tablehead{
\colhead{Galaxy} & \colhead {ID} & \colhead{Ref.} & \colhead{RA} &
\colhead{DEC} & \colhead{\sloanr} & \colhead{\sloangr} & 
\colhead{$\mathrm{z} \pm \Delta \mathrm{z}$} & \colhead{R/\#l} & \colhead{Flag} & 
\colhead{$\overline{\mathrm{Z}}\pm \Delta \overline{\mathrm{Z}}$} & \colhead{Member?} \\
\colhead{(1)} & \colhead{(2)} & \colhead{(3)} & \colhead{(4)} &  
\colhead{(5)} & \colhead{(6)} & \colhead{(7)} &
\colhead{(8)} & \colhead{(9)} & \colhead{(10)} & \colhead{(11)} & \colhead{(12)}}
\startdata
 MS044308.5+021152 &    284 & M & 04 43 08.55 & +02 11 52.4 & 22.27 & 1.15 & 0.19461$\pm$0.00019 &  6.0/---  & 1 & 0.19461$\pm$0.00027 & Yes  \\
                   &     K9 & G &             &             &       &      & 0.16575$\pm$0.00200 &  0.0/---  & 0 &                     &      \\
 MS044308.7+020839 &   1347 & M & 04 43 08.66 & +02 08 39.0 & 19.90 & 1.09 & 0.15994$\pm$0.00019 &  5.5/---  & 1 & 0.15994$\pm$0.00027 & No   \\
                   &     35 & G &             &             &       &      & 0.19866$\pm$0.00030 &  0.0/---  & 0 &                     &      \\
 MS044308.9+021142 &    371 & M & 04 43 08.85 & +02 11 42.3 & 22.37 & 1.36 & 0.35688$\pm$0.00009 &---/10     & 1 & 0.35688$\pm$0.00027 & No   \\
 MS044309.0+021208 &    231 & M & 04 43 08.97 & +02 12 08.9 & 21.19 & 1.22 & 0.19716$\pm$0.00009 & 13.7/---  & 1 & 0.19716$\pm$0.00027 & Yes  \\
 MS044309.1+020845 &   1267 & M & 04 43 09.10 & +02 08 45.9 & 21.89 & 1.13 & 0.19104$\pm$0.00018 &  6.5/---  & 1 & 0.19104$\pm$0.00027 & Yes  \\
 MS044309.2+021053 &    592 & M & 04 43 09.20 & +02 10 53.0 & 21.32 & 1.27 & 0.19546$\pm$0.00013 & 11.3/---  & 1 & 0.19546$\pm$0.00027 & Yes  \\
                   &     05 & G &             &             &       &      & 0.19492$\pm$0.00030 &  0.0/---  & 0 &                     &      \\
 MS044309.3+021350 & 101153 & C & 04 43 09.34 & +02 13 50.0 & 19.12 & 1.18 & 0.39989$\pm$0.00050 &  5.0/---  & 1 & 0.39939$\pm$0.00050 & No   \\
 MS044309.4+021037 &    619 & M & 04 43 09.45 & +02 10 36.7 & 21.38 & 1.12 & 0.19369$\pm$0.00014 &  6.9/---  & 1 & 0.19369$\pm$0.00027 & Yes  \\
                   &     04 & G &             &             &       &      & 0.19665$\pm$0.00051 &  0.0/---  & 0 &                     &      \\
 MS044309.6+021156 &    279 & M & 04 43 09.58 & +02 11 56.3 & 22.91 & 0.91 & 0.38943$\pm$0.00003 &---/4      & 1 & 0.38943$\pm$0.00027 & No   \\
 MS044309.7+021027 &599 (D) & M & 04 43 09.67 & +02 10 27.4 & 18.58 & 1.28 & 0.19906$\pm$0.00012 & 12.2/---  & 1 & 0.19906$\pm$0.00027 & Yes  \\
                   &     33 & G &             &             &       &      & 0.19860$\pm$0.00034 &  0.0/---  & 0 &                     &      \\
 MS044309.7+021004 &    195 & M & 04 43 09.68 & +02 10 03.9 & 18.54 & 1.23 & 0.19223$\pm$0.00014 &  9.1/---  & 1 & 0.19221$\pm$0.00027 & Yes  \\
                   &     30 & G &             &             &       &      & 0.19279$\pm$0.00018 &  0.0/---  & 1 &                     &      \\
 MS044309.7+020815 &   1359 & M & 04 43 09.70 & +02 08 15.9 & 22.92 & 1.11 & 0.43362$\pm$0.00012 &---/5      & 1 & 0.43362$\pm$0.00027 & No   \\
 MS044309.7+021023 &894 (C) & M & 04 43 09.74 & +02 10 22.6 & 18.64 & 1.21 & 0.18780$\pm$0.00013 & 10.4/---  & 1 & 0.18781$\pm$0.00027 & No   \\
                   & 100610 & C &             &             &       &      & 0.18833$\pm$0.00032 &  3.6/---  & 1 &                     &      \\
                   &     13 & G &             &             &       &      & 0.18842$\pm$0.00017 &  0.0/---  & 1 &                     &      \\
 MS044309.9+020947 &    844 & M & 04 43 09.89 & +02 09 46.8 & 21.40 & 0.97 & 0.32986$\pm$0.00026 &  5.5/---  & 1 & 0.32986$\pm$0.00026 &  No \\
                   &     32 & G &             &             &       &      & 0.20317$\pm$0.00026 &  0.0/---  & 0 &                     &      \\
 MS044309.9+021019 &896 (B) & M & 04 43 09.92 & +02 10 19.2 & 17.73 & 1.32 & 0.19873$\pm$0.00021 &  8.0/---  & 1 & 0.19831$\pm$0.00027 & Yes  \\
                   & 100599 & C &             &             &       &      & 0.19909$\pm$0.00025 &  5.9/---  & 1 &                     &      \\
                   &     03 & G &             &             &       &      & 0.19868$\pm$0.00013 &  0.0/---  & 1 &                     &      \\
 MS044310.0+021027 &616 (E) & M & 04 43 10.04 & +02 10 27.0 & 18.99 & 1.28 & 0.19600$\pm$0.00016 &  8.7/---  & 1 & 0.19624$\pm$0.00041 & Yes  \\
                   &     37 & G &             &             &       &      & 0.19737$\pm$0.00023 &  0.0/---  & 1 &                     &      \\
 MS044310.1+021104 &    407 & M & 04 43 10.07 & +02 11 04.0 & 20.20 & 1.34 & 0.19929$\pm$0.00014 &  6.3/---  & 1 & 0.19929$\pm$0.00027 & Yes  \\
                   &     10 & G &             &             &       &      & 0.19855$\pm$0.00022 &  0.0/---  & 0 &                     &      \\
 MS044310.1+020941 &    820 & M & 04 43 10.09 & +02 09 41.3 & 22.75 & 0.70 & 0.57586$\pm$0.00009 &---/7      & 1 & 0.57586$\pm$0.00027 & No   \\
 MS044310.1+021021 &100606 (A)& C & 04 43 10.09 & +02 10 21.2 & 17.70 & 1.26 & 0.18908$\pm$0.00036 &  2.7/---  & 1 & 0.18857$\pm$0.00027 & Yes  \\
                   &     09 & G &             &             &       &      & 0.18797$\pm$0.00014 &  0.0/---  & 0 &                     &      \\
 MS044310.4+021039 &    535 & M & 04 43 10.36 & +02 10 39.1 & 20.46 & 1.21 & 0.18921$\pm$0.00008 & 12.6/---  & 1 & 0.18921$\pm$0.00027 & Yes  \\
 MS044311.0+021114 &    419 & M & 04 43 11.00 & +02 11 13.7 & 21.44 & 1.21 & 0.18957$\pm$0.00023 &  4.2/---  & 1 & 0.18957$\pm$0.00027 & Yes  \\
                   &     24 & G &             &             &       &      & 0.19105$\pm$0.00032 &  0.0/---  & 0 &                     &      \\
 MS044311.1+021206 &    124 & M & 04 43 11.08 & +02 12 06.5 & 21.11 & 1.14 & 0.19340$\pm$0.00005 & 19.5/---  & 1 & 0.19340$\pm$0.00027 & Yes  \\
 MS044311.2+021157 &    139 & M & 04 43 11.19 & +02 11 57.1 & 19.78 & 1.22 & 0.19616$\pm$0.00020 &  6.5/---  & 1 & 0.19612$\pm$0.00027 & Yes  \\
                   & 100902 & C &             &             &       &      & 0.19655$\pm$0.00030 &  4.2/---  & 1 &                     &      \\
                   &     28 & G &             &             &       &      & 0.19833$\pm$0.00027 &  0.0/---  & 0 &                     &      \\
\enddata
\tablecomments{The meaning of the columns are the following: (1) - Galaxy name; 
(2) - Internal ID; (3) - Redshift source; C: CFHTMOS, M - Gemini GMOS, G - Gioia
et al. (1998); (4) and (5): Right Ascension and Declination (J2000.0). The units
of Right Ascension are hours, minutes and seconds, and the units of Declination
are degrees, arcminutes and arcseconds; (6) and (7) - Total \sloanr~ magnitudes
and \sloangr~ colors measured inside a fixed aperture of 1\farcs2 corrected by
galactic extinction ($A(g^{\prime})=0.587$, $A(r^{\prime})=0.406$ mag) using the
reddening maps of \citet{schlafly2011} and assuming a reddening law of
$R_{v}=3.1$ \citep{fitz1999}.; (8) -  Individual redshift and associated errors, 
corrected by the zero point; (9): $R$ values (Tonry \& Davis  1979 - real numbers) or
the number of emission lines (integer values) used to calculate redshift; (10)
Flag for galaxies with multiple measurements included in the final weighted redshift: 
`1'' $-$ the individual redshift is included, ``0'' $-$ the individual redshift is discarded; 
(11) - final redshift and the associated error used in the analysis, corrected to the 
heliocentric system. For galaxies with multiple measurements, these values are the 
mean weighted value; (12) -  Membership flag.\\ (The table is available in its entirety 
machine readable form in the online journal. A portion is show here for guidance regarding 
its form and content.)}
\end{deluxetable*}

The redshifts of 195 out of 204 galaxies from the original sample, 
corrected to the heliocentric reference frame, the corresponding errors, the 
magnitudes and other relevant parameters are shown in Table \ref{tab:galcat}.
Our final catalog contains the redshift of 185 galaxies observed with GMOS and 
CFHTMOS and in common with \citetalias{gioia1998} and 10 galaxies at the redshift 
of the cluster from \citetalias{gioia1998} not covered by our observations. The 
redshifts of these galaxies, corrected by the zero-point shift, are listed at 
the end of the Table \ref{tab:galcat}. We have excluded from the final catalog 
9 foreground and background galaxies from \citetalias{gioia1998}. 
The final sample contains 30 galaxies with more than one redshift estimation and
with differences in velocity $|\Delta V| \le 250$ \kms~: 9 galaxies with 3 measurements 
and 21 galaxies with 2 measurements. There are 107 galaxies located inside the  
GMOS area and 135 galaxies have their spectroscopic redshifts determined for the first 
time (49 with GMOS and 86 with CHFTMOS). Of these, 44 are located at redshift of the 
cluster, increasing by a factor of $\gtrsim 2$ the number of cluster members 
with known redshifts.

\begin{figure}[t!]
\centering
\includegraphics[width=0.98\hsize]{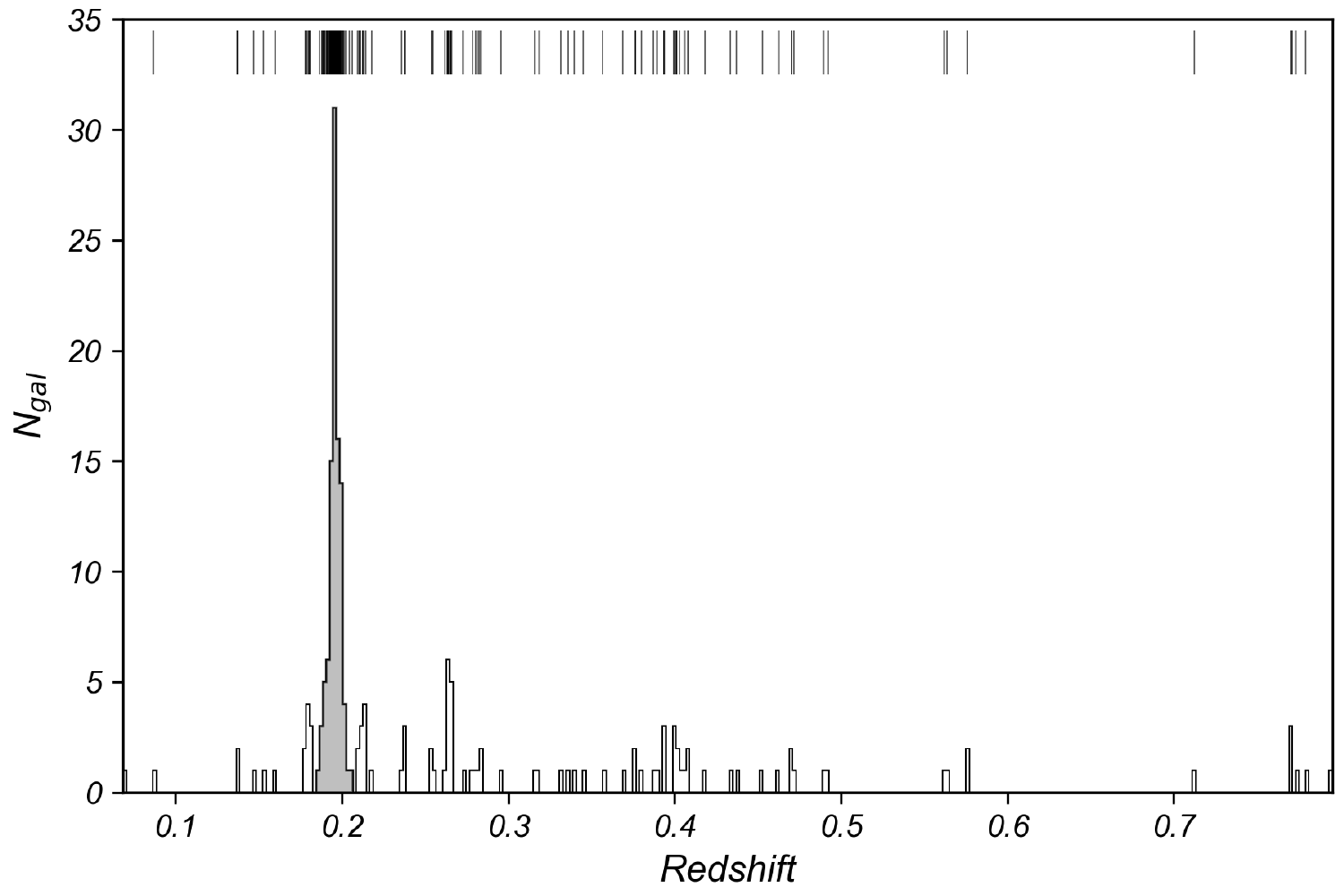}
\caption{Redshift distribution of the 195 galaxies with secure spectroscopic
redshifts in the field of \object{MS\,0440.5$+$0204} (bin width of $\Delta(z)=0.002$). The
gray region indicates the galaxies at the redshift of the cluster ( $z \thickapprox
0.196$) within $\pm 4000$ \kms ($0.183 \lesssim z \lesssim 0.209$). The tick marks at the top 
of the figure represent  the redshift of individual galaxies. \label{fig:histoall}}
\end{figure}

\section{Lensing modeling}\label{sec:lensmod}

The detailed lensing analysis of MS\,0440.5+0204 is presented in \citetalias{verdugo2020}. 
In this section we provide a brief explanation of the method and highlight the main results.

\subsection{Weak lensing}\label{sec:weaklen}

Weak lensing was performed using the methodology described in
\cite{foex12}. 
The lensed sources were selecting after removing galaxies located in the red
sequence, estimated in the \sloanr~$-$ \sloangr~ diagram, down to a magnitude of
\sloanr$=23\,\mathrm{mag}$. Galaxies brighter than \sloanr$=21\,\mathrm{mag}$
and fainter than \sloanr$=25\,\mathrm{mag}$ were also removed. These cuts led to
a number density of $\sim12\,\mathrm{arcmin^{-2}}$.

The shape of the background galaxies was estimated with the software {\sc
Im2shape} \citep{bridle02}, as done in \cite{foex12,foex13}. The lensing
strength (i.e. the factor needed to convert shear into mass), was obtained using
the photometric redshifts from the T0004 release of the CFHTLS-DEEP survey that
were computed with {\sc HyperZ} \citep{bolzonella2000}. We found an averaged
lensing strength $\beta=<D_{ls}/D_{os}>=0.66$, where $D_{ls}$ and $D_{os}$ are
the angular-diameter distances lens-source and observer-source, respectively
(see \citealt{foex13} for a detailed discussion on the $\beta$ factor).

The shear profile was computed in logarithmically-spaced annuli, fitted from 100
kpc to 2.5 Mpc, and using two parametric mass models: a singular isothermal
sphere (SIS), and a Navarro-Frenk-White \citep[NFW,][]{nfw96, nfw1997} model. 
Both models provide a good fit, with a reduced
$\chi^2=9.1/8$ and $\chi^2=8.3/7$ for the SIS and NFW profiles, respectively.
For the SIS model we obtain a velocity dispersion of $\sigma_{v}=894^{+52}_{-61}$ \kms and
for the NFW profile we obtain a $M_{200}$ = 3.9$^{+1.4}_{-0.8}$
$\times10^{14}\,M_{\odot}$ at a radius $r_{200}=1.46\pm0.13$~\hval~Mpc, and a
concentration parameter of $c_{200}=9.3^{+4.8}_{-3.5}$. The concentration
parameter is much larger than expected for a cluster having a mass of
$M_{200}\sim4\times10^{14}\,M_{\odot}$ when the value is compared to the 
$c(M)$ simulation predictions  at $z=0.2$ \citep[e.g][]{Duff08, klypin2016}. 
However, the derived value is in agreement with the strong lensing results (see next
section), and in fairly agreement, within the errors, with the $c(M)$ relation of 
\cite{foex14} obtained by combining stacks of strong-lensing galaxy groups and
clusters.

Finally, to characterize the 2D shape of the cluster from the weak lensing data,
we followed the same approach presented in \citet{soucail2015}, , which is based 
in the software {\sc LENSENT2} \citep{marshall2002}, that uses the shape of each 
background galaxy as a local estimator of the reduced shear. The method, employing 
an entropy-regularized maximum-likelihood allow us to generate a projected mass map 
of the cluster. 

\subsection{Strong lensing}\label{sec:stronglen}

The three strong lensing models presented in \citetalias{verdugo2020} are
performed using the last version of the LENSTOOL ray-tracing code and the
Bayesian  Markov chain Monte Carlo (MCMC) method \citep{jullo2007}. We model the
dark matter component of MS\,0440.5+0204 using first a single large scale clump,
and then adding smaller-scale clumps, as galaxy scale perturbation that are
associated with individual cluster galaxies (the thirteen central galaxies, see
inset in Fig. \ref{fig:colorfig}).  We use a single clump description given the
lack of a second luminous sub-clump and the elliptical pattern of the lensed
arcs. In addition, the single cluster model was accurate enough. For the main
large scale clump we adopt a NFW density profile \citep{nfw96,nfw1997}, which is
characterized by seven parameters: the center position, ($X,Y$), the ellipticity
$\epsilon$, the position angle $\theta$, the velocity $\sigma_s$ and the scale
radius $\mathrm{r}_s$.

The velocity dispersion in the line-of-sight is related to the mass and the
density profile  through the expression

\begin{equation}\label{eq:sigmaNFW_pro}
\tilde{\sigma_s}^{2}   = \frac{2G}{\Sigma(r_s)}\int_{r_s}^{\infty}  (\sqrt{1-(r_s/r)^2}) M(r)\rho_s(r) \frac{dr}{r},
\end{equation}

\noindent where $\rho_s$ is the characteristic density, and $\Sigma(r_s)$ is the
characteristic surface mas density. In case of an isotropic velocity,
Eq.\,\ref{eq:sigmaNFW_pro} is equivalent to the $\sigma^{2}_{p}$ (Eq.\,7) used
in \citet{ver2011}, and represents the velocity dispersion at radius $r_s$
calculated analytically from the mass density profile (see
\citetalias{verdugo2020}). In our model, the velocity $\sigma_s$ is compared
with the velocity dispersion obtained in section\S, \ref{sec:redcluster}, adding
an extra constraint.

The smaller-scale clumps associated with the galaxies are modeled using a pseudo
isothermal elliptical mass distribution (PIEMD). A clump modeled with this
profile is characterized by the seven following parameters: the center position,
($X,Y$), the ellipticity $\epsilon$, the position angle $\theta$, and the
parameters, $\sigma_0$, $\mathrm{r}_{core}$, and $\mathrm{r}_{cut}$
\citep[see][]{Limousin2005, elr07}. The parameters $\sigma_0$ and
$\mathrm{r}_{cut}$ are scaled as a function of their galaxy luminosities; using
as a scaling factor the luminosity $\mathrm{L}_{*}$, associated with the
$\mathrm{r}'$ magnitude of the central galaxy 100606 (A) (see inset in
Figure\,\ref{fig:colorfig} and Table\,\ref{tab:galcat}). While the parameter
$\mathrm{r}^{*}_{core}$ is fixed at 0.15 kpc.

Besides, some parameters describing the dark matter halos associated with two
individual galaxies, namely 896 (B) and 599 (D) (see insets in
Figure\,\ref{fig:colorfig}), were allowed to vary in the optimization procedure,
as they perturb some arclets. Our models are computed and optimized in the image
plane with 17 free parameters: \{$X$,  $Y$, $\epsilon$,  $\theta$, $\mathrm{r}_s$,
$\sigma_{s}$\} for the main halo, \{$\mathrm{r}_{cut}^{*}$, $\sigma^*_0$\}  for
the smaller-scale clumps, \{$\epsilon$,  $\theta$, $\mathrm{r}_{cut}$,
$\sigma_0$\} for galaxies 896 (B) and 599 (D), and \{$\mathrm{z}_{M5}$\} for the arc
system M5. All the parameters are allowed to vary with uniform priors. As we
discuss in \citetalias{verdugo2020}, we tried to include system M4 in the
calculations, leaving the redshift as a free parameter, but we were
unsuccessfully to reproduce the configuration. Therefore, we exclude this system
from our final models.

The three models, \textit{$\mathrm{M}_{\mathrm{lens}}$}, which uses only arc's
positions, \textit{$\mathrm{M}_{\mathrm{lens}-\sigma_s}$}, where we add the
velocity dispersion of the cluster as additional constraints (see section \S \ref{sec:redcluster}), 
and \textit{$\mathrm{M}_{\mathrm{lens}-\sigma_s-\mathrm{mass}}$}, which include the
weak lensing mass at $r_s$, reproduce equally good the image positions of the
arcs, with similar $\chi^{2}/_{DOF}$ and \textit{rmsi}). However, the inclusion
of the additional constraints such as the velocity dispersion and the mass
\textit{$\mathrm{M}_{\mathrm{lens}-\sigma_s-\mathrm{mass}}$} model) gives more
stringent constraints than those provided by the other two models. Despite that
the three models  reproduce well the strong lensing features presented in
MS\,0440$+$0204, at large scale the performance of
\textit{$\mathrm{M}_{\mathrm{lens}-\sigma_s-\mathrm{mass}}$} model is slightly
better (see section \S 5.1 in \citetalias{verdugo2020}). For our best model,
\textit{$\mathrm{M}_{\mathrm{lens}-\sigma_s-\mathrm{mass}}$} we obtain a mass of
$\mathrm{M}_{200}$ = 3.1$^{+0.6}_{-0.6} \times 10^{14}$ \hvaltwo~\Msol~and a 
concentration parameter of $\mathrm{c}_{200}=9.3^{+2.2}_{-1.4}$, consistent with 
the values obtained from the weak lensing analysis.

\section{Data analysis and discussion}\label{sec:analysis}


\subsection{Completeness of the spectroscopic sample}\label{sec:completeness}

We used the photometric and spectroscopic catalogs constructed from 
the observations inside the $\sim 0.45$\degr $\times$ $0.14$\degr~region, 
defined by the CFHTMOS observations, to determine the completeness of the 
spectroscopic sample. Overall the completeness fractions inside the observed 
area are $\sim 50$\% for galaxies brighter than \sloanr~$=20.5$ mag and 
drop to $\sim 19$\% for \sloanr~$=21.5$ and to $\sim 13$\% for \sloanr~$=22.5$ 
The completeness fractions in the fainter magnitude bins are much higher in the 
inner 3\farcm5 region of the cluster ($\lesssim$ 0.66 \hval~Mpc, roughly 
the GMOS field of view). Indeed, of the galaxies selected for spectroscopy, 
we were able to determine the redshifts for $\sim 77$\% of those 
brighter than \sloanr~$=20.5$ and for $\sim 56$\% and $\sim 38$\% at 
\sloanr~$=21.5$ and at \sloanr~$=22.5$, respectively.

In addition to the above completeness estimation, we also compared the radial 
distribution of the galaxies with measured redshifts, to the galaxies selected 
for spectroscopy based on the color–magnitude relation (bottom panel in Fig. \ref{fig:histmem}).
We calculated the completeness fraction comparing the ratio of the number of galaxies 
with measured redshifts, to the number of galaxy candidates without, for six different 
radial bins of 1\farcm5 wide ($\sim 292$ \hval~kpc) and centered at the cluster center. 
Table \ref{tab:completeness} shows the spectroscopic completeness fraction for the six 
radial bins and for different magnitude intervals. The radial distribution 
shows that the completeness fraction decrease toward the outskirt of the cluster 
as expected. In particular, for the 4\farcm5 bin, corresponding to a physical radius 
of $\sim 0.88$ \hval~Mpc (roughly $\sim0.5\times$\rtwo, see section \S \ref{sec:mass} 
below), the completeness fractions are $\sim 70$\% at \sloanr~$=20.5$, $\sim 50$\% at 
\sloanr~$=21.5$ and $\sim 38$\% at \sloanr~$=22.5$.
Based on the above results, we believe that our spectroscopic sample represents well 
the galaxy population in \object{MS\,0440.5$+$0204}.

\begin{deluxetable}{ccccccc}[t!] 
\tablefontsize{\footnotesize}
\tablecolumns{7}
\tablecaption{Completeness fraction\label{tab:completeness}}
\tablehead{
\colhead{Magnitude} & \multicolumn{6}{c}{\textit{f(\sloanr)}} \\
\colhead{(bin)} & \colhead{1\farcm5} & \colhead{3\farcm0} & \colhead{4\farcm5} &
\colhead{6\farcm0} & \colhead{7\farcm5} & \colhead{9\farcm0} \\
\colhead{(1)} & \colhead{(2)} & \colhead{(3)} & \colhead{(4)} & \colhead{(5)} & \colhead{(6)}
& \colhead{(7)}}
\startdata
16.5 - 17.5 & 1.00 & 1.00 & 1.00 & 1.00 & 1.00 & 1.00 \\
17.5 - 18.5 & 1.00 & 1.00 & 1.00 & 1.00 & 1.00 & 1.00 \\
18.5 - 19.5 & 0.83 & 0.88 & 0.86 & 0.89 & 0.83 & 0.80 \\
19.5 - 20.5 & 0.89 & 0.94 & 0.70 & 0.66 & 0.66 & 0.62 \\
20.5 - 21.5 & 0.74 & 0.74 & 0.48 & 0.37 & 0.32 & 0.28 \\
21.5 - 22.5 & 0.54 & 0.48 & 0.34 & 0.27 & 0.21 & 0.16 \\
\enddata
\end{deluxetable}

\subsection{Redshifts distribution}\label{sec:redcluster}

The histogram of the redshift distribution for the 195 galaxies in the field of 
\object{MS0440.5$+$0204} ($\sim$ 0.45\degr $\times$ 0.14\degr) 
is shown in Figure~\ref{fig:histoall}. Ninety six galaxies lie in the 
$0.183 \lesssim z \lesssim 0.209$ redshift interval (within $\pm\, 4000$ \kms~ 
around $z \thickapprox0.196$, gray shaded region in the figure).  We 
estimated the average redshift and the one dimensional line-of-sight 
velocity dispersion and the number of member galaxies of the cluster using 
the robust bi-weight estimators of central location ($C_{BI}$) and scale 
($S_{BI}$) of \citet{beers1990}. We calculated the average redshift and 
the one dimensional line-of-sight velocity dispersion and the number of 
member galaxies of the cluster using the robust bi-weight estimators of 
central location ($C_{BI}$) and scale ($S_{BI}$) of \citet{beers1990}. 
An iterative procedure was used to determine the location and scale 
with the program ROSTAT and applying a 3-$\sigma$ clipping algorithm 
to remove outliers. The procedure was repeated until the velocity dispersion 
converged to a constant value (after two iterations). The best estimates of
the location ($\overline Z$) and scale (\slos) are shown in Table \ref{tab:dympar}. 
The table also shows the number of member galaxies, $N_{\mathrm{mem}}$, the maximum 
radius at which we have spectroscopic members, $r_{\mathrm{max}}$, and other relevant 
dynamical parameters derived in section \S \ref{sec:mass}. It is worth noting that 
the calculated \slos~ is lower than the value reported by \citetalias{gioia1998} 
($872^{+124}_{-90}$ \kms)  and the value derived in \citetalias{verdugo2020} 
(see section \S \ref{sec:weaklen}), but agrees well, within the quoted uncertainties.

\begin{figure}[ht!]
\centering
\includegraphics[width=0.8\hsize]{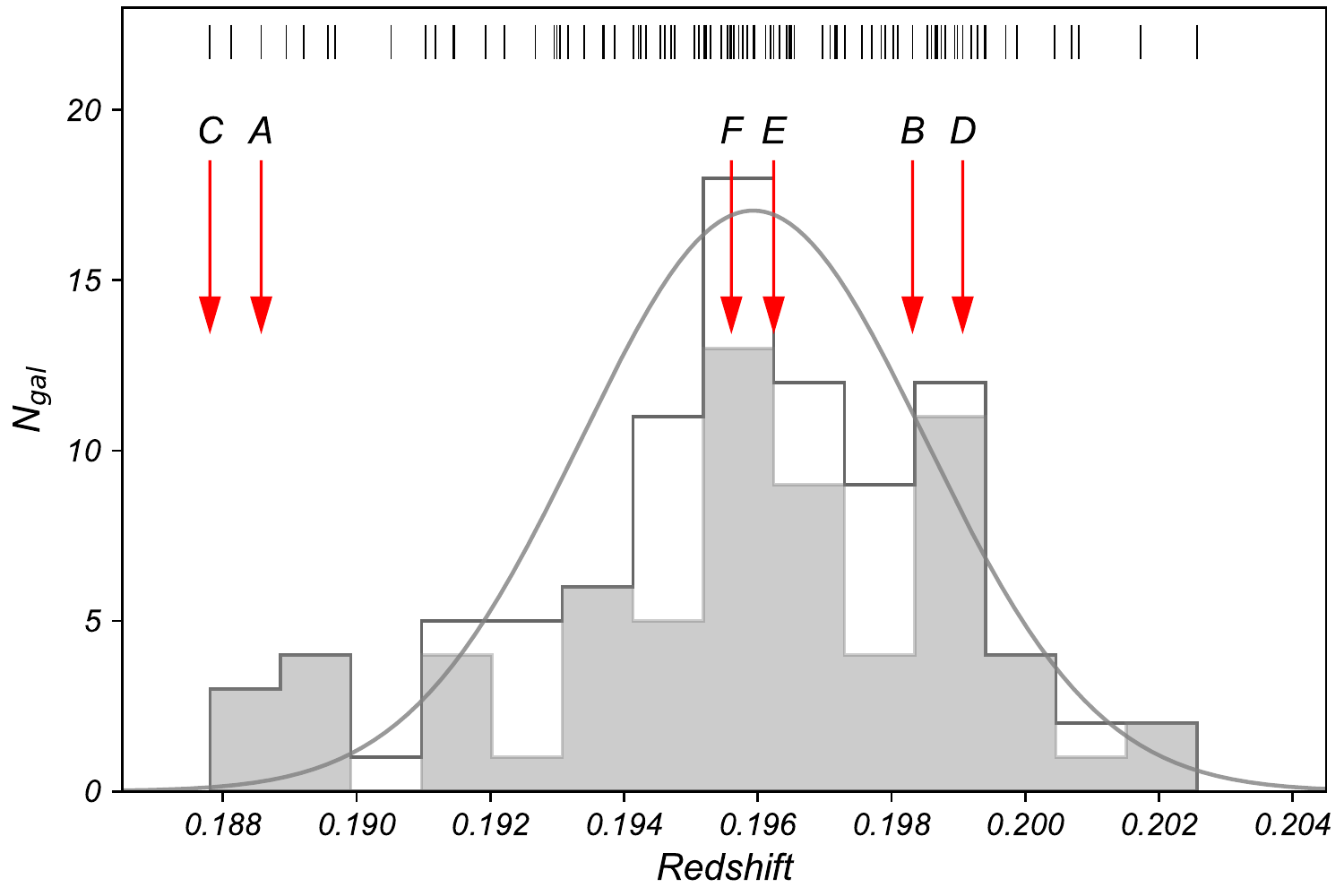}
\includegraphics[width=0.8\hsize]{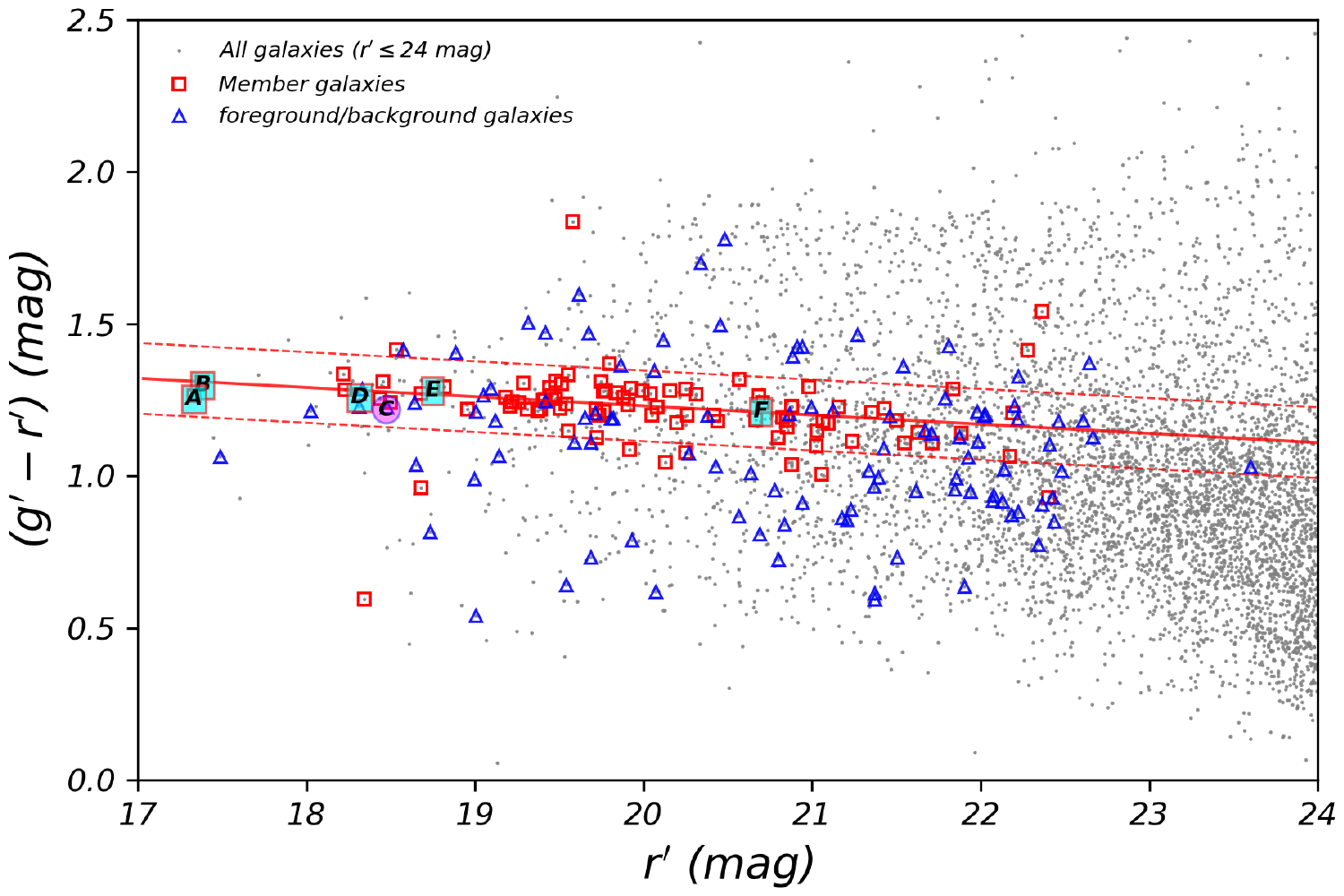}
\caption{\textit{Top panel:} Open histogram - redshift distribution of the 94
confirmed member galaxies. Gray shaded histogram - redshift distribution of 
67 member galaxies located inside a radius $0.5 \times r_{200}$. The arrows
indicate the location of the six central elliptical galaxies hosted by the BCG.
\textit{Bottom panel:} Color-magnitude diagram for all galaxies with \sloanr $\leq24$
mag (small gray dots) detected on the CFHT/Megacam images within an area of
$0.5\degr \times 0.5\degr$. The open squares (red) represent the member galaxies
and the open triangles (blue) the foreground/background field galaxies. The
location of the six central elliptical galaxies is represented by a letter with cyan
background. The solid line shows the best fit for member galaxies lying in the 
red cluster sequence and the dashed lines the 1-$\sigma$ errors of the fit.
\label{fig:histmem}}
\end{figure}

The histogram of the redshift distribution of member galaxies is shown in the
upper panel of Fig.\ref{fig:histmem}. The open histogram shows the distribution
of spectroscopic confirmed member galaxies inside the surveyed area.
Sixty seven member galaxies ($\sim 68$\% of the sample) are located within $0.5
\times r_{200}$ radius (gray histogram), roughly the region covered by the GMOS
observations. Note that the central elliptical galaxy ``C'' (894) is formally
rejected as a cluster member by the 3-sigma clipping algorithm. The galaxy is at
the limit of the redshift distribution in Figure \ref{fig:histmem} and it is
close to the bright elliptical galaxy ``A'' (100606), and could be a  member of
a small group of galaxies located behind the cluster.  Although it is not
formally a member of the cluster,  galaxy ``C''(894) has a line-of-sight
velocity (see Table \ref{tab:centgal}) below the escape velocity of the cluster
of $v_{\mathrm{esc}} = 2334^{+443}_{-350}$\kms, calculated using the
\masstwo~and \rtwo~determined in section \S \ref{sec:mass}\footnote{A
cluster member can be defined as a galaxy with line-of-sight
velocity lower than the escape velocity, where $v_{\mathrm{esc}} \simeq
(\mathrm{M}_{200}/10^{14} \mathrm{h}^{-1} \mathrm{M}_{\odot})^{1/2}\,
(\mathrm{r}_{200}/\mathrm{h}^{-1} \mathrm{Mpc})^{-1/2}$ \citep{diaferio1999}}.
Therefore, hereinafter, galaxy ``C''(894) is included in the analysis, thus
extending to 94 the number of cluster galaxies.

The lower panel in Fig. \ref{fig:histmem} shows the color-magnitude diagram
(CMD) of all galaxies brighter than \sloanr~$=24$ mag detected inside an area of
$0.5\degr \times 0.5\degr$ ($\sim 5.8 \times 5.8$ \hvaltwo~${\rm Mpc}^{2}$). The
majority of member galaxies are located in the region defined by the linear CMD
relation for early-type galaxies in clusters; the Red Cluster Sequence (RCS)
\citep{gladders1998, gladders2000}. Using a standard linear regression plus an
iterative 3-$\sigma$ clipping algorithm to remove outliers, we have determined a
RCS slope of $-0.030\pm0.006$, with an intercept of $1.835\pm0.116$ (red solid
line in CMD in Fig. \ref{fig:histmem}). The number of galaxies following the RCS
and within $\pm$1-$\sigma$~ (dashed lines) is 81 (80\% of the sample), while the
number of galaxies bluer and redder than the slope are 8 (9\%) and 5
(5\%), respectively. This is not surprised since the result is a consequence of
how the galaxies were selected for spectroscopic follow-up. We used the
information provided by the CMD to analyze the spatial distribution of the
member galaxies and the possible connection of \object{MS\,0440.5$+$0204} with
other nearby structures. This point is discussed in detail in section \S
\ref{sec:environment}.

\subsection{Cluster mass}\label{sec:mass}

We have computed the dynamical mass of the cluster using the $\sigma$-\masstwo~
scaling relation of \citet{munari2013} obtained from zoomed-in hydro-dynamical
simulations of Dark Matter (DM) halos calibrated using DM particles and taking 
into account prescriptions for cooling, star formation, and Active Galactic 
Nuclei (AGN) feedback:

\begin{equation}
\sigma_{1D} = A_{1D}  \left[\frac{h(z)\,M_{200}}{10^{15} M_{\odot}}\right]^{\alpha}
\label{eq2}
\end{equation}

\noindent where $\sigma_{1D}$ is the one-dimensional (1D) velocity dispersion,
$A_{1D}=1177\pm4.2$ \kms, $\alpha = 0.364\pm0.002$, $h(z) = H(z)/70$
\kms~Mpc$^{-1}$, and \masstwo~ is the mass within \rtwo\footnote{\rtwo~is the
radius where the over-density is 200 times the critical density of the universe
and is defined as \rtwo$= (\sqrt{3}\,\sigma_{\rm los})/ (10\,H(z))$
\citep{carlberg1997}}. For our calculation, we assume the \stwo~computed
within the  \rtwo~as a proxy for the 1D velocity dispersion. The values for the
mass \masstwo, the radius \rtwo,  the number of member galaxies inside \rtwo,
$\mathrm{N}_{200}$, and  the \stwo~are listed in Table \ref{tab:dympar}.


\begin{deluxetable}{lc}[ht!]
	\tabletypesize{\scriptsize}
	\tablewidth{0pc}
	\tablecolumns{2}
	\tablecaption{Cluster dynamical parameters\label{tab:dympar}}
	\tablehead{
		\colhead {Parameter} & \colhead {Values}}
	\startdata
	$RA\,(2000)$ & 04\hh43\hm10\fs04  \\
	$DEC\,(2000)$ & $+$02\degr10\arcmin20\farcs21 \\
	$\overline{Z}$  & $0.195929^{+0.00033}_{-0.00031}$ \\
	$N_{mem}$\tablenotemark{a} & 94 \\
	$N_{200}$\tablenotemark{a} & 86 \\
	\slos~(\kms) & $771^{+63}_{-71}$ \\ 
	$r_{max}$ (\hval~Mpc) & 2.59  \\
	\stwo~(\kms) &  $807^{+56}_{-81}$ \\
	\rtwo~(\hval~Mpc) & $1.73^{+0.14}_{-0.16}$   \\
	\masstwo ($10^{14}$ \hvaltwo~\Msol) &  $3.66^{+0.67}_{-0.59}$  \\
	$\mathrm{M}_{x} (< r_{200})$ ($10^{14}$ \hvaltwo~\Msol) & $2.99^{+0.61}_{-0.46}$  \\
\enddata
\tablecomments{All quoted errors are at the 68\% confidence level (1-$\sigma$).}
\tablenotetext{a}{includes the galaxy ``C''(894) (see section \S \ref{sec:redcluster})}
\end{deluxetable}


We also estimated the total X-ray mass at \rtwo~radius assuming that the gas is 
isothermal and in hydrostatic equilibrium with a density distribution that 
follows a  circular single $\beta$-model. The X-ray mass at any radius can
be computed using the following relation \citep{evrard1996}:

\begin{equation}
\label{eq:xraymass}
\mathrm{M}_{X} (< \mathrm{r}) = 1.13 \times 10^{14} \beta \frac{\mathrm{T}_X}{\mathrm{keV}} \frac{\mathrm{r}}{{\mathrm{Mpc}}}  \frac{(r/r_c)^{2}} {[1 + (r/r_c)^2]}\,h_{70}^{-1}\,\mathrm{M}_{\odot}
\end{equation}

\noindent where $\mathrm{T}_{X}$ is the isothermal  X-ray cluster temperature,
$\mathrm{r}_c$ is the core radius in the $\beta$-model, and $\beta$ is the model
parameter. For the X-ray mass calculation we assume an average values of 
$\beta=0.45\pm0.03$, $\mathrm{r}_{c}=0.03\pm0.01$ \hval~Mpc and 
$\mathrm{T}_{X} = 3.4\pm0.4$ keV \citep{shan2010, mahdavi2013}. 
The resulting  X-ray mass is given in Table \ref{tab:dympar}. 

We also computed the virial mass following the prescription of
\citet{heisler1985}. Using the classical estimator for the mass and the radius,
we obtained a virial mass of $\mathrm{M}_{vir} = 3.75^{+0.38}_{-0.25} \times
10^{14}$ \hvaltwo~\Msol~at a radius of \rvirial$=1.34^{+0.11}_{-0.05}$ \hval~Mpc.
It is important to remember that the virial mass calculation assumes that the
system is self-gravitating, the galaxies have equal masses, and relies on the
velocities of the galaxies, the mean velocity of the system, and the projected
separation between the galaxies \citep[eq. (4) in][]{heisler1985}.
 
The estimated \masstwo~mass is in a good agreement, within the uncertainties, with 
other mass estimates. The mass is $\sim18$\% larger that the X-ray mass and the
mass derived from our best-fit lensing model \textit{$\mathrm{M}_{\mathrm{lens}-\sigma_{s}-\mathrm{mass}}$}, 
$\sim 6$\% smaller than the \masstwo~derived from weak lensing and $\sim 8$\% smaller 
than the virial mass (see also section \S 5.4 in \citetalias{verdugo2020}).

\subsection{Substructures}\label{sec:substructure}

In this section we analyze the dynamical state of \object{MS\,0440.5$+$0204} by 
examining the redshifts and the projected distributions of the member galaxies.

\subsubsection{Line-of-sight substructures}\label{sec:1Dsub}

\begin{figure*}[ht!]
\centering \includegraphics[width=0.95\hsize]{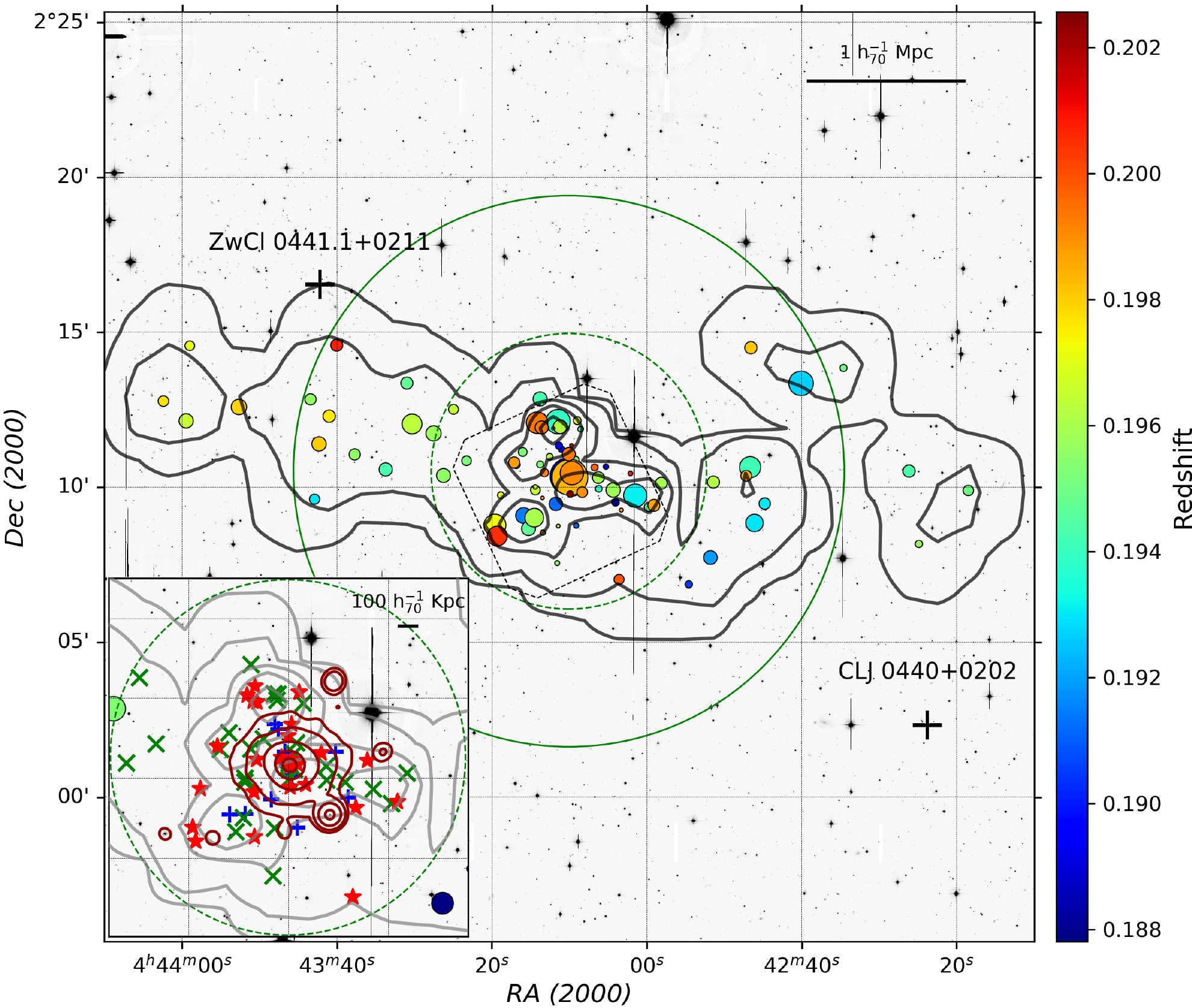}
\caption{Projected galaxy density map of the 94 cluster members. The adaptive
kernel density map is superimposed on the \sloanr\ band CFHT image
(black contours). The size of the image is 0.5\degr $\times$ 0.5\degr ($\sim 5.2 \times 5.2$ 
\hvaltwo\,Mpc$^2$ at the cluster distance). The colored circles and
the bar denote the positions and redshifts of the member galaxies. The size of
each circle is proportional to the luminosity of the galaxies. The GMOS field of view
is represented by a dashed square at the center of the image. The green solid and
dashed large circles represent the \rtwo~and 0.5$\times$\rtwo~radii,
respectively. The big pluses indicate the location of the two known clusters
inside the surveyed area. The inset in the lower left corner shows the central 
0.5$\times$\rtwo~region of the cluster with the Chandra X-ray emission
(brown contours) and the adaptive kernel density map (gray contours)
overlaid. The different symbols show the location of the galaxies linked to the 
structures  S1 (blue pluses), S2 (green crosses) and S3 (red stars), respectively. 
\label{fig:denmap_mem}}
\end{figure*}

The histogram in Fig.~\ref{fig:histmem} shows two prominent peaks in the
distribution, one located at $z \sim 0.196$ and another located at $z \sim
0.199$. The histogram also show a prominent tail toward the lower redshifts
(between $0.0188 < z < 0.192$). This is more evident for the 67 galaxies (gray 
histogram in Fig. \ref{fig:histmem}) located inside the $0.5 \times$\rtwo~radius
(the central overdensity in Fig. \ref{fig:denmap_mem}). A first indication of the existence 
of a multi-modal distribution in the redshift space using all member galaxies is provided 
by the different statistical tests implemented in the {\rm ROSTAT} program. All
statistical tests, such as the Watson U$^2$ and Anderson-Darling A$^2$, reject
the hypothesis of a single Gaussian distribution at the significant level of
94\% in average. Note that some member galaxies distributed toward the
East and outside the $0.5\times$\rtwo~region, seem to be located in a
filament (see Fig.~\ref{fig:denmap_mem}). Thus, this structure, as well as others 
located West and South-West (see section \ref{sec:2Dprojected}) could bias the 
1D-analysis and hence provide a wrong interpretation of the results. To have a more 
realistic picture of the  dynamical state of the cluster, we have restricted the 
1D-analysis to central $0.5 \times$\rtwo~region of the cluster 
(i.e. within $\mathrm{r}\le0.866$ \hval~Mpc in radius).

We used the Gaussian mixture modeling analysis implemented in the GMM code by
\citet{muratov2010} to investigate in detail the structures identified in 
Fig.\ref{fig:histmem} and quantify the significance of the multi-modal distribution.
The GMM code fit different Gaussian mixtures modes and compare them to an
unimodal distribution. The robustness in the multimodal distribution provided by
GMM  is based on three parameters: the kurtosis of the distribution
(\textit{k}), the maximum log-likelihood of the model convergence ($\log
\lambda$), and the separation between the means relatively to their widths $DD$.
A deviation from an unimodal distribution is statistical significant when
$DD>2$, $k<0$ and the log-likelihood values are grater than that for a 
unimodal fit. The errors of the output parameters are estimated using 
non-parametric bootstrapping, and the confidence intervals at which an 
unimodal distribution can be rejected are calculated using parametric 
bootstrapping.

\begin{deluxetable}{cccc}[t!]
\tabletypesize{\scriptsize}
\tablewidth{0pc}
\tablecolumns{4}
\tablecaption{Redshifts and \slos~of the line-of-sight structures\label{tab:1Dsub}}
\tablehead{
\colhead{Structure} & \colhead{N$_{gal}$} & \colhead{$z$} & \colhead{$\sigma$}  \\
\colhead{(1)} & \colhead{(2)} & \colhead{(3)} & \colhead{(4)}}
\startdata
S1 & 12 & $0.189811\pm0.001189$ & $379\pm186$ \\
S2 & 30 & $0.195350\pm0.000964$ & $397\pm172$  \\
S3 & 25 & $0.198903\pm0.001084$ & $416\pm172$  \\
\enddata
\tablecomments{The meaning of the columns are: col. (1) - Structure
identification; col2 (2) - number of associated galaxies with assigned probability
$P>0.6$; col (3) - average redshift and the associated error; col (4) $\sigma$ 
and the associated error in \kms.}
\end{deluxetable}

\begin{figure}[h!]
\centering
\includegraphics[width=0.9\hsize]{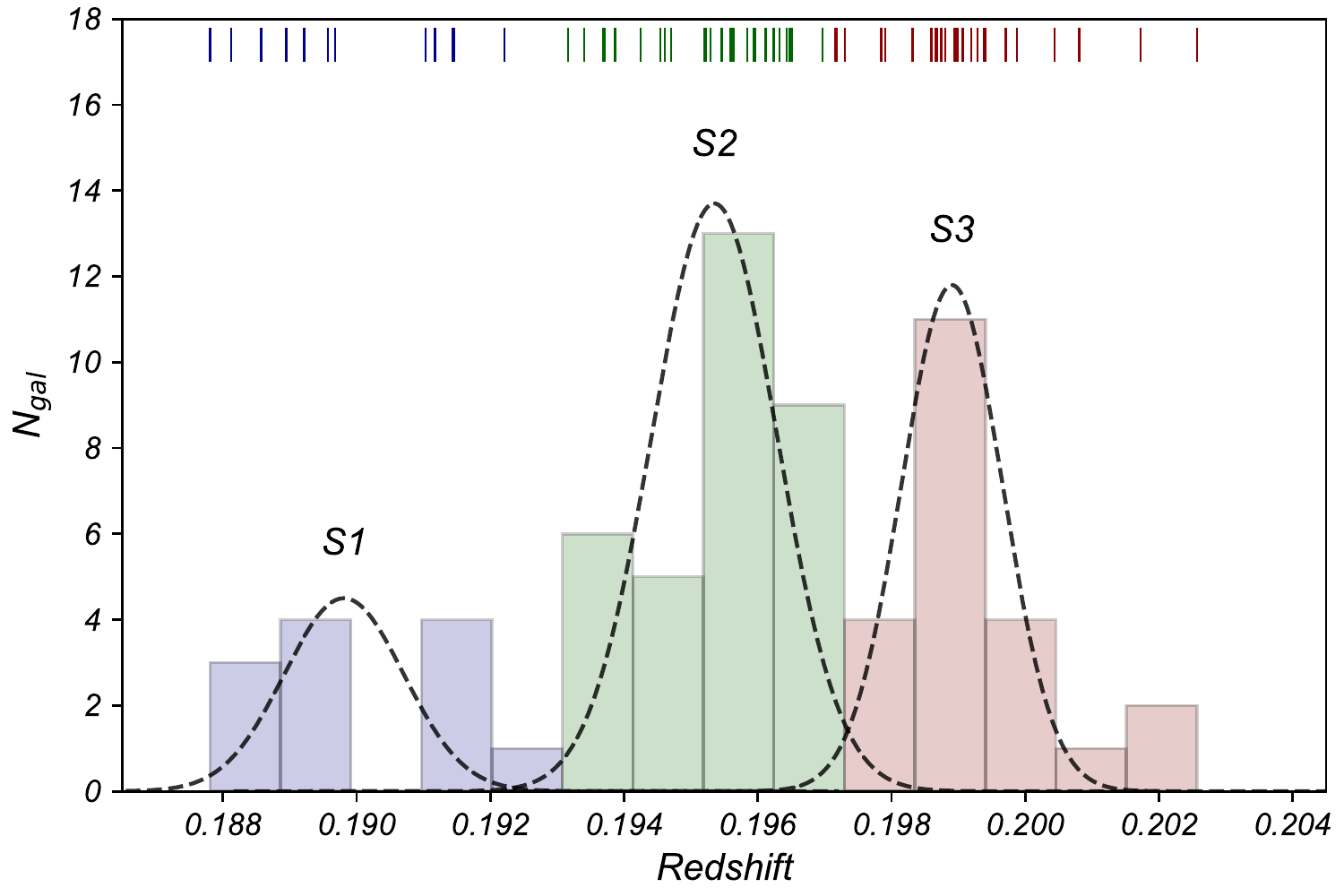}
\caption{Redshift distribution of the 67 member galaxies in the central region 
of \object{MS\,0440.5$+$0204}. The different distributions corresponds to the three 
line-of-sight structures (S1, S2 and S3, respectively) found by GMM with
a confidence level of 98.5\%. The dashed lines indicate the Gaussian fits to the 
respective sub-structures.
\label{fig:1Dsub}}
\end{figure}

We considered whether the redshift distribution of the 67 galaxies is consistent with 
two and three components. In the first case, an unimodal distribution is marginally 
rejected, at the confidence level of 87\% (p-value of 0.13), with no clear separations
($DD=2.93\pm1.17$). If we assume that the redshift distribution in Fig.\ref{fig:histmem} 
is indeed tri-modal, the GMM test rejects an unimodal distribution at the confidence 
level of 98.5\% (p-value 0.015), with a clear separation between the peaks 
($DD=4.26\pm1.27$). Therefore, a model with three components is statistically more 
significant than a model with two components. Table~\ref{tab:1Dsub} shows the 
results from the GMM test considering a tri-modal distribution in the redshift space.
The histogram of the redshift distribution for the galaxies assigned to each of the 
three  line-of-sight structure found by GMM is shown in Fig. \ref{fig:1Dsub}.

Another interesting feature is the location of the six components of the BCG
(red arrows in Fig.~\ref{fig:histmem}). These galaxies are grouped in
pairs and close, in redshift, to the mean redshifts derived for the structures S1,
S2 and S3. 
The main parameters of the six central ellipticals (nuclei) are summarized in
Table \ref{tab:centgal}. The relative position (cols. 4 and 5) are measured
respect to the center of the X-ray emission located at 04\hh43\hm09\fs85,
$+$02\degr10\arcmin20\farcs3. The absolute \sloanr\, magnitudes (col. 8) 
are corrected by galactic extinction and K-corrected. The line-of-sight velocity,
i.e. the radial velocity offset from the median redshift of the cluster, for each 
central elliptical galaxy shown in col. (3) is calculated as follow:

\begin{equation}
\Delta V_{i} =  \frac{(z_{i} - \overline{z})}{(1 + \overline{z})}\, \rm{c}
\label{eq:pecvel} 
\end{equation}

\noindent where $z_{i}$ is the redshift of the $i$ galaxy, $\overline{z}$ is the
average redshift of the cluster and $c$ is the speed of light. The two brightest components, 
``A'' and ``B'', have a difference in velocity of $\sim 2600$ \kms. Moreover, the 
pairs ``A - C'', ``E - F'' and ``B - D'' have differences in velocity between them
of $192$ \kms, $159$ \kms~and $188$ \kms, respectively. The small line-of-sight velocity 
separation between the three pairs and the location of the galaxies in the 
histogram indicates that these galaxies (in pairs) are associated to the line-of-sight 
structures S1, S2 and S3, respectively. Interestingly, the two faintest components
``E'' and ``F'' are located only at $\pm 100$ \kms\, from the average redshift of the 
cluster.

\begin{deluxetable*}{lcccrrrc}[t!]
\tabletypesize{\scriptsize}
\tablewidth{0pc}
\tablecolumns{8}
\tablecaption{Principal parameters of the central galaxies \label{tab:centgal}}
\tablehead{
\colhead{Gal.} & \colhead{$V_{hel}$} &\colhead{$\Delta V$} & \colhead{X} & \colhead{Y} & \colhead{X} & \colhead{Y} & $M_{r'}$  \\
\colhead{(1)} & \colhead{(2)} & \colhead{(3)} & \colhead{(4)} &  \colhead{(5)} & \colhead{(6)} & \colhead{(7)} & \colhead{(8)}}
\startdata
A (100606) & 56533$\pm$80 &$-$1845 & $-$3.69 & $+$1.01 & $-$11.99 &  $+$3.27 & $-$22.84 \\
B (896)  & 59453$\pm$80 & $+$597 & $-$1.09 & $-$1.08 &  $-$3.53 &  $-$3.49 & $-$22.80 \\
C (894)  & 56303$\pm$80 &$-$2037 & $+$1.69 & $+$2.26 &  $+$5.48 &  $+$7.35 & $-$21.89 \\
D (599)  & 59678$\pm$80 & $+$785 & $+$2.70 & $+$7.15 &  $+$8.77 & $+$23.23 & $-$21.95 \\
E (616)  & 58831$\pm$123 &  $+$77 & $-$2.76 & $+$6.71 &  $-$8.96 & $+$21.80 & $-$21.55 \\
F (K1)  & 58641$\pm$89 &  $-$82 & $-$1.36 & $+$1.54 &  $-$4.42 &  $+$4.99 & $-$20.29 \\
\enddata
\tablecomments{col. (1) - central component identification; col. (2) and (3) - heliocentric radial velocity 
and line-of-sight velocity in units of \kms; cols. (4) to (7) - relative positions of the galaxies
respect to the X-ray center in arcseconds and kiloparsecs; col. (8) - absolute magnitude 
in \sloanr~corrected by galactic extinction ($A(r^{\prime})=0.406$ mag) and K-corrected.}
\end{deluxetable*}

\subsubsection{Galaxy projected distribution}\label{sec:2Dprojected}

Figure \ref{fig:denmap_mem} shows an adaptive-kernel density map \citep{sil86}
of the cluster members. The member galaxies are predominately distributed along 
the East-West direction. The projected distribution shows at least four overdensities.
The galaxies in the North-East over-density are distributed in a filament that extends 
for $\sim 2$ \hval~Mpc in the direction of another structure, the poor cluster of 
galaxies \object{ZwCL\,0441.1$+$0211} \citep[RA$=$04\hh43\hm42\fs2,
DEC=$+$02\degr16\arcmin33\arcsec~][]{zwicky1965}. This distribution is striking, 
as it suggests that the two structures might be connected. We return to this point 
in section \S \ref{sec:environment}. Another cluster in the area is CLJ\,0440$+$0202. 
However this cluster is a background structure at redshift $z=1.1$ \citep{stern2003}.

Figure \ref{fig:denmap_mem} shows three other overdensities located
$\sim7$\arcmin~SW,  $\sim10$\arcmin~NW and $\sim10$\arcmin~W  from the cluster
center. These overdensities are small groups of 3-9 galaxies that might be 
falling onto the cluster. Our interpretation is based in the location of the 
galaxies belonging to these groups in the projected phase diagram (see Fig. 
\ref{fig:phasediam}) and in the analysis presented in section 
\S \ref{sec:phasediam}.  

The central region of the cluster (lower-left inset in Fig. \ref{fig:denmap_mem}) 
shows at least three other overdensities.
The galaxies linked to the line-of-sight structures S1 and S2 seem to be
distributed randomly and apparently they are not associated to any of these
overdensities. However, there is a weak evidence that the distribution of
the galaxies linked to the  line-of-sight structure S3 (red stars) could be
associated to these overdensities.

\begin{figure}[ht!]
\centering
\includegraphics[width=0.95\hsize]{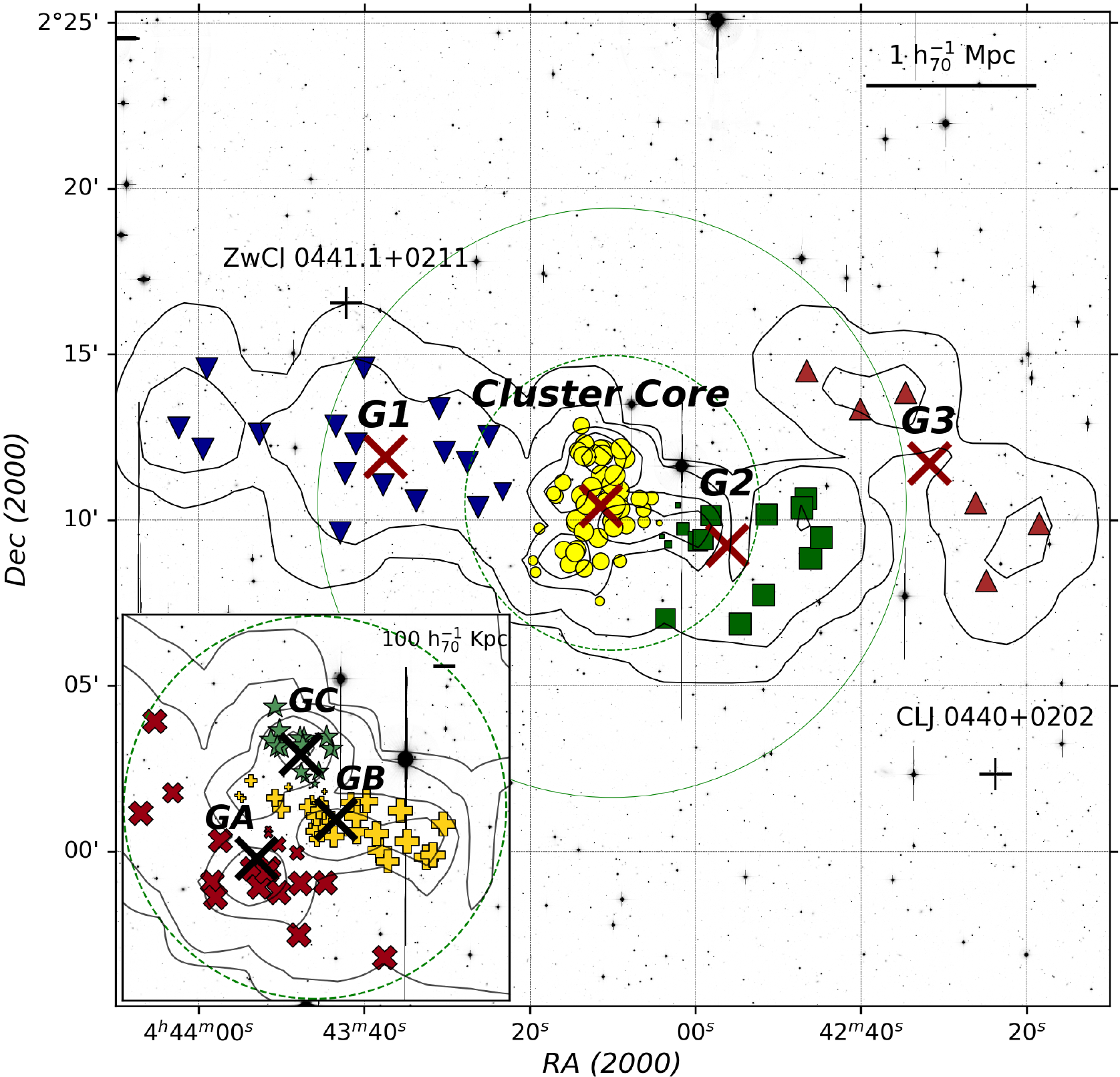}
\caption{Results from the MDGMM test. The contours and the size of the image are the
same as Fig. \ref{fig:denmap_mem}. The different colors and symbols show the
galaxies assigned to each sub-component by MDGMM. The lower-left inset shows the
distribution of the 67 galaxies in the central region of the cluster. The size
of each symbol is proportional to the probability of a galaxy to be a member of
one of this sub-components. The big red and black crosses show the mean location
of each sub-component.  \label{fig:mdgmm}}
\end{figure}

To further investigate the overdensities in Fig.\ref{fig:denmap_mem}, we use the
Multi-Dimensional Gaussian Mixture Model (MDGMM) implemented in the
sklearn.mixture python package \citep[GaussianMixture,][]{scikit-learn}. MDGMM
implements the expectation-maximization (EM) algorithm for fitting mixture of
Gaussian models. The output of MDGMM provides the number of sub-components, the
mean value of the location for each sub-component, predict the labels for the
data sample and provides two different criteria for the number of sub-components
found: the Bayesian information criterion (BIC), and the Akaike information
criterion (AIC). The optimal number of sub-components is the value that minimizes
the AIC or BIC. We run the test assuming initially 1 to 8 components in the
projected distribution of the member galaxies in  Fig. \ref{fig:denmap_mem}. We
looked at the AIC and BIC values as a function of the number of components, and
both BIC and AIC converged  to an identical minimum number of four components.
Figure \ref{fig:mdgmm} shows the location of the galaxies associated to each
sub-component. We can see that there are four sub-components (hereinafter groups)
associated to the overdensities shown in Fig. \ref{fig:denmap_mem}: the group G1 
located NE with 17 galaxies, the cluster core with 67 galaxies, and the groups
G2 and G3 located W with 15 and 6 galaxies, respectively.

\begin{figure}[ht!] 
\centering
\includegraphics[width=1.0\hsize]{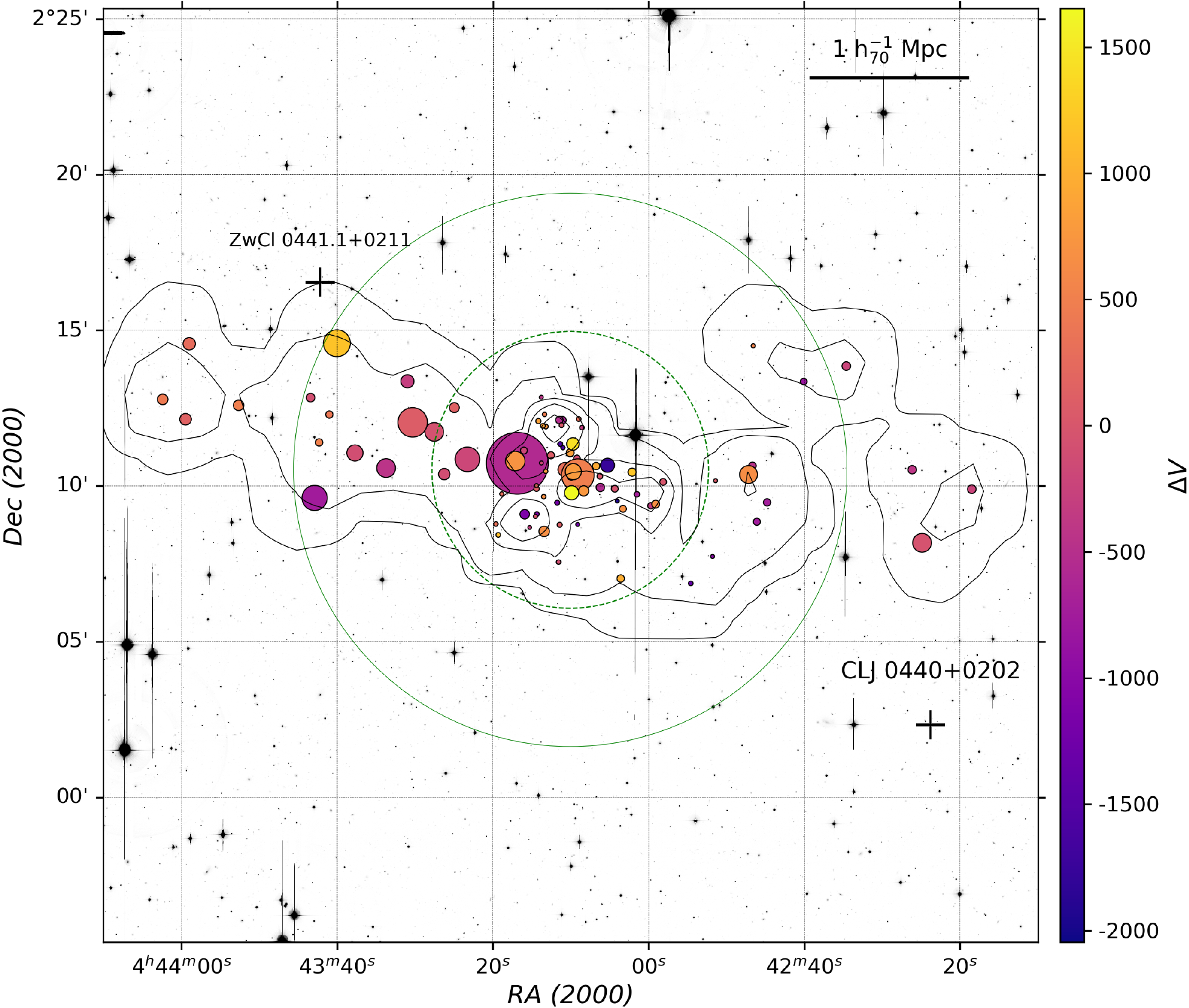} 
\caption{Graphical representation of the DS-test results using the 94 member
galaxies. The size of the image is the same as Fig. \ref{fig:denmap_mem}. The
colored circles and the bar denote the positions and the line-of-sight velocities
in \kms~obtained using equation (\ref{eq:pecvel}), respectively. 
The size of each circle is proportional to $e^{\delta}$, where $\delta$ is the 
DS-test measure of the local deviation (the larger the symbol, the higher is 
the significance of substructure). \label{fig:dstest}}
\end{figure}

We repeated the MDGMM test using the 67 galaxies located in the central
region of the cluster and assigned to the line-of-sight structures S1, S2 and S3, 
assuming the same number of initial components. In this case, both BIC and AIC 
converged to the same minimum of three components. The lower-left inset in Fig. 
\ref{fig:mdgmm}  shows the location and the assigned galaxies to these three 
sub-components. It is remarkable to see how well the galaxies associated to 
these sub-components follow the galaxy density map in the inner region of the 
cluster and at large scale. 

Motivated by the results in the redshift distribution (Fig. \ref{fig:1Dsub}) and
by the existence of different sub-components in the projected distribution 
(Figs. \ref{fig:denmap_mem} and \ref{fig:mdgmm}, we used the Dressler–Shectman 
(DS-) test \citep{dressler1988} to statistically evaluate the likelihood of 
sub-structures using the information about the spatial and kinematical positions 
of galaxies. The DS-test takes a number of $\mathrm{N}_{nn}$ neighbours from each 
galaxy in the projection, estimates the local mean velocity and velocity dispersion 
of the sub-sample, and compares this with the global mean velocity and velocity 
dispersion of the whole sample. The final values are then summed to obtain 
$\Delta=\sum_{i}^{N} \delta_{i}$, where $\Delta/N > 1$ is an indicator of sub-structure. 
A graphical representation of DS-test (Figure \ref{fig:dstest}) reveals two 
substructures, the main cluster at the center of the figure and the eastern substructure
(the largest circle in the figure) corresponding to the galaxies in group G1 in Fig. 
\ref{fig:mdgmm} . Using the ten nearest neighbors ($N_{nn} \approx \sqrt{N}$), we obtained 
$\Delta/N=1.14$. We used the Monte Carlo technique to find  the significance of the 
DS-test. We re-calculated the statistic by randomly shuffling 10,000 times the velocities 
of the galaxies among the cluster members, while holding the positions fixed. The result 
of the statistic shows a $\Delta/N>1$ for the $\sim 60$\% of the trials, implying that 
there are some evidences for substructures that are perturbing the cluster, in agreement 
with 1D- and 2D- results presented above. Apart from the eastern G1 group, the 
DS-test does not reveal the substructures at the center of the cluster. Therefore, 
the strongest evidences of a merging event are coming from three sources: the
redshift distribution (Fig. \ref{fig:1Dsub}), the galaxy projected distribution 
(Figs. \ref{fig:denmap_mem} and \ref{fig:mdgmm}) and the location of the different 
structures in the projected phase-space diagram (Fig. \ref{fig:phasediam}).

The cluster shows a complex structure and appears to be dynamically active,
accreting galaxies and groups from their neighborhoods. This is not surprised,
since there are numerous evidences that clusters are still accreting substructures 
at intermediate and low redshifts \citep[e.g.,][]{gonzalez2005,carrasco2007,dawson2015}.

\begin{figure}[ht!]
\centering
\includegraphics[width=0.95\hsize]{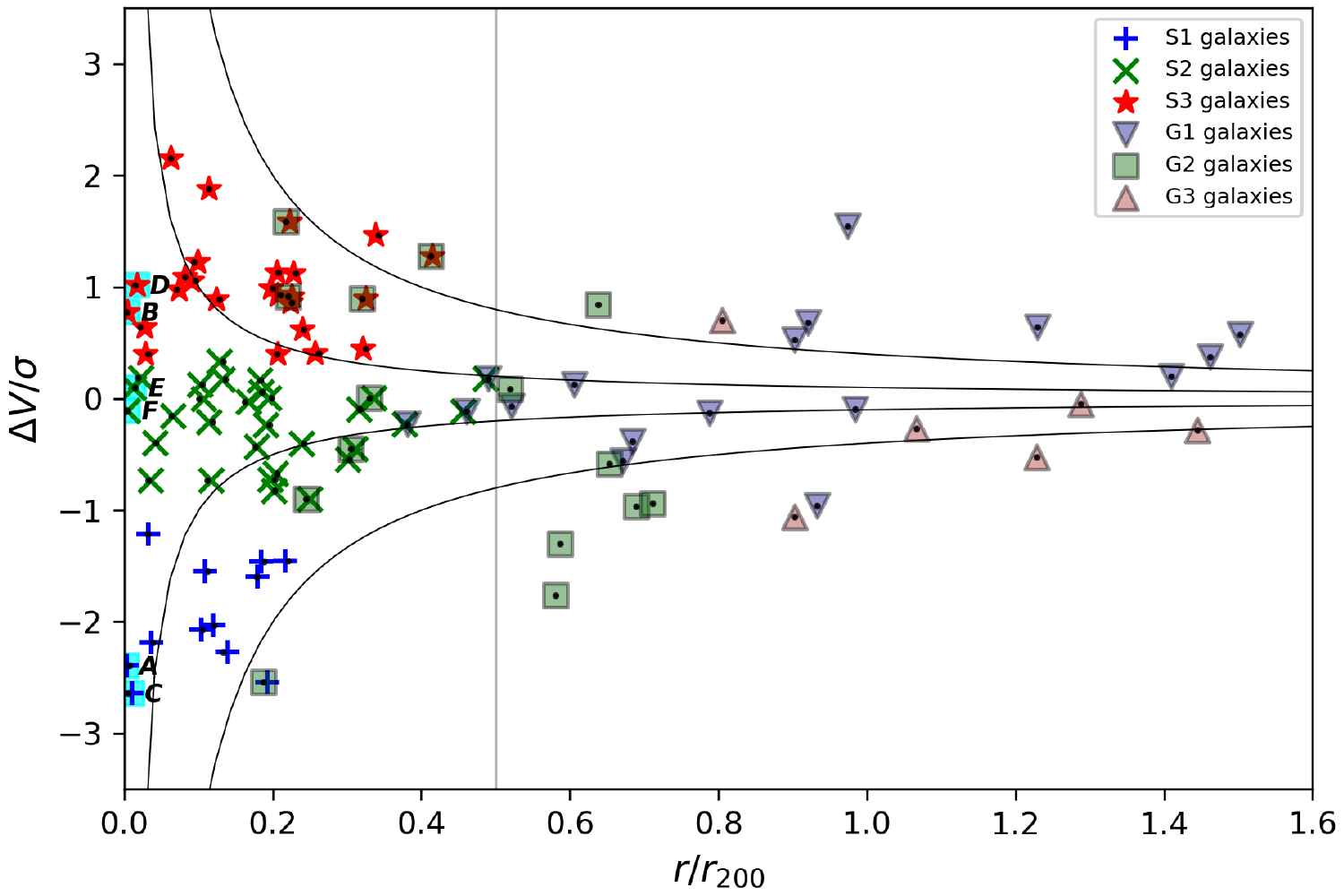} 
\caption{Phase-space diagram constructed from confirmed cluster members. The plot
shows the radial velocity offsets from the mean cluster velocity (line-of-sight
velocity from equation (\ref{eq:pecvel})) divided by the line-of-sight velocity
dispersion as a function of the projected clustercentric distance, measured from
the center of the X-ray emission, divided by \rtwo. The colored symbols are
the same as in the inset of Fig. \ref{fig:denmap_mem} (line-of-sight
structures S1, S2 and S3) and Fig.  \ref{fig:mdgmm} (overdensities G1, G2
and G3). The curves are the caustic profiles-lines of constant $ (r/r_{200})
\times (\Delta V/\sigma)$ at $\pm$0.1 and $\pm$0.4, respectively. \label{fig:phasediam}}
\end{figure}

\subsubsection{Projected phase-space diagram}\label{sec:phasediam}

The analysis of the phase-space diagram has been proven to be a useful
complementary method to investigate the dynamical state of a cluster and
differentiate between the virialized, from recently or early accreted, and
in-falling cluster galaxies \citep[e.g. ][]{gill2005,mahajan2011,haines2012,
noble2013,noble2016,rhee2017}. In a relaxed system, the line-of-sight velocities
of those galaxies that were accreted during the cluster formation have large
values in velocities and occupy a narrow band in the spatial direction, mostly 
in the central regions of the cluster. Galaxies in in-falling groups tend to 
populate separate regions in phase-space diagram. 

Here we investigate the kinematics  of the cluster from  the projected
phase-space diagram, using caustic profiles, i.e. constant velocity-radial
lines, following the approach described in \citet{noble2013}. In our analysis we
use values of constant velocity-radial lines of $\pm 0.1$ and $\pm 0.4$. The two
curves define three different regions in the phase-space diagram: $0 < (r/r_{200})
\times (\Delta V/\sigma) < 0.1$ - the central region where virialized galaxies
are located; $0.1 < (r/r_{200}) \times (\Delta V/\sigma) < 0.4$ - in the
intermediate region, where the recently accreted or early accreted (i.e. the
backsplash population) reside; $(r/r_{200}) \times (\Delta V/\sigma) > 0.4$ -
in-falling region, where infall galaxies or group of galaxies are preferentially
located.

Figure \ref{fig:phasediam} shows the phase-space diagram for all spectroscopic
confirmed member galaxies of \object{MS\,0440$+$0204}. We can see that the different
overdensities/groups and line-of-sight structures detected in Figures 
\ref{fig:1Dsub} - \ref{fig:mdgmm} occupy a distinct location in the phase-space diagram.

The galaxies lying in the lower region and at $r/r_{200}>0.5$ in Fig. \ref{fig:phasediam} 
(brown triangles) are part of the group G3 located West. On the other hand, galaxies 
located in the upper region and at $r/r_{200}>0.5$ 
(blue inverted triangles) are part of the group G1 located East. Roughly 
50\% of galaxies associated to the G2 group (green squares) lie in the virialized and
intermediate regions and at $r/r_{200}<0.5$, where the backsplash population reside, 
indicating that these galaxies could have been recently accreted or they could have 
been accreted at an earlier time and then rebounded outward 
\citep[e.g.][]{balogh2000,gill2005,rhee2017}. The remaining galaxies in the G2 group 
and galaxies in the G1 and G3 groups are distributed mainly in the infalling 
region ($(r/r_{200}) \times  (\Delta V/\sigma) > 0.4$), suggesting that these 
galaxies has been recently accreated.

The galaxies associated to the line-of-sight structures S1, S2 and S3 show 
larges $\Delta V/\sigma$ values as expected based on the analysis and presented in 
section \S \ref{sec:1Dsub}. The galaxies in the structure S2 occupy the inner central region
in the phase-space diagram (inside the $0 < (r/r_{200}) \times (\Delta V/\sigma) < 0.1$ bin)
with a narrow $\Delta V/\sigma$ values, indicating that the galaxies associated to the 
S2 structures were accreted during the cluster formation \citep[e.g.][]{haines2012,noble2013}.
A large fraction of the galaxies in the S1 and S3 structures occupy the intermediate region 
in the phase-space diagram ($0.1 < (r/r_{200}) \times (\Delta V/\sigma) < 0.4$ bin, 
the backsplash region), but with large values of $\Delta V/\sigma$. This can be 
interpreted as that the galaxies in S1 and S3 have already passed through the core 
of the cluster, but have not yet had time to join the virialized population.
Based on these results, it is likely that the cluster is experiencing 
a strong merging process along the line-of sight. Moreover, the strong
merging event could explain the presence of the diffuse extended symmetric 
envelope  \citep[e.g][]{dubinski1998} and the high concentration value 
obtained in our lensing analysis \citepalias{verdugo2020}.

\begin{figure*}[!t]
\centering\includegraphics[width=0.98\hsize]{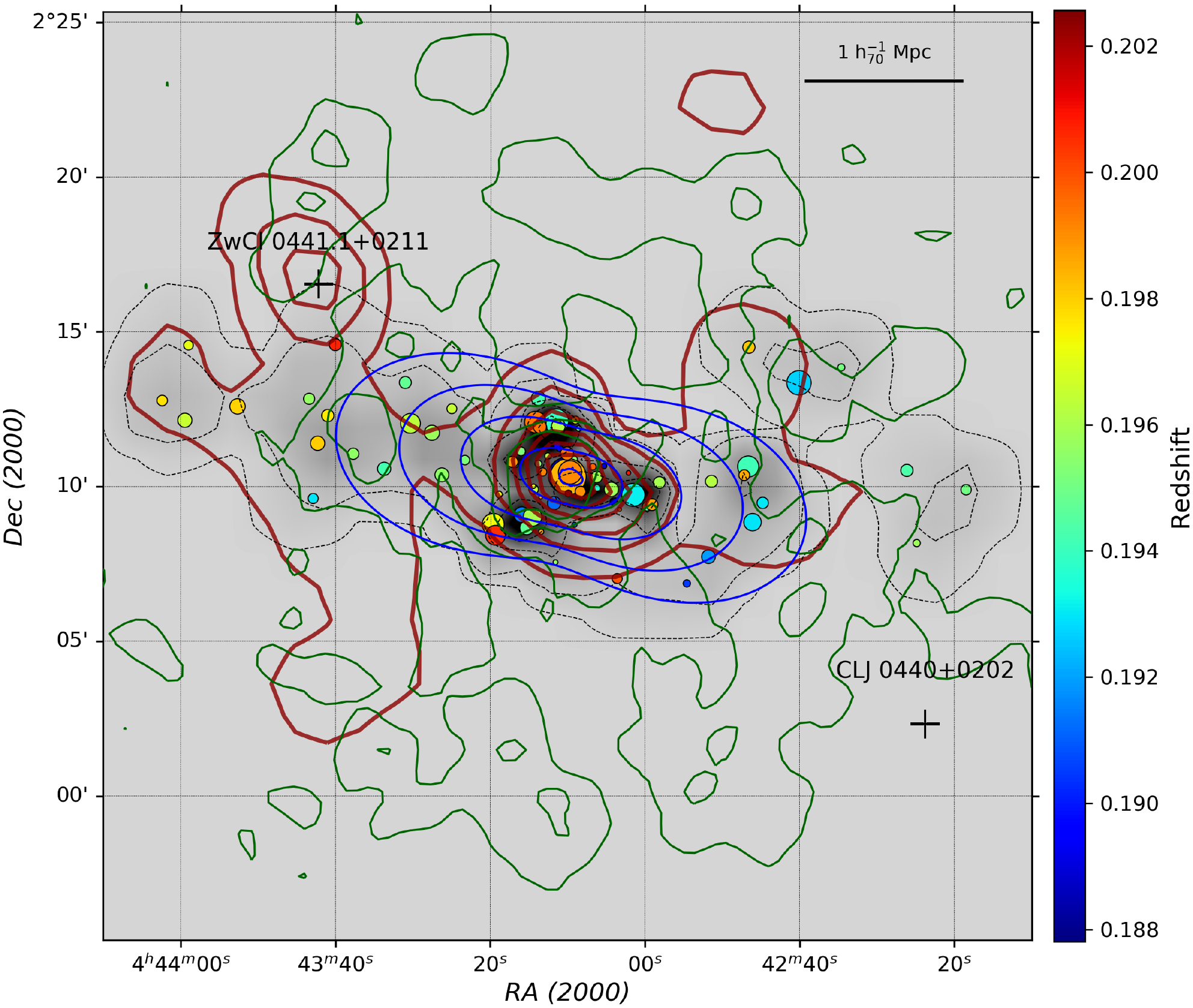}
\caption{Projected galaxy density map of 1328 galaxies brighter than
\sloanr~$=24$ mag and distributed in the RCS in Fig. \ref{fig:histmem} 
(dark red contours) overlay over a gray scale image of the member galaxy
density map (dashed black contours). The size of the image and the meaning
of the colored circles and the bar are the same as in Fig. \ref{fig:denmap_mem}.
The dark green and blue contours are from \citepalias{verdugo2020} and
correspond to the weak lensing 2D mass and the strong lensing mass contours,
respectively. \label{fig:RCS_proj_dist}}
\end{figure*}

\subsection{The environment around MS\,0440.5$+$0204 and the connection with other structures}\label{sec:environment}

The analysis presented in section \S \ref{sec:2Dprojected} suggests that 
\object{MS\,0440.5$+$0204} might be connected to the poor cluster of galaxies
\object{ZwCL\,0441.1$+$0211} through the galaxies located East from the
cluster (group G1 in Fig.\ref{fig:mdgmm}). To further investigate this 
possibility, we have used the NASA/IPAC Extragalactic Database
\footnote{http://ned.ipac.caltech.edu/} (NED) to search for information
about the redshift of the \object{ZwCL\,0441.1$+$0211} cluster and the
galaxies belonging to it. Unfortunately, the NED database does not provide any 
information on the redshift of the cluster and its member galaxies. The cluster 
is listed as a X-ray source in the ASCA Medium Sensitive Survey \citep{ueda2001},
but the redshift of the cluster is not provided. We also used the Sloan 
Digital Sky Survey Data Release 12\footnote{http://skyserver.sdss.org/dr12/en/tools/search/sql.aspx} 
database to search for any information about this cluster, but unfortunately the region 
is not in the SDSS footprints.

In absence of a redshift information for the galaxies in \object{ZwCL\,0441.1$+$0211} 
cluster, we used an alternative approach to investigate if there is a potential 
connection between the two clusters. If we assume that the galaxies in
\object{ZwCL\,0441.1$+$0211} are located close in redshift to 
\object{MS\,0440.5$+$0204}, then one could expect that the early-type galaxies 
in \object{ZwCL\,0441.1$+$0211} share the same location in the 
color-magnitude relation of Fig. \ref{fig:histmem}. Based on this assumption, 
we have analyzed the projected density distribution of all galaxies lying 
in the RCS and within $\pm 1 \sigma$ from the best-fit shown in the
bottom panel of Fig. \ref{fig:histmem}.

Figure \ref{fig:RCS_proj_dist} shows the adaptive kernel density map of
a sample of 1328 galaxies located in the RCS and brighter than \sloanr$=24$.
The galaxy density map (thick dark red contours) shows to prominent peaks,
one coincident with the center of \object{MS\,0440.5$+$0204}, and another near
the position of \object{ZwCL\,0441.1$+$0211}. Remarkably, the contours of 
the density map follow the same distribution pattern than the contours of 
the density map of member galaxies in Fig. \ref{fig:denmap_mem} 
(thin dashed black contours), suggesting that the two structures might
be connected. The figure also shows the 2D mass map from the strong lensing 
model (blue lines) and the 2D mass map from  weak lensing (green lines) 
from \citetalias{verdugo2020}. The direction of the 
contours in the strong lensing 2D mass map is consistent with the direction of 
the contours of the 2D weak lensing signal and with the direction of the contours 
in the density maps for member galaxies and the galaxies lying in the RCS. 
In addition, the elongation and position angle of the 2D strong lensing  mass 
map point into the NE direction, where \object{ZwCL\,0441.1$+$0211} 
cluster is located. As it is discussed in \citetalias{verdugo2020}, this elongation is 
likely produced by the influence of \object{ZwCL\,0441.1$+$0211} cluster. 
We refer the reader to \citetalias{verdugo2020} for a detailed discussion about 
the 2D mass maps obtained from the lensing models.

\begin{figure}[ht!]
\includegraphics[width=0.90\hsize]{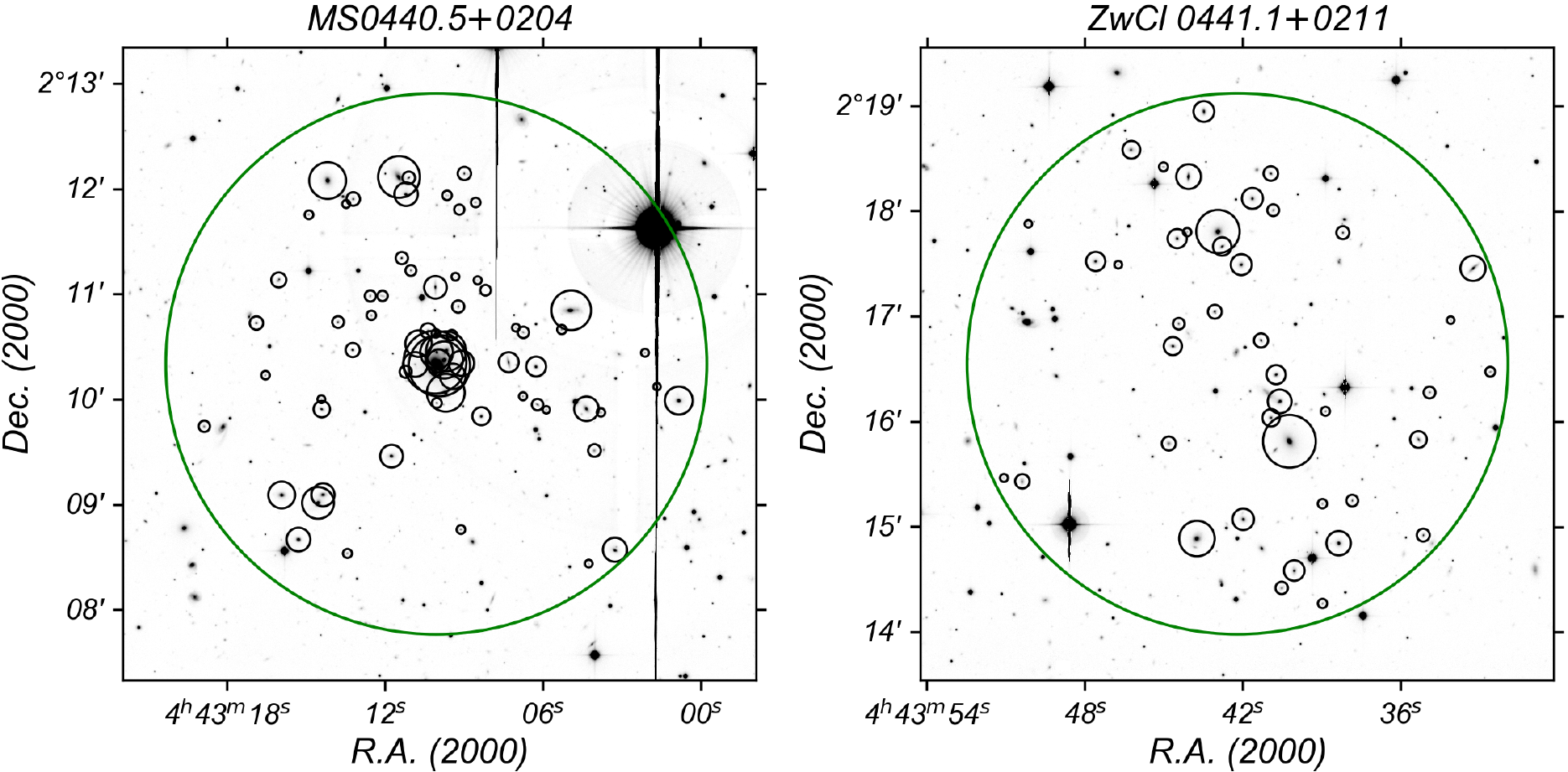}
\includegraphics[width=0.90\hsize]{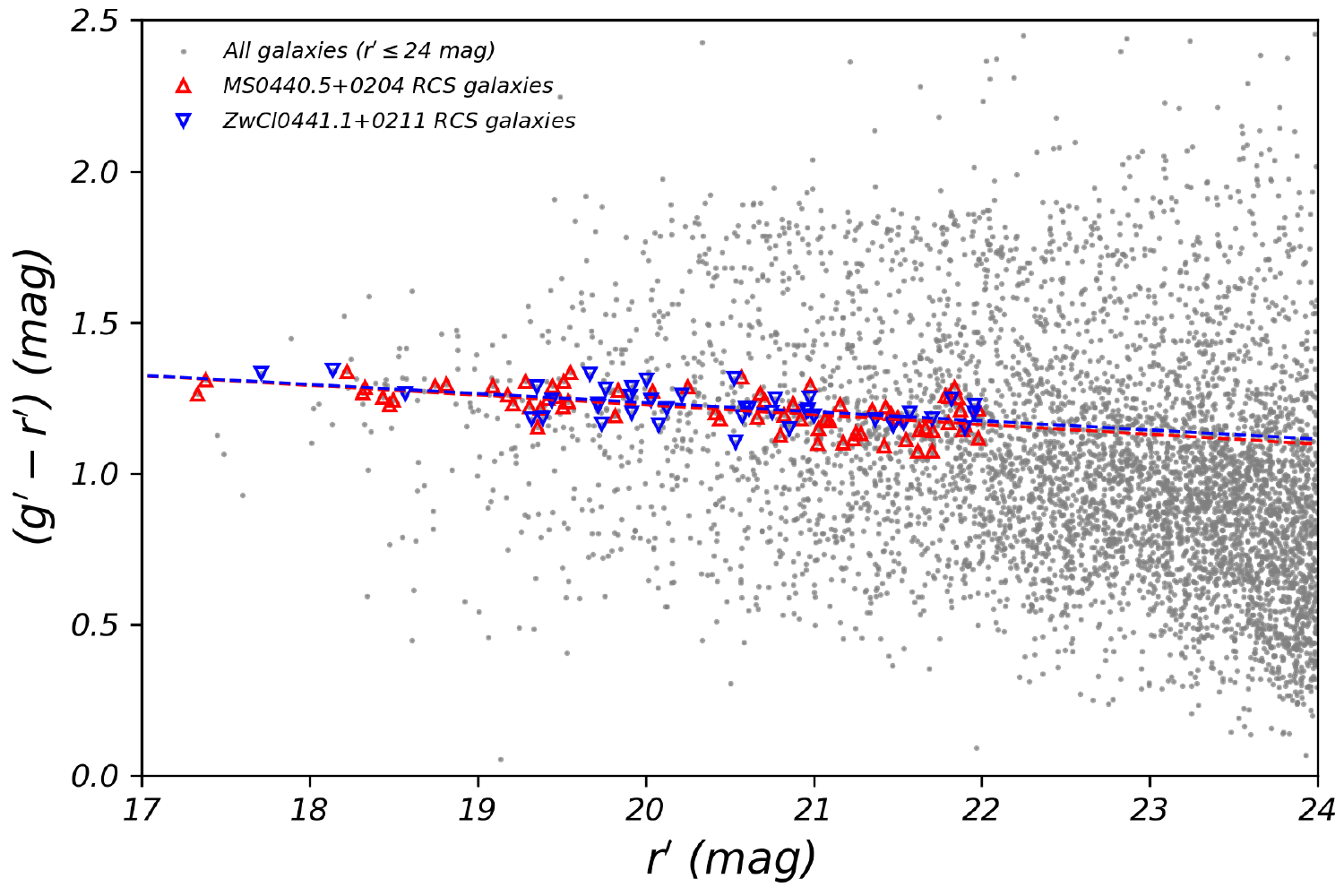}
\caption{\textit{Top:} \sloanr-band stamp images showing the selected galaxies
located in the RCS of Fig. \ref{fig:histmem} (bottom panel) and
inside the 3\arcmin~radius (green circle) from the center of the high-density
peaks in MS\,0440$+$0204 (left panel) and ZwCL\,0441.1$+$0211 (right panel),
respectively.  \textit{Bottom:} The color-magnitude diagram of all galaxies
brighter than \sloanr$=24$ mag (gray dots). The dashed (red and blue) lines show
the best-fit for the galaxies brighter than \sloanr$=22.5$ mag located in the 
RCS and inside the 3\arcmin~radius from the center of MS\,0440$+$0204 (red
triangles) and ZwCL\,0441.1$+$0211 (blue inverted triangles) clusters,
respectively.
\label{fig:ms0440Zwcl}}
\end{figure}

To confirm the above results, we have used the magnitudes and colors of the
galaxies located in the RCS and inside a circle of 3\arcmin~ in radius ($\sim 585$
\hval~kpc at the redshifts of \object{MS\,0440.5$+$0204}) from the high-density
peak in Fig. \ref{fig:ms0440Zwcl} to analyze the magnitude-color relation of
these galaxies. The selected galaxies are shown in the top panel of Fig. 
\ref{fig:ms0440Zwcl}. It is important to mention that the selection of the
galaxies in \object{MS\,0440.5$+$0204} is independent of its membership. We used
a standard linear regression plus an iterative 3-$\sigma$ clipping algorithm to
determine the slope and the intercept of the two populations. For
\object{MS\,0440.5$+$0204}, we obtained a slope of $-0.032\pm0.005$, with an
intercept of $1.875\pm0.107$.  In the case of ZwCL\,0441.1$+$0211, we obtained a
slope of $-0.030\pm0.007$, and an intercept of $1.838\pm0.147$. The best-fits
from the linear regression are shown in the CMD at the bottom panel of Fig.
\ref{fig:ms0440Zwcl} (red and blue dashed lines). We can see that values for the
intercepts and the slopes are similar, within the errors. This result reinforces
our finding that \object{ZwCL\,0441.1$+$0211} and \object{MS\,0440.5$+$0204}
are practically at the same distance. Even more, the distribution of the
galaxies inside the filament (G1 group) and the elongation seen in the 2D strong
lensing mass map, strongly support the idea that these two clusters might be
connected. However, to probe this statement it is necessary to determine
the redshifts of the galaxies in the area of \object{ZwCL\,0441.1$+$0211}.

\section{Summary and conclusions}\label{sec:summary}

In this work we have presented a detailed kinematical and dynamical analysis of
the strong lensing galaxy cluster \object{MS\,0440.5$+$0204}, a massive
structure located at $z=0.1959$. The analysis is  based on new spectroscopic
data obtained for the most prominent arc systems in the cluster core
and hundreds of galaxies observed in a wider area on the sky. Our main results and 
conclusion of this study are summarized in the next paragraphs. 

We have determined, for the first time, the spectroscopic redshifts for the
gravitational arc systems M1, M2 and M3, and confirmed the redshift for the arc
system S1, which is in a good agreement with the redshift obtained by
\citetalias{gioia1998}. We have measured the spectroscopic redshifts for 
185 galaxies observed with GMOS and CFHTMOS inside an area of $\sim0.45\degr\,\times\,0.14\degr$ 
in RA/Dec ($\sim 5.3\,\times\,1.6$ \hvaltwo~Mpc$^2$ at the distance of the cluster). 
The redshift catalog was supplemented with additional 10 galaxies located at redshift 
the cluster from \citetalias{gioia1998} and not covered by our observations, increasing 
the sample to 195 galaxies with confirmed redshifts (section \S \ref{sec:galred}). 
We have estimated new redshifts for 135 galaxies, of which 44 are at the distance 
of \object{MS\,0440.5$+$0204}, increasing the number of cluster members by a factor 
of $\gtrsim 2$. Using the robust bi-weight estimators of \citet{beers1990}, we determined 
an average redshift for the cluster of $0.195929^{+0.00033}_{-0.00031}$ and a 
line-of-sight velocity dispersion of $771^{+63}_{-71}$ \kms\, for 94 confirmed member 
galaxies. The velocity dispersion is slightly lower, but in agreement within the uncertainties, 
with the values reported by \citetalias{gioia1998} (sections \S \ref{sec:redcluster})
and derived by \citetalias{verdugo2020} using the SIS model profile
(section \S \ref{sec:weaklen}).  We have computed the mass of the cluster using
the $\sigma$-\masstwo~scaling relation of \citet{munari2013}. We determined a
dynamical mass \masstwo$=3.66^{+0.67}_{-0.59} \times 10^{14}$ \hvaltwo~\Msol~at
\rtwo $=1.73^{+0.14}_{-0.16}$ \hval~Mpc (section \S \ref{sec:mass}). The dynamical
mass is in a very good agreement with the total mass inferred from the weak-lensing 
analysis \citepalias[\masstwo$=3.9^{+1.4}_{-0.8} \times 10^{14}$\hvaltwo~\Msol,][]{verdugo2020}
and from X-ray ($\mathrm{M}_{x} = 3.80^{+0.62}_{-0.46} \times 10^{14}$ \hvaltwo~\Msol). 

The cluster shows several structures along the line-of-sight and in the spatial
direction. We find that the redshift distribution of the galaxies in the inner
$r\sim 0.86$ \hval~Mpc of the cluster ($0.5 \times$\rtwo) is at least
tri-modal at the 98.5\%~confidence level, and with three well defined
structures. These structures, named S1, S2 and S3 basically form the cluster core 
(section \S \ref{sec:1Dsub}). The six central elliptical galaxies 
located at the center of the cluster have different line-of-sight 
velocities. The galaxies pairs ``A - C'', ``E - F'', and ``B - D''  have a very 
small differences in their line-of-sight velocities ($\sim 160 - 190$ \kms). Each
pair is associated to the structures S1 (``A - C''),  S2 (``E - F''), and S3 (``B - D''). 
The presence of these structures along the line-of-sight suggests that the cluster 
core is likely experiencing a strong merging process. This statement is supported by the results 
obtained in the analysis of the phase-space diagram (section \S \ref{sec:phasediam}), 
where the galaxies show large values in the line-of-sight velocity and within a narrow band 
in the spatial direction (see Fig. \ref{fig:phasediam}). A strong merging event
scenario could explain also the presence of the diffuse extended symmetric envelope 
\citep[e.g][]{dubinski1998} and the high concentration value of $c_{200} \sim 9$ 
obtained in our lensing analysis \citepalias{verdugo2020}. Strong-lensing is sensitive 
to the projected mass along the line of vision, therefore the presence of several 
structures along the line-of-sight can bias the lensing results 
\citep[e.g.][,\citetalias{verdugo2020}]{bayliss2014,li2019}.

We have identified four different high-density regions in the projected density
map of member galaxies (Fig. \ref{fig:denmap_mem}). These high density regions
are (see Fig. \ref{fig:mdgmm}): the cluster core formed by the line-of-sight structures 
S1, S2 and S3, with 67 galaxies, the group G1 at $\sim 8$\arcmin~NE  from the 
cluster core with 17 galaxies, the group G2 at $\sim 7$\arcmin~SW from 
the cluster core with 15 galaxies, and the group G3 located $\sim 13$\arcmin~NW 
from the cluster core with 6 galaxies. All these substructures are also presented in the
projected density map of 1328 galaxies located in the linear CMD relation for
early-type galaxies (the Red Cluster Sequence) and in the 2D weak lensing mass
map of Fig. \ref{fig:ms0440Zwcl}. The distribution of the galaxies in the phase-space 
diagram of galaxies assigned to the G1 and G3 suggests that these groups are likely
falling into the cluster core. The galaxies assigned to the G2 group are mainly 
located in the  intermediate region in the phase-space diagram, indicating that 
the galaxies in this group has been recently accreted. 

The projected distribution of the member galaxies (Fig. \ref{fig:denmap_mem}) 
and the distribution of the galaxies in the G1 group along a filament suggest a
connection between {MS\,0440.5$+$0204} and \object{ZwCL\,0441.1$+$0211}. 
In the absence of redshift information for the galaxies that could belong to 
\object{ZwCL\,0441.1$+$0211}, we have used the magnitudes and colors of the 
galaxies inside a 3\arcmin~radius around the peak of the overdensities in 
Fig. \ref{fig:ms0440Zwcl} to investigate if  \object{ZwCL\,0441.1$+$0211} 
and \object{MS\,0440.5$+$0204} are somehow linked. The result shows that 
the magnitudes and colors of the galaxies in \object{ZwCL\,0441.1$+$0211} 
share the same linear CMD relation than the galaxies in \object{MS\,0440.5$+$0204} 
within the same radius. Based on analysis presented in section 
\S \ref{sec:analysis}, the clusters \object{ZwCL\,0441.1$+$0211} and 
\object{MS\,0440.5$+$0204} might be connected.
It is importan to point out that \object{ZwCL\,0441.1$+$0211} is not detected in the
2D weak lensing map. There is only a weak detection near the cluster position (Fig.
\ref{fig:ms0440Zwcl}). The absence of detection could be an argument against our
conclusion. However, fake lensing peaks are expected in weak lensing analysis
and could be caused by sampling and shape noise, effect that could explain the
week lensing peak near the position of \object{ZwCL\,0441.1$+$0211}
\citepalias[see][for details] {verdugo2020}.

The complex redshift distribution of member galaxies in the cluster core and
along the the line-of-sight, and the presence of several structures in the 
galaxy density maps and in the weak-lensing mass map reveal that 
\object{MS\,0440.5$+$0204} is dynamically active. The central region of  
the cluster might be experiencing a merging process along the line-of-sight.
The central elliptical galaxies might be eventually merge to form a massive galaxy 
at the cluster core. Additional evidence of dynamical activities is given by the groups 
G1 and G3, which are likely falling into the cluster core. The presence of a bridge of matter
connecting the \object{MS\,0440.5$+$0204} with the \object{ZwCL\,0441.1$+$0211}
poor cluster suggest that both structures might be linked. However, to
confirm this hypothesis, spectroscopic observations of galaxies in the area of
the \object{ZwCL\,0441.1$+$0211} cluster are required. Also, it would be necessary to
extend the observations to a larger area on the sky to analyze in detail 
the dynamical state of the \object{MS\,0440.5$+$0204} and \object{ZwCL\,0441.1$+$0211} 
clusters in a merging scenario.

\acknowledgments

We would like to thank the anonymous referee for constructive and helpful comments. 
ERC acknowledges the hospitality of the Observatorio Astron\'omico Nacional
of the Universidad Nacional Aut\'onoma de M\'exico in Ensenada, where this work
was partially done.  This research has been carried out thanks to PROGRAMA 
UNAM-DGAPA-PAPIIT IA102517. VM acknowledges the partial support from 
Centro de Astrof\'{\i}sica de Valpara\'{\i}so. This work is based on observations 
obtained at the international Gemini Observatory, a program of NSF’s NOIRLab, 
which is managed by the Association of
Universities for Research in Astronomy (AURA) under a cooperative agreement with
the National Science Foundation. on behalf of the Gemini Observatory
partnership: the National Science Foundation (United States), National Research
Council (Canada), Agencia Nacional de Investigaci\'{o}n y Desarrollo (Chile),
Ministerio de Ciencia, Tecnolog\'{i}a e Innovaci\'{o}n (Argentina),
Minist\'{e}rio da Ci\^{e}ncia, Tecnologia, Inova\c{c}\~{o}es e
Comunica\c{c}\~{o}es (Brazil), and Korea Astronomy and Space Science Institute
(Republic of Korea). The paper uses data obtained with MegaPrime/MegaCam, a
joint project of CFHT and CEA/DAPNIA, at  the Canada-France-Hawaii Telescope
(CFHT) which is operated by the National Research Council (NRC) of Canada, the
Institut National des Science de l'Univers of the Centre National de la
Recherche Scientifique (CNRS) of France, and the University of Hawaii. This
research also used the facilities of the Canadian Astronomy Data Centre operated
by the National Research Council of Canada with the support of the Canadian
Space Agency. 

%

\vspace{5mm}
\facilities{Gemini South: GMOS-S; CFHT: Megacam, MOS/SIS}


\software{SExtractor \citep{ber96}, THELI \citep{esd05,schirmer13}, Astropy \citep{astropy:2013, astropy:2018}}

\clearpage

\bibliography{myreferences}

\begin{thebibliography}{}
\expandafter\ifx\csname natexlab\endcsname\relax\def\natexlab#1{#1}\fi
\providecommand{\url}[1]{\href{#1}{#1}}
\providecommand{\dodoi}[1]{doi:~\href{http://doi.org/#1}{\nolinkurl{#1}}}
\providecommand{\doeprint}[1]{\href{http://ascl.net/#1}{\nolinkurl{http://ascl.net/#1}}}
\providecommand{\doarXiv}[1]{\href{https://arxiv.org/abs/#1}{\nolinkurl{https://arxiv.org/abs/#1}}}

\bibitem[{{Astropy Collaboration} {et~al.}(2013){Astropy Collaboration},
  {Robitaille}, {Tollerud}, {Greenfield}, {Droettboom}, {Bray}, {Aldcroft},
  {Davis}, {Ginsburg}, {Price-Whelan}, {Kerzendorf}, {Conley}, {Crighton},
  {Barbary}, {Muna}, {Ferguson}, {Grollier}, {Parikh}, {Nair}, {Unther},
  {Deil}, {Woillez}, {Conseil}, {Kramer}, {Turner}, {Singer}, {Fox}, {Weaver},
  {Zabalza}, {Edwards}, {Azalee Bostroem}, {Burke}, {Casey}, {Crawford},
  {Dencheva}, {Ely}, {Jenness}, {Labrie}, {Lim}, {Pierfederici}, {Pontzen},
  {Ptak}, {Refsdal}, {Servillat}, \& {Streicher}}]{astropy:2013}
{Astropy Collaboration}, {Robitaille}, T.~P., {Tollerud}, E.~J., {et~al.} 2013,
  \aap, 558, A33, \dodoi{10.1051/0004-6361/201322068}

\bibitem[{{Balestra} {et~al.}(2016){Balestra}, {Mercurio}, {Sartoris},
  {Girardi}, {Grillo}, {Nonino}, {Rosati}, {Biviano}, {Ettori}, {Forman},
  {Jones}, {Koekemoer}, {Medezinski}, {Merten}, {Ogrean}, {Tozzi}, {Umetsu},
  {Vanzella}, {van Weeren}, {Zitrin}, {Annunziatella}, {Caminha}, {Broadhurst},
  {Coe}, {Donahue}, {Fritz}, {Frye}, {Kelson}, {Lombardi}, {Maier},
  {Meneghetti}, {Monna}, {Postman}, {Scodeggio}, {Seitz}, \&
  {Ziegler}}]{balestra2016}
{Balestra}, I., {Mercurio}, A., {Sartoris}, B., {et~al.} 2016, \apjs, 224, 33,
  \dodoi{10.3847/0067-0049/224/2/33}

\bibitem[{{Balogh} {et~al.}(2000){Balogh}, {Navarro}, \& {Morris}}]{balogh2000}
{Balogh}, M.~L., {Navarro}, J.~F., \& {Morris}, S.~L. 2000, \apj, 540, 113,
  \dodoi{10.1086/309323}

\bibitem[{{Bayliss} {et~al.}(2014){Bayliss}, {Johnson}, {Gladders}, {Sharon},
  \& {Oguri}}]{bayliss2014}
{Bayliss}, M.~B., {Johnson}, T., {Gladders}, M.~D., {Sharon}, K., \& {Oguri},
  M. 2014, \apj, 783, 41, \dodoi{10.1088/0004-637X/783/1/41}

\bibitem[{{Beers} {et~al.}(1990){Beers}, {Flynn}, \& {Gebhardt}}]{beers1990}
{Beers}, T.~C., {Flynn}, K., \& {Gebhardt}, K. 1990, \aj, 100, 32,
  \dodoi{10.1086/115487}

\bibitem[{{Bertin}(2006)}]{ber06}
{Bertin}, E. 2006, in ASP Conf. Ser., Vol. 351, ADASS XV, ed. C.~{Gabriel},
  C.~{Arviset}, D.~{Ponz}, \& S.~{Enrique}, 112

\bibitem[{{Bertin}(2010)}]{ber2010a}
{Bertin}, E. 2010, {SWarp: Resampling and Co-adding FITS Images Together},
  Astrophysics Source Code Library.
\newblock \doeprint{1010.068}

\bibitem[{Bertin \& Arnouts(1996)}]{ber96}
Bertin, E., \& Arnouts, S. 1996, A\&AS, 117, 393

\bibitem[{{Bolzonella} {et~al.}(2000){Bolzonella}, {Miralles}, \&
  {Pell{\'o}}}]{bolzonella2000}
{Bolzonella}, M., {Miralles}, J.~M., \& {Pell{\'o}}, R. 2000, \aap, 363, 476.
\newblock \doarXiv{astro-ph/0003380}

\bibitem[{{Bonoli} {et~al.}(2020){Bonoli}, {Mar{\'\i}n-Franch}, {Varela},
  {V{\'a}zquez Rami{\'o}}, {Abramo}, {Cenarro}, {Dupke}, {V{\'\i}lchez},
  {Crist{\'o}bal-Hornillos}, {Gonz{\'a}lez Delgado},
  {Hern{\'a}ndez-Monteagudo}, {L{\'o}pez-Sanjuan}, {Muniesa}, {Civera},
  {Ederoclite}, {Hern{\'a}n-Caballero}, {Marra}, {Baqui}, {Cortesi},
  {Cypriano}, {Daflon}, {de Amorim}, {D{\'\i}az-Garc{\'\i}a}, {Diego},
  {Mart{\'\i}nez-Solaeche}, {P{\'e}rez}, {Placco}, {Prada}, {Queiroz},
  {Alcaniz}, {Alvarez-Candal}, {Cepa}, {Maroto}, {Roig}, {Siffert}, {Taylor},
  {Benitez}, {Moles}, {Sodr{\'e}}, {Carneiro}, {Mendes de Oliveira}, {Abdalla},
  {Angulo}, {Aparicio Resco}, {Balaguera-Antol{\'\i}nez}, {Ballesteros},
  {Brito-Silva}, {Broadhurst}, {Carrasco}, {Castro}, {Cid Fernandes}, {Coelho},
  {de Melo}, {Doubrawa}, {Fernandez-Soto}, {Ferrari}, {Finoguenov},
  {Garc{\'\i}a-Benito}, {Iglesias-P{\'a}ramo}, {Jim{\'e}nez-Teja}, {Kitaura},
  {Laur}, {Lopes}, {Lucatelli}, {Mart{\'\i}nez}, {Maturi}, {Quartin},
  {Pigozzo}, {Rodr{\`\i}guez-Mart{\`\i}n}, {Salzano}, {Tamm}, {Tempel},
  {Umetsu}, {Valdivielso}, {von Marttens}, {Zitrin}, {D{\'\i}az-Mart{\'\i}n},
  {L{\'o}pez-Alegre}, {L{\'o}pez-Sainz}, {Yanes-D{\'\i}az}, {Rueda-Teruel},
  {Rueda-Teruel}, {Abril Iba{\~n}ez}, {Ant{\'o}n Bravo}, {Bello Ferrer},
  {Bielsa}, {Casino}, {Castillo}, {Chueca}, {Cuesta}, {Garzar{\'a}n Calderaro},
  {Iglesias-Marzoa}, {{\'I}niguez}, {Lamadrid Gutierrez}, {Lopez-Martinez},
  {Lozano-P{\'e}rez}, {Ma{\'\i}cas Sacrist{\'a}n}, {Molina-Ib{\'a}{\~n}ez},
  {Moreno-Signes}, {Rodr{\'\i}guez Llano}, {Royo Navarro}, {Tilve Rua},
  {Andrade}, {Alfaro}, {Akras}, {Arnalte-Mur}, {Ascaso}, {Barbosa},
  {Beltr{\'a}n Jim{\'e}nez}, {Benetti}, {Bengaly}, {Bernui}, {Blanco-Pillado},
  {Borges Fernandes}, {Bregman}, {Bruzual}, {Calderone}, {Carvano}, {Casarini},
  {Chies-Santos}, {Coutinho de Carvalho}, {Dimauro}, {Duarte Puertas},
  {Figueruelo}, {Gonz{\'a}lez-Serrano}, {Guerrero}, {Gurung-L{\'o}pez},
  {Herranz}, {Huertas-Company}, {Irwin}, {Izquierdo-Villalba}, {Kanaan},
  {Kehrig}, {Kirkpatrick}, {Lim}, {Lopes}, {Lopes de Oliveira},
  {Marcos-Caballero}, {Mart{\'\i}nez-Delgado}, {Mart{\'\i}nez-Gonz{\'a}lez},
  {Mart{\'\i}nez-Somonte}, {Oliveira}, {Orsi}, {Overzier}, {Penna-Lima},
  {Reis}, {Spinoso}, {Tsujikawa}, {Vielva}, {Vitorelli}, {Xia}, {Yuan},
  {Arroyo-Polonio}, {Dantas}, {Galarza}, {Gon{\c{c}}alves}, {Gon{\c{c}}alves},
  {Gonzalez}, {Gonzalez}, {Greisel}, {Land im}, {Lazzaro}, {Magris},
  {Monteiro-Oliveira}, {Pereira}, {Rebou{\c{c}}as}, {Rodriguez-Espinosa},
  {Santos da Costa}, \& {Telles}}]{miniJPAS2020}
{Bonoli}, S., {Mar{\'\i}n-Franch}, A., {Varela}, J., {et~al.} 2020, arXiv
  e-prints, arXiv:2007.01910.
\newblock \doarXiv{2007.01910}

\bibitem[{{Bridle} {et~al.}(2002){Bridle}, {Kneib}, {Bardeau}, \&
  {Gull}}]{bridle02}
{Bridle}, S.~L., {Kneib}, J.~P., {Bardeau}, S., \& {Gull}, S.~F. 2002, in The
  Shapes of Galaxies and their Dark Halos, ed. P.~{Natarajan}, 38--46,
  \dodoi{10.1142/9789812778017_0006}

\bibitem[{{Carlberg} {et~al.}(1994){Carlberg}, {Yee}, {Ellingson}, {Pritchet},
  {Abraham}, {Smecker-Hane}, {Bond}, {Couchman}, {Crabtree}, {Crampton},
  {Davidge}, {Durand}, {Eales}, {Hartwick}, {Hesser}, {Hutchings}, {Kaiser},
  {Mendes de Oliveira}, {Myers}, {Oke}, {Rigler}, {Schade}, \&
  {West}}]{calberg1994}
{Carlberg}, R.~G., {Yee}, H.~K.~C., {Ellingson}, E., {et~al.} 1994, \jrasc, 88,
  39.
\newblock \doarXiv{astro-ph/9311020}

\bibitem[{{Carlberg} {et~al.}(1997){Carlberg}, {Yee}, {Ellingson}, {Morris},
  {Abraham}, {Gravel}, {Pritchet}, {Smecker-Hane}, {Hartwick}, \&
  {Hesser}}]{carlberg1997}
---. 1997, \apjl, 485, L13, \dodoi{10.1086/310801}

\bibitem[{{Carrasco} {et~al.}(2007){Carrasco}, {Cypriano}, {Neto}, {Cuevas},
  {Sodr{\'e}}, {de Oliveira}, \& {Ramirez}}]{carrasco2007}
{Carrasco}, E.~R., {Cypriano}, E.~S., {Neto}, G.~B.~L., {et~al.} 2007, \apj,
  664, 777, \dodoi{10.1086/518925}

\bibitem[{{Carrasco} {et~al.}(2006){Carrasco}, {Mendes de Oliveira}, \&
  {Infante}}]{carrasco2006}
{Carrasco}, E.~R., {Mendes de Oliveira}, C., \& {Infante}, L. 2006, \aj, 132,
  1796, \dodoi{10.1086/507447}

\bibitem[{{Carrasco} {et~al.}(2010){Carrasco}, {Gomez}, {Verdugo}, {Lee},
  {Diaz}, {Bergmann}, {Turner}, {Miller}, \& {West}}]{carrasco2010}
{Carrasco}, E.~R., {Gomez}, P.~L., {Verdugo}, T., {et~al.} 2010, \apjl, 715,
  L160, \dodoi{10.1088/2041-8205/715/2/L160}

\bibitem[{{Dawson} {et~al.}(2015){Dawson}, {Jee}, {Stroe}, {Ng}, {Golovich},
  {Wittman}, {Sobral}, {Br{\"u}ggen}, {R{\"o}ttgering}, \& {van
  Weeren}}]{dawson2015}
{Dawson}, W.~A., {Jee}, M.~J., {Stroe}, A., {et~al.} 2015, \apj, 805, 143,
  \dodoi{10.1088/0004-637X/805/2/143}

\bibitem[{{Devillard}(1997)}]{devillard1997}
{Devillard}, N. 1997, The Messenger, 87, 19

\bibitem[{{Diaferio}(1999)}]{diaferio1999}
{Diaferio}, A. 1999, \mnras, 309, 610, \dodoi{10.1046/j.1365-8711.1999.02864.x}

\bibitem[{{Diego} {et~al.}(2015){Diego}, {Broadhurst}, {Benitez}, {Umetsu},
  {Coe}, {Sendra}, {Sereno}, {Izzo}, \& {Covone}}]{diego2015}
{Diego}, J.~M., {Broadhurst}, T., {Benitez}, N., {et~al.} 2015, \mnras, 446,
  683, \dodoi{10.1093/mnras/stu2064}

\bibitem[{{Dressler} \& {Shectman}(1988)}]{dressler1988}
{Dressler}, A., \& {Shectman}, S.~A. 1988, \aj, 95, 985, \dodoi{10.1086/114694}

\bibitem[{{Dubinski}(1998)}]{dubinski1998}
{Dubinski}, J. 1998, \apj, 502, 141, \dodoi{10.1086/305901}

\bibitem[{{Duffy} {et~al.}(2008){Duffy}, {Schaye}, {Kay}, \& {Dalla
  Vecchia}}]{Duff08}
{Duffy}, A.~R., {Schaye}, J., {Kay}, S.~T., \& {Dalla Vecchia}, C. 2008,
  \mnras, 390, L64, \dodoi{10.1111/j.1745-3933.2008.00537.x}

\bibitem[{{El{\'{\i}}asd{\'o}ttir} {et~al.}(2007){El{\'{\i}}asd{\'o}ttir},
  {Limousin}, {Richard}, {Hjorth}, {Kneib}, {Natarajan}, {Pedersen}, {Jullo},
  \& {Paraficz}}]{elr07}
{El{\'{\i}}asd{\'o}ttir}, {\'A}., {Limousin}, M., {Richard}, J., {et~al.} 2007,
  astro-ph/0710.5636.
\newblock \doarXiv{0710.5636}

\bibitem[{Erben {et~al.}(2005)Erben, Schirmer, Dietrich, {et~al.}}]{esd05}
Erben, T., Schirmer, M., Dietrich, J., {et~al.} 2005, AN, 326, 432

\bibitem[{{Evrard} {et~al.}(1996){Evrard}, {Metzler}, \&
  {Navarro}}]{evrard1996}
{Evrard}, A.~E., {Metzler}, C.~A., \& {Navarro}, J.~F. 1996, \apj, 469, 494,
  \dodoi{10.1086/177798}

\bibitem[{{Fitzpatrick}(1999)}]{fitz1999}
{Fitzpatrick}, E.~L. 1999, \pasp, 111, 63, \dodoi{10.1086/316293}

\bibitem[{{Fo{\"e}x} {et~al.}(2014){Fo{\"e}x}, {Motta}, {Jullo}, {Limousin}, \&
  {Verdugo}}]{foex14}
{Fo{\"e}x}, G., {Motta}, V., {Jullo}, E., {Limousin}, M., \& {Verdugo}, T.
  2014, \aap, 572, A19, \dodoi{10.1051/0004-6361/201424706}

\bibitem[{{Fo{\"e}x} {et~al.}(2013){Fo{\"e}x}, {Motta}, {Limousin}, {Verdugo},
  {More}, {Cabanac}, {Gavazzi}, \& {Mu{\~n}oz}}]{foex13}
{Fo{\"e}x}, G., {Motta}, V., {Limousin}, M., {et~al.} 2013, \aap, 559, A105,
  \dodoi{10.1051/0004-6361/201321112}

\bibitem[{{Fo{\"e}x} {et~al.}(2012){Fo{\"e}x}, {Soucail}, {Pointecouteau},
  {Arnaud}, {Limousin}, \& {Pratt}}]{foex12}
{Fo{\"e}x}, G., {Soucail}, G., {Pointecouteau}, E., {et~al.} 2012, \aap, 546,
  A106, \dodoi{10.1051/0004-6361/201218973}

\bibitem[{{Geller} {et~al.}(2014){Geller}, {Hwang}, {Diaferio}, {Kurtz}, {Coe},
  \& {Rines}}]{geller2014}
{Geller}, M.~J., {Hwang}, H.~S., {Diaferio}, A., {et~al.} 2014, \apj, 783, 52,
  \dodoi{10.1088/0004-637X/783/1/52}

\bibitem[{{Gill} {et~al.}(2005){Gill}, {Knebe}, \& {Gibson}}]{gill2005}
{Gill}, S. P.~D., {Knebe}, A., \& {Gibson}, B.~K. 2005, \mnras, 356, 1327,
  \dodoi{10.1111/j.1365-2966.2004.08562.x}

\bibitem[{{Gioia} {et~al.}(1990){Gioia}, {Maccacaro}, {Schild}, {Wolter},
  {Stocke}, {Morris}, \& {Henry}}]{gioia1990}
{Gioia}, I.~M., {Maccacaro}, T., {Schild}, R.~E., {et~al.} 1990, \apjs, 72,
  567, \dodoi{10.1086/191426}

\bibitem[{{Gioia} {et~al.}(1998){Gioia}, {Shaya}, {Le F{\`e}vre}, {Falco},
  {Luppino}, \& {Hammer}}]{gioia1998}
{Gioia}, I.~M., {Shaya}, E.~J., {Le F{\`e}vre}, O., {et~al.} 1998, \apj, 497,
  573, \dodoi{10.1086/305471}

\bibitem[{{Gladders} {et~al.}(1998){Gladders}, {L{\'o}pez-Cruz}, {Yee}, \&
  {Kodama}}]{gladders1998}
{Gladders}, M.~D., {L{\'o}pez-Cruz}, O., {Yee}, H.~K.~C., \& {Kodama}, T. 1998,
  \apj, 501, 571, \dodoi{10.1086/305858}

\bibitem[{Gladders \& Yee(2000)}]{gladders2000}
Gladders, M.~D., \& Yee, H. K.~C. 2000, The Astronomical Journal, 120, 2148,
  \dodoi{10.1086/301557}

\bibitem[{{Glazebrook} \& {Bland-Hawthorn}(2001)}]{glaz2001}
{Glazebrook}, K., \& {Bland-Hawthorn}, J. 2001, \pasp, 113, 197,
  \dodoi{10.1086/318625}

\bibitem[{{Gonzalez} {et~al.}(2005){Gonzalez}, {Tran}, {Conbere}, \&
  {Zaritsky}}]{gonzalez2005}
{Gonzalez}, A.~H., {Tran}, K.-V.~H., {Conbere}, M.~N., \& {Zaritsky}, D. 2005,
  \apjl, 624, L73, \dodoi{10.1086/430518}

\bibitem[{{Gwyn}(2008)}]{gwyn08}
{Gwyn}, S.~D.~J. 2008, \pasp, 120, 212, \dodoi{10.1086/526794}

\bibitem[{{Haines} {et~al.}(2012){Haines}, {Pereira}, {Sanderson}, {Smith},
  {Egami}, {Babul}, {Edge}, {Finoguenov}, {Moran}, \& {Okabe}}]{haines2012}
{Haines}, C.~P., {Pereira}, M.~J., {Sanderson}, A.~J.~R., {et~al.} 2012, \apj,
  754, 97, \dodoi{10.1088/0004-637X/754/2/97}

\bibitem[{{Heisler} {et~al.}(1985){Heisler}, {Tremaine}, \&
  {Bahcall}}]{heisler1985}
{Heisler}, J., {Tremaine}, S., \& {Bahcall}, J.~N. 1985, \apj, 298, 8,
  \dodoi{10.1086/163584}

\bibitem[{{Hicks} {et~al.}(2006){Hicks}, {Ellingson}, {Hoekstra}, \&
  {Yee}}]{hicks2006}
{Hicks}, A.~K., {Ellingson}, E., {Hoekstra}, H., \& {Yee}, H.~K.~C. 2006, \apj,
  652, 232, \dodoi{10.1086/508138}

\bibitem[{{Hoekstra} {et~al.}(2015){Hoekstra}, {Herbonnet}, {Muzzin}, {Babul},
  {Mahdavi}, {Viola}, \& {Cacciato}}]{hoekstra2015}
{Hoekstra}, H., {Herbonnet}, R., {Muzzin}, A., {et~al.} 2015, \mnras, 449, 685,
  \dodoi{10.1093/mnras/stv275}

\bibitem[{{Hoekstra} {et~al.}(2012){Hoekstra}, {Mahdavi}, {Babul}, \&
  {Bildfell}}]{hoekstra2012}
{Hoekstra}, H., {Mahdavi}, A., {Babul}, A., \& {Bildfell}, C. 2012, \mnras,
  427, 1298, \dodoi{10.1111/j.1365-2966.2012.22072.x}

\bibitem[{{Hook} {et~al.}(2004){Hook}, {J{\o}rgensen}, {Allington-Smith},
  {Davies}, {Metcalfe}, {Murowinski}, \& {Crampton}}]{hook2004}
{Hook}, I.~M., {J{\o}rgensen}, I., {Allington-Smith}, J.~R., {et~al.} 2004,
  \pasp, 116, 425, \dodoi{10.1086/383624}

\bibitem[{{Jauzac} {et~al.}(2015){Jauzac}, {Richard}, {Jullo}, {Cl{\'e}ment},
  {Limousin}, {Kneib}, {Ebeling}, {Natarajan}, {Rodney}, {Atek}, {Massey},
  {Eckert}, {Egami}, \& {Rexroth}}]{Jauzac2015}
{Jauzac}, M., {Richard}, J., {Jullo}, E., {et~al.} 2015, \mnras, 452, 1437,
  \dodoi{10.1093/mnras/stv1402}

\bibitem[{{Jullo} {et~al.}(2007){Jullo}, {Kneib}, {Limousin},
  {El{\'{\i}}asd{\'o}ttir}, {Marshall}, \& {Verdugo}}]{jullo2007}
{Jullo}, E., {Kneib}, J.-P., {Limousin}, M., {et~al.} 2007, New Journal of
  Physics, 9, 447, \dodoi{10.1088/1367-2630/9/12/447}

\bibitem[{{Klypin} {et~al.}(2016){Klypin}, {Yepes}, {Gottl{\"o}ber}, {Prada},
  \& {He{\ss}}}]{klypin2016}
{Klypin}, A., {Yepes}, G., {Gottl{\"o}ber}, S., {Prada}, F., \& {He{\ss}}, S.
  2016, \mnras, 457, 4340, \dodoi{10.1093/mnras/stw248}

\bibitem[{{Le Fevre} {et~al.}(1994){Le Fevre}, {Crampton}, {Felenbok}, \&
  {Monnet}}]{lefevre1994}
{Le Fevre}, O., {Crampton}, D., {Felenbok}, P., \& {Monnet}, G. 1994, \aap,
  282, 325

\bibitem[{{Li} {et~al.}(2019){Li}, {Gladders}, {Heitmann}, {Rangel}, {Child},
  {Florian}, {Bleem}, {Habib}, \& {Finkel}}]{li2019}
{Li}, N., {Gladders}, M.~D., {Heitmann}, K., {et~al.} 2019, \apj, 878, 122,
  \dodoi{10.3847/1538-4357/ab1f74}

\bibitem[{{Limousin} {et~al.}(2005){Limousin}, {Kneib}, \&
  {Natarajan}}]{Limousin2005}
{Limousin}, M., {Kneib}, J.-P., \& {Natarajan}, P. 2005, \mnras, 356, 309,
  \dodoi{10.1111/j.1365-2966.2004.08449.x}

\bibitem[{{Limousin} {et~al.}(2013){Limousin}, {Morandi}, {Sereno},
  {Meneghetti}, {Ettori}, {Bartelmann}, \& {Verdugo}}]{Limousin2013}
{Limousin}, M., {Morandi}, A., {Sereno}, M., {et~al.} 2013, \ssr, 177, 155,
  \dodoi{10.1007/s11214-013-9980-y}

\bibitem[{{Limousin} {et~al.}(2007){Limousin}, {Richard}, {Jullo}, {Kneib},
  {Fort}, {Soucail}, {El{\'{\i}}asd{\'o}ttir}, {Natarajan}, {Ellis}, {Smail},
  {Czoske}, {Smith}, {Hudelot}, {Bardeau}, {Ebeling}, {Egami}, \&
  {Knudsen}}]{lim2007}
{Limousin}, M., {Richard}, J., {Jullo}, E., {et~al.} 2007, \apj, 668, 643,
  \dodoi{10.1086/521293}

\bibitem[{{Luppino} {et~al.}(1993){Luppino}, {Gioia}, {Annis}, {Le Fevre}, \&
  {Hammer}}]{luppino1993}
{Luppino}, G.~A., {Gioia}, I.~M., {Annis}, J., {Le Fevre}, O., \& {Hammer}, F.
  1993, \apj, 416, 444, \dodoi{10.1086/173249}

\bibitem[{{Mahajan} {et~al.}(2011){Mahajan}, {Mamon}, \&
  {Raychaudhury}}]{mahajan2011}
{Mahajan}, S., {Mamon}, G.~A., \& {Raychaudhury}, S. 2011, \mnras, 416, 2882,
  \dodoi{10.1111/j.1365-2966.2011.19236.x}

\bibitem[{{Mahdavi} {et~al.}(2013){Mahdavi}, {Hoekstra}, {Babul}, {Bildfell},
  {Jeltema}, \& {Henry}}]{mahdavi2013}
{Mahdavi}, A., {Hoekstra}, H., {Babul}, A., {et~al.} 2013, \apj, 767, 116,
  \dodoi{10.1088/0004-637X/767/2/116}

\bibitem[{{Marshall} {et~al.}(2002){Marshall}, {Hobson}, {Gull}, \&
  {Bridle}}]{marshall2002}
{Marshall}, P.~J., {Hobson}, M.~P., {Gull}, S.~F., \& {Bridle}, S.~L. 2002,
  \mnras, 335, 1037, \dodoi{10.1046/j.1365-8711.2002.05685.x}

\bibitem[{{Monna} {et~al.}(2015){Monna}, {Seitz}, {Zitrin}, {Geller}, {Grillo},
  {Mercurio}, {Greisel}, {Halkola}, {Suyu}, {Postman}, {Rosati}, {Balestra},
  {Biviano}, {Coe}, {Fabricant}, {Hwang}, \& {Koekemoer}}]{Monna2015}
{Monna}, A., {Seitz}, S., {Zitrin}, A., {et~al.} 2015, \mnras, 447, 1224,
  \dodoi{10.1093/mnras/stu2534}

\bibitem[{{Moorwood} {et~al.}(1998){Moorwood}, {Cuby}, \&
  {Lidman}}]{moorwood1998}
{Moorwood}, A., {Cuby}, J.~G., \& {Lidman}, C. 1998, The Messenger, 91, 9

\bibitem[{{Munari} {et~al.}(2013){Munari}, {Biviano}, {Borgani}, {Murante}, \&
  {Fabjan}}]{munari2013}
{Munari}, E., {Biviano}, A., {Borgani}, S., {Murante}, G., \& {Fabjan}, D.
  2013, \mnras, 430, 2638, \dodoi{10.1093/mnras/stt049}

\bibitem[{{Muratov} \& {Gnedin}(2010)}]{muratov2010}
{Muratov}, A.~L., \& {Gnedin}, O.~Y. 2010, \apj, 718, 1266,
  \dodoi{10.1088/0004-637X/718/2/1266}

\bibitem[{Navarro {et~al.}(1996)Navarro, Frenk, \& White}]{nfw96}
Navarro, J., Frenk, C., \& White, S. 1996, ApJ, 462, 563

\bibitem[{{Navarro} {et~al.}(1997){Navarro}, {Frenk}, \& {White}}]{nfw1997}
{Navarro}, J.~F., {Frenk}, C.~S., \& {White}, S.~D.~M. 1997, \apj, 490, 493

\bibitem[{{Newman} {et~al.}(2013){Newman}, {Treu}, {Ellis}, {Sand}, {Nipoti},
  {Richard}, \& {Jullo}}]{new13}
{Newman}, A.~B., {Treu}, T., {Ellis}, R.~S., {et~al.} 2013, \apj, 765, 24,
  \dodoi{10.1088/0004-637X/765/1/24}

\bibitem[{{Noble} {et~al.}(2013){Noble}, {Webb}, {Muzzin}, {Wilson}, {Yee}, \&
  {van der Burg}}]{noble2013}
{Noble}, A.~G., {Webb}, T.~M.~A., {Muzzin}, A., {et~al.} 2013, \apj, 768, 118,
  \dodoi{10.1088/0004-637X/768/2/118}

\bibitem[{{Noble} {et~al.}(2016){Noble}, {Webb}, {Yee}, {Muzzin}, {Wilson},
  {van der Burg}, {Balogh}, \& {Shupe}}]{noble2016}
{Noble}, A.~G., {Webb}, T.~M.~A., {Yee}, H.~K.~C., {et~al.} 2016, \apj, 816,
  48, \dodoi{10.3847/0004-637X/816/2/48}

\bibitem[{Pedregosa {et~al.}(2011)Pedregosa, Varoquaux, Gramfort, Michel,
  Thirion, Grisel, Blondel, Prettenhofer, Weiss, Dubourg, Vanderplas, Passos,
  Cournapeau, Brucher, Perrot, \& Duchesnay}]{scikit-learn}
Pedregosa, F., Varoquaux, G., Gramfort, A., {et~al.} 2011, Journal of Machine
  Learning Research, 12, 2825

\bibitem[{{Price-Whelan} {et~al.}(2018){Price-Whelan}, {Sip{\H{o}}cz},
  {G{\"u}nther}, {Lim}, {Crawford}, {Conseil}, {Shupe}, {Craig}, {Dencheva},
  {Ginsburg}, {VanderPlas}, {Bradley}, {P{\'e}rez-Su{\'a}rez}, {de Val-Borro},
  {Paper Contributors}, {Aldcroft}, {Cruz}, {Robitaille}, {Tollerud},
  {Coordination Committee}, {Ardelean}, {Babej}, {Bach}, {Bachetti}, {Bakanov},
  {Bamford}, {Barentsen}, {Barmby}, {Baumbach}, {Berry}, {Biscani}, {Boquien},
  {Bostroem}, {Bouma}, {Brammer}, {Bray}, {Breytenbach}, {Buddelmeijer},
  {Burke}, {Calderone}, {Cano Rodr{\'\i}guez}, {Cara}, {Cardoso}, {Cheedella},
  {Copin}, {Corrales}, {Crichton}, {D{\textquoteright}Avella}, {Deil},
  {Depagne}, {Dietrich}, {Donath}, {Droettboom}, {Earl}, {Erben}, {Fabbro},
  {Ferreira}, {Finethy}, {Fox}, {Garrison}, {Gibbons}, {Goldstein}, {Gommers},
  {Greco}, {Greenfield}, {Groener}, {Grollier}, {Hagen}, {Hirst}, {Homeier},
  {Horton}, {Hosseinzadeh}, {Hu}, {Hunkeler}, {Ivezi{\'c}}, {Jain}, {Jenness},
  {Kanarek}, {Kendrew}, {Kern}, {Kerzendorf}, {Khvalko}, {King}, {Kirkby},
  {Kulkarni}, {Kumar}, {Lee}, {Lenz}, {Littlefair}, {Ma}, {Macleod},
  {Mastropietro}, {McCully}, {Montagnac}, {Morris}, {Mueller}, {Mumford},
  {Muna}, {Murphy}, {Nelson}, {Nguyen}, {Ninan}, {N{\"o}the}, {Ogaz}, {Oh},
  {Parejko}, {Parley}, {Pascual}, {Patil}, {Patil}, {Plunkett}, {Prochaska},
  {Rastogi}, {Reddy Janga}, {Sabater}, {Sakurikar}, {Seifert}, {Sherbert},
  {Sherwood-Taylor}, {Shih}, {Sick}, {Silbiger}, {Singanamalla}, {Singer},
  {Sladen}, {Sooley}, {Sornarajah}, {Streicher}, {Teuben}, {Thomas},
  {Tremblay}, {Turner}, {Terr{\'o}n}, {van Kerkwijk}, {de la Vega}, {Watkins},
  {Weaver}, {Whitmore}, {Woillez}, {Zabalza}, \& {Contributors}}]{astropy:2018}
{Price-Whelan}, A.~M., {Sip{\H{o}}cz}, B.~M., {G{\"u}nther}, H.~M., {et~al.}
  2018, \aj, 156, 123, \dodoi{10.3847/1538-3881/aabc4f}

\bibitem[{{Quintana} {et~al.}(2000){Quintana}, {Carrasco}, \&
  {Reisenegger}}]{quintana2000}
{Quintana}, H., {Carrasco}, E.~R., \& {Reisenegger}, A. 2000, \aj, 120, 511,
  \dodoi{10.1086/301476}

\bibitem[{{Rhee} {et~al.}(2017){Rhee}, {Smith}, {Choi}, {Yi}, {Jaff{\'e}},
  {Candlish}, \& {S{\'a}nchez-J{\'a}nssen}}]{rhee2017}
{Rhee}, J., {Smith}, R., {Choi}, H., {et~al.} 2017, \apj, 843, 128,
  \dodoi{10.3847/1538-4357/aa6d6c}

\bibitem[{{Rines} {et~al.}(2013){Rines}, {Geller}, {Diaferio}, \&
  {Kurtz}}]{rines2013}
{Rines}, K., {Geller}, M.~J., {Diaferio}, A., \& {Kurtz}, M.~J. 2013, \apj,
  767, 15, \dodoi{10.1088/0004-637X/767/1/15}

\bibitem[{{Schirmer}(2013)}]{schirmer13}
{Schirmer}, M. 2013, \apjs, 209, 21, \dodoi{10.1088/0067-0049/209/2/21}

\bibitem[{{Schlafly} \& {Finkbeiner}(2011)}]{schlafly2011}
{Schlafly}, E.~F., \& {Finkbeiner}, D.~P. 2011, \apj, 737, 103,
  \dodoi{10.1088/0004-637X/737/2/103}

\bibitem[{{Shan} {et~al.}(2010){Shan}, {Qin}, \& {Zhao}}]{shan2010}
{Shan}, H.~Y., {Qin}, B., \& {Zhao}, H.~S. 2010, \mnras, 408, 1277,
  \dodoi{10.1111/j.1365-2966.2010.17209.x}

\bibitem[{{Silverman}(1986)}]{sil86}
{Silverman}, B.~W. 1986, {Density estimation for statistics and data analysis}

\bibitem[{{Soucail} {et~al.}(2015){Soucail}, {Fo{\"e}x}, {Pointecouteau},
  {Arnaud}, \& {Limousin}}]{soucail2015}
{Soucail}, G., {Fo{\"e}x}, G., {Pointecouteau}, E., {Arnaud}, M., \&
  {Limousin}, M. 2015, \aap, 581, A31, \dodoi{10.1051/0004-6361/201322689}

\bibitem[{{Stern} {et~al.}(2003){Stern}, {Holden}, {Stanford}, \&
  {Spinrad}}]{stern2003}
{Stern}, D., {Holden}, B., {Stanford}, S.~A., \& {Spinrad}, H. 2003, \aj, 125,
  2759, \dodoi{10.1086/374229}

\bibitem[{{Tonry} \& {Davis}(1979)}]{tonry1979}
{Tonry}, J., \& {Davis}, M. 1979, \aj, 84, 1511, \dodoi{10.1086/112569}

\bibitem[{{Ueda} {et~al.}(2001){Ueda}, {Ishisaki}, {Takahashi}, {Makishima}, \&
  {Ohashi}}]{ueda2001}
{Ueda}, Y., {Ishisaki}, Y., {Takahashi}, T., {Makishima}, K., \& {Ohashi}, T.
  2001, \apjs, 133, 1, \dodoi{10.1086/319189}

\bibitem[{{Verdugo} {et~al.}(2020){Verdugo}, {Carrasco}, {Fo{\"e}x}, {Motta},
  {Gomez}, {Limousin}, {Maga{\~n}a}, \& {de Diego}}]{verdugo2020}
{Verdugo}, T., {Carrasco}, E.~R., {Fo{\"e}x}, G., {et~al.} 2020, \apj, 897, 4,
  \dodoi{10.3847/1538-4357/ab9635}

\bibitem[{{Verdugo} {et~al.}(2011){Verdugo}, {Motta}, {Mu{\~n}oz}, {Limousin},
  {Cabanac}, \& {Richard}}]{ver2011}
{Verdugo}, T., {Motta}, V., {Mu{\~n}oz}, R.~P., {et~al.} 2011, \aap, 527, A124,
  \dodoi{10.1051/0004-6361/201014965}

\bibitem[{{Verdugo} {et~al.}(2016){Verdugo}, {Limousin}, {Motta}, {Mamon},
  {Fo{\"e}x}, {Gastaldello}, {Jullo}, {Biviano}, {Rojas}, {Mu{\~n}oz},
  {Cabanac}, {Maga{\~n}a}, {Fern{\'a}ndez-Trincado}, {Adame}, \& {De
  Leo}}]{verdugo2016}
{Verdugo}, T., {Limousin}, M., {Motta}, V., {et~al.} 2016, \aap, 595, A30,
  \dodoi{10.1051/0004-6361/201628629}

\bibitem[{{Yee} {et~al.}(1996){Yee}, {Ellingson}, \& {Carlberg}}]{yee1996}
{Yee}, H.~K.~C., {Ellingson}, E., \& {Carlberg}, R.~G. 1996, \apjs, 102, 269,
  \dodoi{10.1086/192259}

\bibitem[{{Zitrin} {et~al.}(2013){Zitrin}, {Meneghetti}, {Umetsu},
  {Broadhurst}, {Bartelmann}, {Bouwens}, {Bradley}, {Carrasco}, {Coe}, {Ford},
  {Kelson}, {Koekemoer}, {Medezinski}, {Moustakas}, {Moustakas}, {Nonino},
  {Postman}, {Rosati}, {Seidel}, {Seitz}, {Sendra}, {Shu}, {Vega}, \&
  {Zheng}}]{zitrin2013}
{Zitrin}, A., {Meneghetti}, M., {Umetsu}, K., {et~al.} 2013, \apjl, 762, L30,
  \dodoi{10.1088/2041-8205/762/2/L30}

\bibitem[{{Zwicky} {et~al.}(1965){Zwicky}, {Karpowicz}, \&
  {Kowal}}]{zwicky1965}
{Zwicky}, F., {Karpowicz}, M., \& {Kowal}, C.~T. 1965, {``Catalogue of Galaxies
  and of Clusters of Galaxies'', Vol. V}

\end{thebibliography}


\end{document}